\documentclass[11pt]{article}

\usepackage[margin=0.75in]{geometry}
\usepackage{amsmath,amssymb,amsthm}
\usepackage{mathtools}
\usepackage{graphicx}
\usepackage{hyperref}
\usepackage{enumerate}
\usepackage{bbm}
\usepackage{float}
\usepackage{color}
\usepackage{braket} 
\usepackage{comment}
\usepackage{tikz}
\usetikzlibrary{decorations.pathreplacing}
\usetikzlibrary{arrows.meta}
\usetikzlibrary{calc}
\usetikzlibrary{positioning}

\newtheorem{theorem}{Theorem}
\newtheorem{lemma}{Lemma}

\theoremstyle{definition}
\newtheorem{definition}{Definition}
\theoremstyle{remark}

\usepackage{amsthm}

\theoremstyle{definition}
\newtheorem{fact}{Fact}

\renewcommand{\ket}[1]{\lvert #1 \rangle}

\newcommand{\cC}{\mathcal{C}}

\begin{document}
\begin{@twocolumnfalse}
\title{Achievability of CSS codes performance limits over Pauli channels under separate ML decoding with bounded graphical complexity }
\author{
  \\ Gulnihal Mercanoglu, \quad Achilleas Anastasopoulos, \quad and \quad S. Sandeep Pradhan \\
  \small Department of Electrical and Computer Engineering \\
  \small University of Michigan, Ann Arbor \\
  \small email: \{gnihalm, anastas, pradhanv\}@umich.edu
}
\date{\today}

\maketitle

\begin{abstract}
Capacity-approaching quantum error correction faces a fundamental tension: the required parity-check matrices need to become increasingly dense, while practical decoding and syndrome extraction favor sparse, bounded-degree structures.
Reconciling these competing requirements remains an important open problem in quantum code design.
We construct Calderbank--Shor--Steane (CSS) code ensembles that approach the CSS rate over memoryless Pauli channels under maximum-likelihood decoding while retaining sparse, bounded-degree graphical representations. The construction separates parity-check density from graphical complexity: although capacity-approaching components may require dense parity-check matrices, they are represented by sparse concatenated factor graphs. We first study a construction combining low-density parity-check (LDPC) and low-density generator-matrix (LDGM) component codes and numerically analyze the average weight distributions of the resulting codes. Motivated by these results, we introduce a second construction based on concatenated LDPC and LDGM codes and generalized irregular repeat--accumulate (GIRA) codes. We derive bounds on the average weight distribution of the resulting sum ensemble and use the Miller--Burshtein framework to prove simultaneous reliability of the two CSS decoding components. Consequently, the proposed ensemble achieves rates arbitrarily close to the CSS hashing bound under separate bit- and phase-error correction, while maintaining bounded graphical complexity, thus providing a foundation for future iterative decoders and indirect syndrome-extraction architectures.
\end{abstract}
\end{@twocolumnfalse}

\clearpage

\section{Introduction}

Calderbank--Shor--Steane (CSS) codes provide one of the most direct connections between classical and quantum error correction~\cite{CalderbankShor1996,steane1996simple,Gottesman2009}. A CSS code is specified by two binary linear codes, $\cC_1, \cC_2 $ satisfying a nesting condition $\cC_2\subseteq \cC_1 $, and the correction of Pauli errors can be separated into two classical decoding problems: one associated with the bit-flips using $\cC_1$ and the other with phase-flips using $\cC_2^\perp$. This structure makes CSS codes attractive both analytically and architecturally, since classical coding constructions and decoding methods can be brought directly into the design of quantum codes~\cite{MacKay2004,PoulinChung2008}.

For a memoryless Pauli channel with error probabilities $p_X,p_Y,p_Z$ for $X,Y,Z$ errors, respectively, separate correction of the bit- and phase-error components gives rise to two binary additive channels whose crossover probabilities are \(p_X+p_Y\) and \(p_Z+p_Y\), respectively~\cite{ShorPreskill2000}. 
Hamada~\cite{Hamada2008} established that the rate $C(W_1)+C(W_2)-1$ is achievable by  random CSS code ensembles over
two memoryless additive channels.
For the two induced binary channels, this yields the achievable CSS rate
$C_{CSS}=1-h_2(p_X+p_Y)-h_2(p_Z+p_Y),$
which we adopt as the benchmark for separate bit- and phase-error correction.
This benchmark can be below the unrestricted hashing bound~\cite{hashingBennett1996}, since separate decoding does not exploit the statistical dependence between the bit- and phase-error components. 
Nevertheless, it is the natural capacity benchmark for the CSS decoding architecture considered in this paper

Random-coding arguments establish the existence of quantum codes, including CSS and stabilizer codes, with strong asymptotic performance under optimal decoding~\cite{CalderbankShor1996,steane1996simple,Hamada2004,hashingBennett1996,Devetak2005}. A fundamental constructive question arises:

\begin{quote}
\emph{
Can nested CSS code families constructed from classical sparse-graph ensembles achieve the separate-decoding CSS rate under maximum-likelihood decoding while retaining bounded graphical complexity?}
\end{quote}

A number of works have investigated quantum code families designed to approach information-theoretic limits, including concatenated quantum codes, sparse-graph quantum codes, and capacity-oriented constructions~\cite{Poulin2006,KovalevPryadko2013,PanteleevKalachev2022,LeverrierZemor2022}. Random stabilizer and CSS constructions provide important existence results, but they generally do not yield bounded-complexity graphical descriptions, practical decoding algorithms, or low-weight syndrome-extraction circuits.
Quantum low-density parity-check (QLDPC) codes, on the other hand, provide sparse stabilizer representations and have undergone substantial recent development~\cite{MacKay2004,TillichZemor2014,PanteleevKalachev2022,LeverrierZemor2022}. 
Recent advances include 
QLDPC ensembles approaching the unrestricted hashing bound~\cite{Mittal2026}.
The simultaneous attainment of capacity, a structured CSS nesting condition, and a bounded-complexity graphical realization nevertheless remains a challenge.

The source of this tension becomes especially clear through the classical representation of a CSS code. 
Under separate decoding, \(\mathcal C_1\) corrects one error component and \(\mathcal C_2^\perp\) corrects the other~\cite{CalderbankShor1996,steane1996simple,Hamada2008}. Thus, approaching the CSS rate requires both \(\mathcal C_1\) and \(\mathcal C_2^\perp\) to approach the capacities of their respective binary marginal channels.
The parity-check matrices relevant to these two classical decoding problems are then subject to a fundamental limitation from classical coding theory. Namely, for sequences of binary linear codes whose rates approach capacity on memoryless binary-input output-symmetric channels, the parity-check density must grow\footnote{The number of ones in the parity check matrix normalized to the code length must grow as $\log(1/\delta)$ for codes with rate $R > (1-\delta)C$.}~\cite{sason2003parity,WiechmanSason2007,RichardsonUrbanke2008}. Consequently, one should not generally expect both classical components of a capacity-approaching CSS family to possess uniformly sparse parity-check matrices. This is not merely a limitation of a particular regular LDPC ensemble, but reflects a fundamental density-versus-performance tradeoff.

This observation has two important consequences in a quantum system. First, the rows of the CSS parity-check matrices determine stabilizer checks. Increasing their row weights therefore leads to higher-weight stabilizer measurements and denser syndrome-extraction circuits. Such circuits may require more physical interactions, greater depth or connectivity, and can create additional opportunities for error propagation~\cite{Gottesman1997,Fowler2012}. 
Second, and crucially, high-degree parity-check graphs are unfavorable for conventional iterative decoding. The computational complexity of belief-propagation decoding is tied to the number of edges and the node degrees in the underlying factor graph~\cite{Kschischang2001,RichardsonUrbanke2008}; increasing those degrees undermines the bounded-complexity motivation for using a sparse-graph code.

These considerations suggest that a capacity-oriented CSS construction must distinguish among three objects that coincide in a conventional QLDPC description:
\begin{enumerate}[(i)]
    \item the parity-check matrix that defines the code;
    \item the graphical representation used for decoding; and
    \item the checks that are physically measured to extract error information.
\end{enumerate}
The first required separation is between \emph{parity-check density} and \emph{graphical complexity}. A code may have a dense parity-check matrix while admitting an extended factor-graph representation assembled from bounded-degree sparse components and auxiliary variables~\cite{Forney2001,Kschischang2001,Pfister2004IRA,HsAn10}. Such a representation preserves the possibility of low-complexity iterative decoding even when no uniformly sparse parity-check matrix exists for the code itself.
The present paper addresses this separation issue rigorously. 

The second separation concerns syndrome extraction. If approaching capacity forces the relevant parity-check matrices to become dense, then an implementation based on directly measuring every row of those matrices cannot retain bounded stabilizer weights~\cite{Gottesman1997,BreuckmannEberhardt2021}. 
A future implementation research program may therefore investigate indirect or extended measurements associated with the sparse graphical realization, rather than identifying the measured checks with the rows of an effective dense parity-check matrix~\cite{Poulin2005,BaconCasaccino2006,BaconFlammiaHarrowShi2017}.
Alternatively, one may consider fully quantum decoders without any required measurement, as those envisioned in~\cite{renes2017belief,Knill_2000,LiNakaharaPoonSzeTomita2012,PremakumarWangSaffmanJoynt2023,HeussenLocherMuller2024,VeroniMullerGiudice2024,ButtLocherBrechtelsbauerBuchlerMuller2025}.
The present paper does not address constructions of  fault-tolerant syndrome-extraction protocols, but its graphical formulation provides a representation on which such a protocol, as well as practical iterative decoding, may subsequently be developed.

Classical coding theory encountered a closely related density-versus-complexity problem, although without the quantum requirement of physically measuring stabilizer syndromes. LDPC codes introduced by Gallager~\cite{Gallager1962,gallager1963ldpc}, showed that sparse parity-check graphs support efficient message-passing algorithms. Suitably designed LDPC ensembles can operate close to capacity, but making the gap to capacity vanish under iterative decoding generally requires increasing node degrees and hence increasing graphical complexity~\cite[Ch.~3.15, pp.~109--112]{RichardsonUrbanke2008}. The parity-check-density lower bounds of Sason and Urbanke show that the corresponding density growth is fundamental~\cite{sason2003parity}.

The resolution of this tension is to distinguish a code's effective parity-check matrix from its graphical realization. Concatenated sparse constructions can define codes with potentially dense parity-check matrices through bounded-degree factor graphs containing hidden variables and sparse constituent transformations. 
The the MacKay–Neal (MN) codes~\cite{MacKay1999}, the irregular repeat--accumulate (IRA) ensembles~\cite{Pfister2004IRA}, and the concatenated LDPC and low-density generator matrix (LDGM) ensembles~\cite{HsAn10}
demonstrated how codes with a sparse representation graph can be capacity achieving (and thus have dense parity check matrices).

The present work imports this classical principle into the CSS setting. The CSS nesting condition
$\cC_2\subseteq\cC_1$ (or equivalently,
$\cC_1^\perp\subseteq\cC_2^\perp$) allows the second decoding code to be represented as the sum $\cC_2^\perp  = \cC_1^\perp + \tilde{\cC}$ of the dual $\cC_1^\perp$ and an auxiliary classical code $\tilde{\cC}$. This representation enforces CSS compatibility by construction and exposes a sparse extended factor graph for the code pair, even when the effective parity-check matrices of the resulting decoding codes are dense. The central challenge is then to select the constituent ensembles so that both \(\cC_1\) and \(\cC_2^\perp\) are simultaneously reliable at rates approaching their respective channel capacities.

Maximum-likelihood decoding is the natural benchmark for establishing the fundamental performance of these ensembles. Its block-error probability can be bounded in terms of the ensemble-average weight distribution. In particular, the Miller--Burshtein framework separates the small-weight region, where codewords must be sufficiently suppressed, from weights bounded away from the endpoints, where the ensemble-average spectrum is compared with that of a random linear code~\cite{Miller2001}. 
Exact ML decoding is computationally intractable in general, so an ML capacity-achievability result does not by itself provide a practical decoder. Nevertheless, retaining a bounded-degree factor graph is important: it provides the structural prerequisite for investigating belief-propagation and related message-passing algorithms~\cite{Kschischang2001,RichardsonUrbanke2008}. Thus, ML decoding serves as a theoretical benchmark, while the sparse extended representation preserves a route toward practical decoding.

The contributions of this paper are as follows.
\begin{enumerate}
    \item \textbf{A sparse-graph framework for nested CSS codes.}
    We express the CSS nesting condition through a sum-of-codes representation of \(\cC_2^\perp\). This representation separates the density of the effective parity-check matrices from the complexity of the extended factor graph and provides a common framework for two constructions.

    \item \textbf{A motivating LDPC-based construction and numerical ML analysis.}
    In Construction~A, \(\cC_1\) is a regular LDPC code, while \(\cC_2^\perp\) is represented as the sum of its LDGM dual and an auxiliary LDPC code. We analyze the two classical components through their ensemble-average weight distributions. The numerical results indicate that the sum-code component can closely reproduce the random-coding spectrum in the linearly scaling weight regime, whereas the regular-LDPC component remains the performance bottleneck.

    \item \textbf{A bounded-graphical-complexity capacity-achieving construction.}
    Motivated by the classical constructions of~\cite{Pfister2004IRA,HsAn10}, Construction~B replaces the regular-LDPC choice for \(\cC_1\) with a concatenated LDPC/LDGM code. Its dual admits a generalized irregular repeat--accumulate (GIRA) representation, so both codes retain sparse extended factor graphs despite having potentially dense effective parity-check matrices.

    \item \textbf{New spectral analysis for the dual and sum ensembles.}
    We derive an ensemble-average weight-distribution characterization for the GIRA component and establish bounds for its small-, linear-, and high-weight regimes. We then extend these results to the sum of an independently drawn GIRA code and an LDPC code.

    \item \textbf{Achievability of the CSS hashing bound under ML decoding.}
    Using the weight-distribution framework of Miller and Burshtein~\cite{Miller2001}, together with the classical LDPC/GM result of~\cite{HsAn10}, we prove that both classical decoding problems can be made simultaneously reliable. Consequently, Construction~B achieves rates arbitrarily close to the CSS hashing bound over memoryless Pauli channels under separate bit- and phase-error ML decoding. Moreover, this is done with fixed regular degree graphs.
\end{enumerate}

The result is an ML achievability theorem, rather than a claim that the proposed ensembles already achieve capacity under belief-propagation decoding or possess a complete fault-tolerant syndrome-extraction procedure. Its significance is that capacity-oriented performance is established without requiring the effective parity-check matrices themselves to remain sparse, while a bounded-degree extended factor graph is retained as a foundation for future iterative decoders and indirect quantum measurement architectures.

Just before the submission of this work, 
we became aware of an independent and parallel work \cite{Kasai2026} that shows similar results. The work uses the concatenated LDPC and LDGM codes of \cite{HsAn10} as $\mathcal{C}_1$ and MN codes as $\mathcal{C}_2$.

The remainder of the paper is organized as follows.
Section~\ref{sec:Preliminaries} introduces CSS codes, Pauli channels, and the classical code ensembles used in the constructions.
Section~\ref{sec:constructionA} presents Construction~A and investigates its ML decoding performance numerically.
Section~\ref{sec:constructionB} develops Construction~B and establishes its weight-distribution and capacity-achievability properties.
Finally, Section~\ref{sec:conclusion} summarizes the main findings and discusses iterative decoding and quantum implementation as directions for future work.
Additional technical results and proofs are provided in the appendices.

\section{Preliminaries}
\label{sec:Preliminaries}

\subsection{ CSS Codes} 
\label{sec:css}
Calderbank--Shor--Steane (CSS) codes constitute a class of quantum
error-correcting codes constructed from pairs of binary
linear codes \cite{CalderbankShor1996,steane1996simple}. Let
$\cC_1$,  $\cC_2$ be classical linear $(n,k_1)$, $(n,k_2)$ codes, respectively, satisfying
\begin{equation}
\cC_2\subseteq \cC_1.
\label{eq:css_inclusion}
\end{equation}

The resulting $[[n,k_1-k_2]]$ CSS code encodes $k_1-k_2$ logical qubits into $n$
physical qubits~\cite{CalderbankShor1996,steane1996simple}.
The encoding can be described in terms of the cosets of
$\cC_2$ in $\cC_1$. For each representative
$u\in\cC_1/\cC_2$, the encoded basis state is the uniform superposition 
\begin{equation}
\ket{u+\cC_2}
=
\frac{1}{\sqrt{|\cC_2|}}
\sum_{u\in\cC_2}
\ket{u+v} .
\end{equation}

An important feature of the CSS construction is that bit-flip ($X$) and
phase-flip ($Z$) errors can be corrected using error-correction procedures. 
In particular, $\cC_1$ is used to correct bit-flip errors, whereas $\cC_2^\perp$ is used to correct phase-flip errors
\cite{CalderbankShor1996,Hamada2008}. This separation follows from
the fact that a phase flip becomes a bit flip after a Hadamard
transformation. Consequently, the corresponding syndromes can be
measured and decoded using the parity-check structures of the two codes without measuring the encoded logical information~\cite{CalderbankShor1996}.

We now consider the performance of such a construction over a
memoryless Pauli channel. Let
$p_I,p_X,p_Y,p_Z$ denote the probabilities of the Pauli operators
$I,X,Y,Z$, respectively, with
$p_I+p_X+p_Y+p_Z=1$.
Since $Y=iXZ$, a $Y$ error contains both a bit-flip and a phase-flip
component. 
Accordingly, when the bit- and phase-error components are corrected separately in the CSS construction, their respective marginal channels can be represented as binary memoryless additive channels 
with crossover probabilities $q_X=p_X+p_Y$ and $q_Z=p_Y+p_Z$.
The hashing bound provides an achievable lower bound on the quantum capacity of memoryless Pauli channels. In this work, we consider CSS codes with separate bit- and phase-error correction, without exploiting possible correlations between the two components. The corresponding achievable CSS rate under separate bit- and phase-error correction is given by
\begin{align}
C_{\mathrm{CSS}}
&=1-h_2(q_X)-h_2(q_Z) 
=1-h_2(p_X+p_Y)-h_2(p_Y+p_Z),
\label{eq:css_achievable_rate}
\end{align}
where $h_2(\cdot)$ denotes the binary entropy function.

We next give a simple representation of the nested pair in \eqref{eq:css_inclusion} that guarantees the inclusion condition by construction and expresses $\cC_2^\perp$ in terms of $\cC_1^\perp$ and an auxiliary linear code. This representation is the common foundation of the two code constructions developed in the following sections.

By duality, \eqref{eq:css_inclusion} is equivalent to $\cC_1^\perp\subseteq\cC_2^\perp$. Since every subspace of a finite-dimensional vector space has a complement \cite[p.~42]{axler2024}, there exists a binary linear $(n,k_1-k_2)$ code $\widetilde{\cC}$ such that
\begin{equation}
\cC_2^\perp
=
\cC_1^\perp + \widetilde{\cC}
\label{eq:css_sum_representation}
\end{equation}

\begin{figure}[H]
    \centering
    \resizebox{0.6\textwidth}{!}{
        \begin{tikzpicture}[
    x=1cm, y=1cm,
    line cap=round,
    line join=round,
    edge/.style={
        draw=black,
        line width=0.9pt
    },
    block/.style={
        draw=black!70,
        fill=gray!10,
        minimum width=5.3cm,
        minimum height=0.65cm,
        line width=0.8pt,
        inner sep=4pt,
        font=\normalsize
    },
    check/.style={
        rectangle,
        draw=black!70,
        fill=gray!80,
        minimum width=0.48cm,
        minimum height=0.48cm,
        line width=0.8pt,
        inner sep=0pt
    },
    var/.style={
        circle,
        draw=black!80,
        fill=white,
        minimum size=0.70cm,
        line width=0.8pt,
        inner sep=1pt,
        font=\normalsize
    },
    state/.style={
        circle,
        draw=black!80,
        fill=blue!15,
        minimum size=0.70cm,
        line width=0.8pt,
        inner sep=1pt,
        font=\normalsize
    }
]

\def\xa{0.0}
\def\xb{2.5}
\def\xc{5.0}
\def\xd{8.3}

\def\xxa{1.05}
\def\xxb{3.55}
\def\xxc{6.05}
\def\xxd{9.35}

\def\yTopBlock{5.5}
\def\yTopVars{4.1}
\def\ySum{2.65}
\def\yBotVars{1.2}
\def\yBotBlock{-0.3}

\node[check, 
      minimum width=5.3cm,
      minimum height=0.65cm,] (C1p) at (4.15,\yTopBlock)
{$1_{\{c_1^\perp\in\mathcal C_1^\perp\}}$};

\node[state] (c1) at (\xa,\yTopVars) {$c_{1,1}^\perp$};
\node[state] (c2) at (\xb,\yTopVars) {$c_{1,2}^\perp$};
\node[state] (c3) at (\xc,\yTopVars) {$c_{1,3}^\perp$};
\node[state] (cn) at (\xd,\yTopVars) {$c_{1,n}^\perp$};
\node[font=\Large] at (7.15,\yTopVars) {$\cdots$};

\draw[edge] ($(C1p.south)+(-2.1,0)$) -- (c1);
\draw[edge] ($(C1p.south)+(-0.7,0)$) -- (c2);
\draw[edge] ($(C1p.south)+(0.7,0)$) -- (c3);
\draw[edge] ($(C1p.south)+(2.1,0)$) -- (cn);

\node[check] (s1) at (\xa,\ySum) {};
\node[check] (s2) at (\xb,\ySum) {};
\node[check] (s3) at (\xc,\ySum) {};
\node[check] (sn) at (\xd,\ySum) {};

\draw[edge] (c1) -- (s1);
\draw[edge] (c2) -- (s2);
\draw[edge] (c3) -- (s3);
\draw[edge] (cn) -- (sn);

\node[var] (x1) at (\xxa,\ySum) {$c_{2,1}^\perp$};
\node[var] (x2) at (\xxb,\ySum) {$c_{2,2}^\perp$};
\node[var] (x3) at (\xxc,\ySum) {$c_{2,3}^\perp$};
\node[var] (xn) at (\xxd,\ySum) {$c_{2,n}^\perp$};
\node[font=\Large] at (7.15,\ySum) {$\cdots$};

\draw[edge] (s1) -- (x1);
\draw[edge] (s2) -- (x2);
\draw[edge] (s3) -- (x3);
\draw[edge] (sn) -- (xn);

\node[state] (ct1) at (\xa,\yBotVars) {$\widetilde c_1$};
\node[state] (ct2) at (\xb,\yBotVars) {$\widetilde c_2$};
\node[state] (ct3) at (\xc,\yBotVars) {$\widetilde c_3$};
\node[state] (ctn) at (\xd,\yBotVars) {$\widetilde c_n$};
\node[font=\Large] at (7.15,\yBotVars) {$\cdots$};

\draw[edge] (ct1) -- (s1);
\draw[edge] (ct2) -- (s2);
\draw[edge] (ct3) -- (s3);
\draw[edge] (ctn) -- (sn);

\node[check, 
      minimum width=5.3cm,
      minimum height=0.65cm,] (Ct) at (4.15,\yBotBlock) {$1_{\{\widetilde c\in\widetilde{\mathcal C}\}}$};

\draw[edge] (ct1) -- ($(Ct.north)+(-2.1,0)$);
\draw[edge] (ct2) -- ($(Ct.north)+(-0.7,0)$);
\draw[edge] (ct3) -- ($(Ct.north)+(0.7,0)$);
\draw[edge] (ctn) -- ($(Ct.north)+(2.1,0)$);

\end{tikzpicture}
    }
    \caption{Factor-graph representation of the sum-code
construction
$\cC_2^\perp
=\cC_1^\perp+\widetilde{\cC}$.
Each sum check enforces
$c_{2,i}^\perp
=c_{1,i}^\perp+\widetilde c_i$
over $\mathbb F_2$.}
    \label{fig:css_const_diagram}
\end{figure}
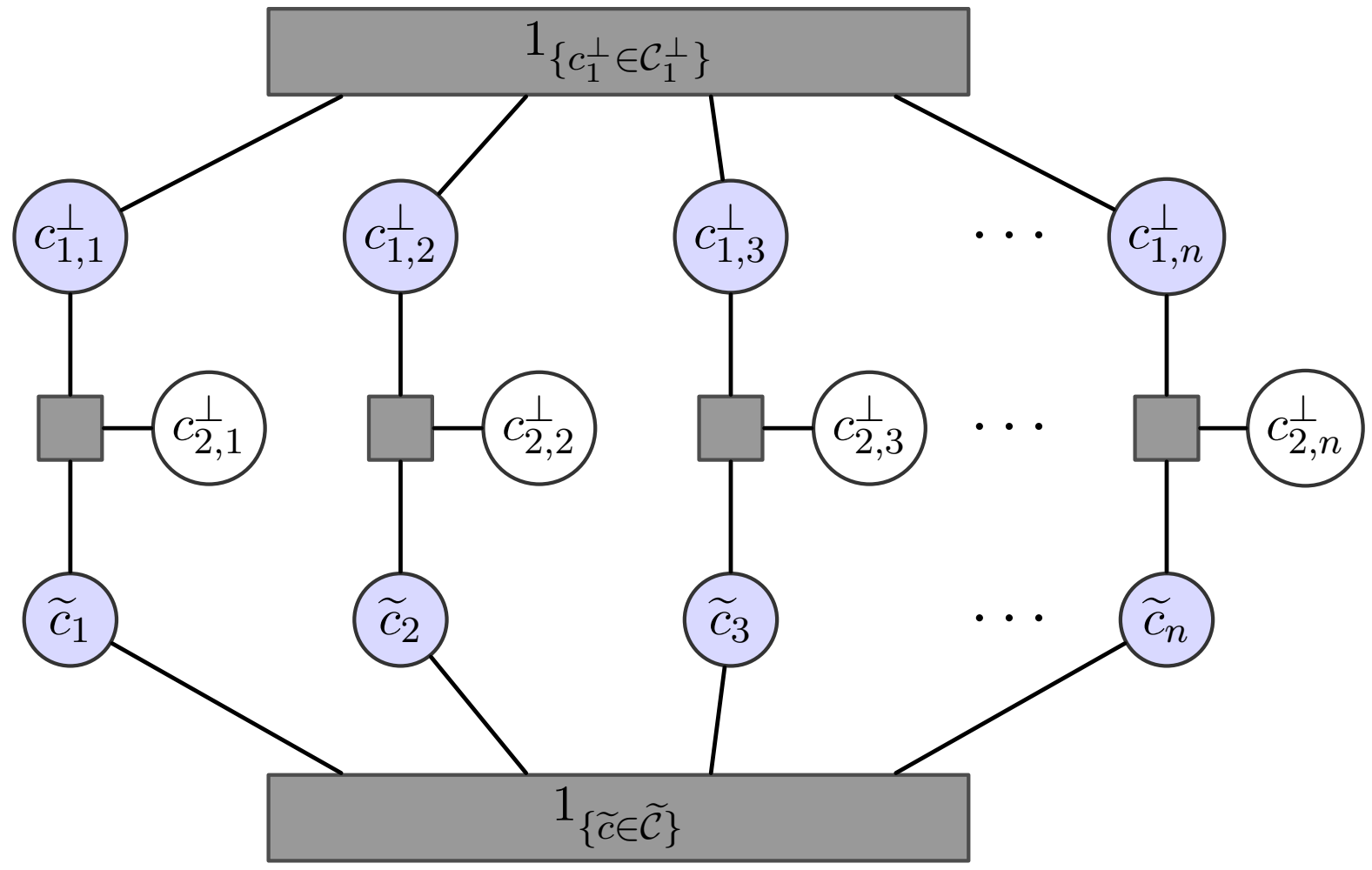

This identity provides a general sum-of-codes representation of the second CSS component. In particular, it shows that \(\cC_2^\perp\) can be constructed by combining the dual of the first component with an additional auxiliary code. The (partial) factor graph of this construction is depicted in Figure~\ref{fig:css_const_diagram}.

To summarize, the above construction reveals that,
any $[[n,k_1-k_2]]$ CSS code can be characterized as a pair $(\cC_1,  \widetilde{\cC})$ of two codes with parameters $(n,k_1)$ and $(n,k_1-k_2)$, respectively.

\subsection{Code-Ensemble Definitions}


\begin{definition}
\label{def:ldpc}
The $(d_v,d_c)$-regular LDPC ensemble of $(n,k)$ codes, denoted by
$\operatorname{LDPC}(n,d_v,d_c)$, consists of bipartite graphs with
$n$ variable nodes of degree $d_v$ and $(n-k)$ check nodes of degree $d_c$, where
$nd_v=(n-k)d_c.$
An ensemble member is obtained by matching the variable-node and
check-node sockets according to a uniformly random permutation. Let $N_{ji}$ denote the number of edges connecting variable node $i$ to check node $j$. The associated binary parity-check matrix $H\in\mathbb{F}_2^{(n-k)\times n}$ is defined by
$H_{ji} = N_{ji}  \pmod 2$, i.e.,  parallel edges are counted modulo 2.
The corresponding code is
\begin{equation}
  \cC_{LDPC}
=
\{c\in\mathbb{F}_2^n: Hc^T=0\},  
\end{equation}
and its design rate is
$ R_{LDPC}
=
\frac{k}{n}
=
1-\frac{d_v}{d_c}. $  
\end{definition}

For every realization $\cC_{LDPC} \in \operatorname{LDPC}(n,d_v,d_c)$ with parity-check matrix
$H\in\mathbb F_2^{(n-k)\times n}$, its dual code
is given by
\begin{equation}
\cC_{LDGM}
=
\cC_{LDPC}^{\perp}
=
\left\{
uH:u\in\mathbb F_2^{n-k}
\right\}.
\end{equation}
Thus, the LDPC ensemble induces a dual
$(d_c,d_v)$-regular LDGM ensemble, denoted by
$\operatorname{LDGM}(n,d_c,d_v)$, with design rate
$R_{LDGM}=d_v/d_c=1-R_{LDPC}$.

\begin{definition}
\label{def:ldpcgm}
Consider an independently drawn graph 
from the
$\operatorname{LDPC}(n,d_v^H,d_c^H)$ ensemble 
(with corresponding parity-check matrix
$H\in\mathbb{F}_2^{(n-k)\times n}$)
and an independently drawn graph
from the $\operatorname{LDGM}(n,d^G,d^G)$ ensemble (with corresponding rate-one generator matrix
$G\in\mathbb{F}_2^{n\times n}$).
The corresponding LDPC/GM code is
\begin{equation}
\cC_{LDPC/GM}
=
\left\{
uG :
u\in\mathbb{F}_2^n,\;
Hu^T=0
\right\}.   
\label{eq:ldpcgm}
\end{equation}
The resulting ensemble is denoted by
$\operatorname{LDPC/GM}(n,d_v^H,d_c^H,d^G),$
and has design rate
$R_{LDPC/GM}
=
1-\frac{d_v^H}{d_c^H}.$
\end{definition}

For every realization 
$\cC_{LDPC/GM}\in \operatorname{LDPC/GM}(n,d_v^H,d_c^H,d^G)$
with constituent parity-check matrix $H\in\mathbb F_2^{(n-k)\times n}$ and rate-one
generator matrix $G\in\mathbb F_2^{n\times n}$, its dual code is given by
\begin{equation}
\cC_{GIRA}
=
\cC_{LDPC/GM}^{\perp}
=
\left\{
c\in\mathbb F_2^n:
xH+cG^T=0
\text{ for some }
x\in\mathbb F_2^{n-k}
\right\}.
\label{eq:def_gira}
\end{equation}
The resulting ensemble has design rate $R_{GIRA}
=
\frac{d_v^H}{d_c^H}.$
Here, $H$ serves as the generator matrix of
the $\operatorname{LDGM}(n,d_c^H,d_v^H)$
ensemble, while $G$ defines a generalized
accumulator. 
The resulting ensemble is denoted by $\operatorname{GIRA}(n,d_c^H,d_v^H,d^G)$, where GIRA stands for generalized irregular repeat--accumulate.
The construction can be viewed as a generalization of the non-systematic IRA ensembles studied in~\cite{Pfister2004IRA}.

\begin{definition}[Classical Capacity-Achieving Code Ensemble]
Let $\{\mathcal{E}(n,\theta)\}_n$ be a sequence of code ensembles indexed by the blocklength $n$, where $\theta$ denotes a collection of fixed parameters and $R(\theta)$ denotes the design rate.
We say that the sequence of ensembles is capacity-achieving under maximum-likelihood (ML) decoding if there exists a fixed choice of $\theta$ such that, for every MBIOS channel $W$ with capacity $C(W)>R(\theta)$, one can find codes $\cC_n\in\mathcal E(n,\theta)$ satisfying
\begin{equation}
\lim_{n\to\infty}
P_B^{\mathrm{ML}}(\cC_n;W)=0,   
\end{equation}
where $P_B^{\mathrm{ML}}(\cC_n;W)$ denotes the block error probability under ML decoding over $W$.
\end{definition}

We remark that, in the context of this paper, the parameters $\theta$ correspond to the degree parameters of the graph-based code ensembles. 
These parameters are fixed independently of the blocklength $n$ and the gap to capacity $C(W)-R(\theta)$.

\begin{definition}[Quantum Capacity-Achieving CSS Code Ensemble with Separate Decoding]
Let $\{\mathcal{E}_1(n,\theta), \tilde{\mathcal{E}}(n,\theta)\}_n$,  be a sequence of code ensemble pairs indexed by the blocklength $n$, where $\theta$ denotes a collection of fixed parameters and $R_1(\theta)$, $\tilde{R}(\theta)$ denotes the design rates of the two ensembles, respectively. 
We say that the sequence of ensemble pairs achieves the quantum capacity~\eqref{eq:css_achievable_rate} of the Pauli channel with error probabilities $(p_I,p_X,p_Y,p_Z)$
with separate bit- and phase-error decoding  if the following is true.

There exists a fixed choice of $\theta$ such that, for every pair of BSC channels with crossover probabilities  $q_X=p_X+p_Y$, $q_Z=p_Z+p_Y$, respectively, having capacities $C_{BSC}(q_X)>R_1(\theta)$, and $C_{BSC}(q_Z)>1-R_1(\theta)+\tilde{R}(\theta)$, respectively, 
one can find codes $\cC_{1,n}\in\mathcal{E}_1(n,\theta)$ and $\tilde{\cC}_n\in\tilde{\mathcal{E}}(n,\theta)$ satisfying
\begin{equation}
\lim_{n\to\infty}
P_B^{\mathrm{ML}}(\cC_{1,n};q_X)=0,  
\qquad
\lim_{n\to\infty}
P_B^{\mathrm{ML}}(\cC_{1,n}^\perp + \tilde{C}_n;q_Z)=0,  
\end{equation}
where $P_B^{\mathrm{ML}}(\cC;q)$ denotes the block error probability of the classical code $\cC$ under ML decoding over $BSC(q)$ and $\cC_{1,n}^\perp + \tilde{C}_n$ denotes the sum code.
\end{definition}

We remark that, for quantum CSS codes, the two classical ML decoding procedures are performed separately after the corresponding stabilizer syndrome measurements. Specifically, the ML decoder associated with $\mathcal C_1$ uses the syndrome obtained from $Z$-type stabilizer measurements to correct bit-flip errors, while the ML decoder associated with $\mathcal C_2^\perp$ uses the syndrome obtained from $X$-type stabilizer measurements, equivalently performed in the Hadamard basis, to correct phase-flip errors.

\section{CSS Code Construction A}
\label{sec:constructionA}

For the constructions considered here, we seek three properties simultaneously. First, the nesting condition~\eqref{eq:css_inclusion} must hold. Second, both $\cC_1$ and $\cC_2^\perp$ should be strong error-correcting codes for binary symmetric channels. Third, both code components should admit graphical representations that permit low-complexity iterative decoding.

A natural starting point is therefore to choose $\cC_1$ to be an LDPC code. This immediately provides both a strong code and a sparse parity-check representation. The nesting condition can then be enforced by adding additional parity checks (sparse or dense) to the parity-check matrix of $\cC_1$. As shown below, this construction also gives a useful representation of $\cC_2^\perp$ as the sum of two codes.
Although $\cC_1$ has a sparse parity-check matrix, the resulting code $\cC_2^\perp$ does not in general have a sparse parity-check matrix. In fact, the matrix used to construct $\cC_2$ becomes a \emph{generator} matrix for $\cC_2^\perp$. Thus, even if this representation is sparse, this does not imply that $\cC_2^\perp$ is itself an LDPC code. This motivates the central idea used throughout this section: rather than requiring every constituent code to have a sparse parity-check matrix, we seek alternative sparse graphical representations that still allow simple iterative decoding.

Construction~A illustrates this idea in its simplest form. Construction~B then extends it by strengthening the second requirement in the construction, namely both $\cC_1$ and $\cC_2^\perp$ should be capacity-achieving codes with ML decoding (not merely strong error-correcting codes), thus necessitating their parity check matrices to be dense~\cite{sason2003parity} while their graphical representation remains sparse.

\subsection{Code Construction} 

We specialize the general CSS framework of Section~\ref{sec:css} by choosing the first code component $\cC_1$ from the $\operatorname{LDPC}(n,d_v^{H_1},d_c^{H_1})$ ensemble, with sparse parity-check matrix $H_1\in\mathbb F_2^{(n-k_1)\times n}$. The corresponding design rate is $R_1=1-d_v^{H_1}/d_c^{H_1}$. Since $H_1$ also serves as a sparse generator matrix for $\cC_1^\perp$, its dual admits an LDGM representation with degrees $(d_c^{H_1},d_v^{H_1})$ .

To construct the second code component, we independently choose an auxiliary code $\widetilde{\cC}$ from the $\operatorname{LDPC}(n,d_v^{\widetilde H},d_c^{\widetilde H})$ ensemble, with dimension $(k_1-k_2)$ and design rate $\widetilde R=1-d_v^{\widetilde H}/d_c^{\widetilde H}$.
As a result, $\cC_2^\perp$ admits a sparse factor-graph representation through the constituent LDGM and LDPC codes and their coordinate wise sum constraints.

Consequently, the code components of Construction~A are summarized as
\begin{equation}
\boxed{
\begin{aligned}
\cC_1
&\sim
\operatorname{LDPC}
(n,d_v^{H_1},d_c^{H_1}),
\
R_1=1-\frac{d_v^H}{d_c^H}\\
\widetilde{\cC}
&\sim
\operatorname{LDPC}
(n,d_v^{\widetilde H},d_c^{\widetilde H}),
\
\tilde{R}=1-\frac{d_v^{\tilde{H}}}{d_c^{\tilde{H}}}\\
\cC_2^\perp &=\cC_1^\perp+\widetilde{\cC}, 
\ R_2^\perp = 1-R_1+\tilde{R}.
\end{aligned}
}
\label{eq:C2dual_construction1_sum}
\end{equation}

The CSS nesting condition $\cC_2\subseteq\cC_1$ is automatically satisfied by construction. Furthermore, the resulting construction retains a sparse graphical representation without requiring a sparse parity-check matrix for $\cC_2^\perp$.

We conclude Construction~A with all three goals realized. Specifically, it automatically satisfies the nesting condition~\eqref{eq:css_inclusion}; it consists of potentially strong error correcting codes (the first is an LDPC code and the second is the sum of an LDGM and an LDPC code);
it has a simple graphical representation (only the sparse matrices $H_1$ and $\tilde{H}$ are used in the description). Construction~A is illustrated in Figs.~\ref{fig:ldpc} and \ref{fig:config1_c2perp}.

\begin{figure}[H]
    \centering
    \resizebox{0.6\textwidth}{!}{
        \begin{tikzpicture}[
    scale=1.0, transform shape,
    line cap=round,
    line join=round,
    edge/.style={draw=black, line width=0.9pt},
    block/.style={
        draw=black!70,
        fill=gray!10,
        line width=0.8pt,
        text=black,
        font=\Large
    },
    check/.style={
        draw=black!70,
        fill=gray!80,
        line width=0.8pt,
        text=black,
        font=\Large,
        minimum width=0.48cm,
        minimum height=0.48cm,
    },
    var/.style={
        circle,
        draw=black!80,
        fill=white,
        line width=0.8pt,
        minimum size=0.50cm,
        inner sep=0pt
    },
    state/.style={
        circle,
        draw=black!80,
        fill=blue!15,
        line width=0.8pt,
        minimum size=0.50cm,
        inner sep=0pt
    }
]


\def\xA{0.7}
\def\xB{2.2}
\def\xC{3.7}
\def\xD{7.3}


\def\yck{4.0}
\node[check] (ck1) at (\xA+0.35,\yck) {};
\node[check] (ck2) at (\xB+0.35,\yck) {};
\node[check] (ckm) at (\xD-0.35,\yck) {};
\node at (4.45,\yck) {\Large $\cdots$};

\def\yHo{3.0}
\node[
    block,
    minimum width=8.0cm,
    minimum height=0.5cm,
    text=black,
    font=\large
] (Ho) at (4,\yHo) {$H_1$};

\def\yx{2.0}
\node[var] (x1) at (\xA,\yx) {$c_1$};
\node[var] (x2) at (\xB,\yx) {$c_2$};
\node[var] (x3) at (\xC,\yx) {$c_3$};
\node[var] (xn) at (\xD,\yx) {$c_n$};
\node at (5.15,\yx) {\Large $\cdots$};



\draw[edge] (ck1) -- ($(ck1.center |- Ho.north)+(0*0.3-0.75,0)$);
\draw[edge] (ck1) -- ($(ck1.center |- Ho.north)+(1*0.3-0.75,0)$);
\draw[edge] (ck1) -- ($(ck1.center |- Ho.north)+(2*0.3-0.75,0)$);
\draw[edge] (ck1) -- ($(ck1.center |- Ho.north)+(3*0.3-0.75,0)$);
\draw[edge] (ck1) -- ($(ck1.center |- Ho.north)+(4*0.3-0.75,0)$);
\draw[edge] (ck1) -- ($(ck1.center |- Ho.north)+(5*0.3-0.75,0)$);

\draw[edge] (ck2) -- ($(ck2.center |- Ho.north)+(0*0.3-0.75,0)$);
\draw[edge] (ck2) -- ($(ck2.center |- Ho.north)+(1*0.3-0.75,0)$);
\draw[edge] (ck2) -- ($(ck2.center |- Ho.north)+(2*0.3-0.75,0)$);
\draw[edge] (ck2) -- ($(ck2.center |- Ho.north)+(3*0.3-0.75,0)$);
\draw[edge] (ck2) -- ($(ck2.center |- Ho.north)+(4*0.3-0.75,0)$);
\draw[edge] (ck2) -- ($(ck2.center |- Ho.north)+(5*0.3-0.75,0)$);

\draw[edge] (ckm) -- ($(ckm.center |- Ho.north)+(0*0.3-0.75,0)$);
\draw[edge] (ckm) -- ($(ckm.center |- Ho.north)+(1*0.3-0.75,0)$);
\draw[edge] (ckm) -- ($(ckm.center |- Ho.north)+(2*0.3-0.75,0)$);
\draw[edge] (ckm) -- ($(ckm.center |- Ho.north)+(3*0.3-0.75,0)$);
\draw[edge] (ckm) -- ($(ckm.center |- Ho.north)+(4*0.3-0.75,0)$);
\draw[edge] (ckm) -- ($(ckm.center |- Ho.north)+(5*0.3-0.75,0)$);


\draw[edge] ($(x1.center |- Ho.south)+(0*0.45-0.45,0)$) -- (x1);
\draw[edge] ($(x1.center |- Ho.south)+(1*0.45-0.45,0)$) -- (x1);
\draw[edge] ($(x1.center |- Ho.south)+(2*0.45-0.45,0)$) -- (x1);
\draw[edge] ($(x2.center |- Ho.south)+(0*0.45-0.45,0)$) -- (x2);
\draw[edge] ($(x2.center |- Ho.south)+(1*0.45-0.45,0)$) -- (x2);
\draw[edge] ($(x2.center |- Ho.south)+(2*0.45-0.45,0)$) -- (x2);
\draw[edge] ($(x3.center |- Ho.south)+(0*0.45-0.45,0)$) -- (x3);
\draw[edge] ($(x3.center |- Ho.south)+(1*0.45-0.45,0)$) -- (x3);
\draw[edge] ($(x3.center |- Ho.south)+(2*0.45-0.45,0)$) -- (x3);
\draw[edge] ($(xn.center |- Ho.south)+(0*0.45-0.45,0)$) -- (xn);
\draw[edge] ($(xn.center |- Ho.south)+(1*0.45-0.45,0)$) -- (xn);
\draw[edge] ($(xn.center |- Ho.south)+(2*0.45-0.45,0)$) -- (xn);



\node[
    right,
    align=center,
    font=\large
] at (8.0,\yck) {$n-k_1$};

\node[
    right,
    align=center,
    font=\large
] at (8.0,\yx)
{$n$};

\end{tikzpicture}
    }
    \caption{Factor Graph of $\cC_1\sim\operatorname{LDPC}(n,3,6)$.}
    \label{fig:ldpc}
\end{figure}
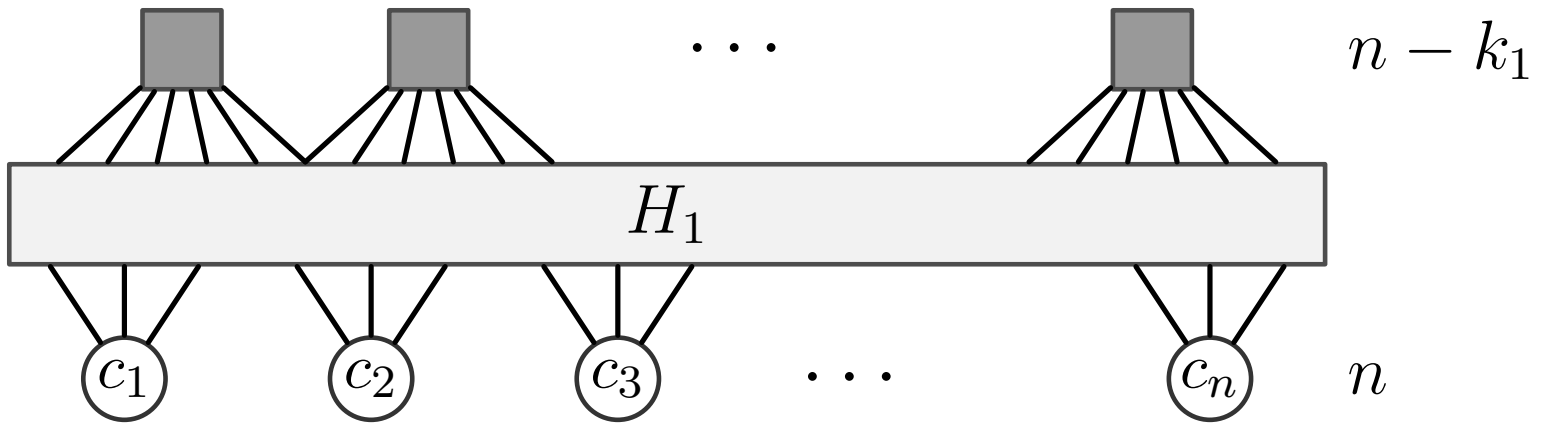

\begin{figure}[H]
    \centering
    \resizebox{0.8\textwidth}{!}{
        \input{figures/constructionA_c2perp_graph.tex}
    } 
    \caption{Factor Graph of $\cC_2^{\perp}\sim \operatorname{LDGM}(n,6,3)+\operatorname{LDPC}(n,3,4)$.}
    \label{fig:config1_c2perp}
\end{figure}

\subsection{ML Performance Analysis}

We next describe how the maximum-likelihood (ML) decoding performance
of Construction~A can be analyzed from the average weight distributions
of its two code components. Reliable decoding of the CSS code
requires both $\cC_1$ and $\cC_2^\perp$ to have vanishing
codeword error probability as the blocklength tends to infinity.

The random-coding argument in \cite{ShulmanFeder1999} shows that
random-coding techniques can be extended to structured code ensembles
by measuring the deviation of their average weight distribution from
that of a random linear-code ensemble. Miller and Burshtein
\cite{Miller2001} refined this approach by separating the weight
spectrum into two regions. A selected set of weights is treated directly
by a Bhattacharyya union bound, whereas the complementary set is handled
by comparing its average weight distribution with that of a random
linear code.


Consider an ensemble $\cC$ of binary linear codes of length $n$
and rate $R$, containing $M=2^{nR}$ codewords and, let
$    N_{\cC}(w)
    \triangleq
    \# \{c\in\cC:\operatorname{wt}(c)=w\}  $
denote the weight distribution. 
For a code ensemble, we denote the ensemble-average  weight
distribution by
$\overline{N_{\cC}}(w)
    \triangleq
    \mathbb E[N_{\cC}(w)], $   
where the expectation is taken over the code ensemble.
For a memoryless binary-input symmetric-output
channel, let $D = 2\sqrt{p(1-p)}$ denote the Bhattacharyya parameter
for binary symmetric channel with crossover probability $p$. Then, for any
$U\subseteq\{1,\ldots,n\}$, the Miller--Burshtein bound on the expected codeword error probability can be expressed as
\begin{equation}
    \mathbb E\left[
P_B^{\mathrm{ML}}(\cC) \right]
    \leq
    \sum_{w\in U}
    \overline{N_{\cC}}(w)D^w
    +
    2^{-nE_r(R+\Delta R_n)},
    \label{eq:two_term_ml_bound}
\end{equation}
where $E_r(\cdot)$ denotes the random-coding error exponent of the underlying channel, which characterizes the exponential decay of the error-probability upper bound for random coding. 
In particular, $E_r(r)$ is strictly positive for rates $r$ below the channel capacity.
The rate adjustment $\Delta R_n$ is defined as
\begin{equation}
    \Delta R_n
    =
    \max_{w\in U^c}
    \left[
        \frac{1}{n}\log_2\overline{N_{\cC}}(w)
        -
        \frac{1}{n}\log_2
        \frac{\binom{n}{w}(M-1)}{2^n}
    \right].
    \label{eq:deltaR_ml_bound}
\end{equation}
Thus, the contribution of weights in $U$ is controlled directly through the union bound, 
whereas the contribution of weights in $U^c$ are controlled through the
deviation of the ensemble-average spectrum from the random linear-code
spectrum. 
Typically the set $U$ is defined as $U=\{w: (w< n \gamma) \lor  (w> n (1-\gamma))\}$ for some $\gamma\in(0,1/2)$. 
Anticipating that sub-linear weights have to be dealt with separately, for a fixed $0<\zeta<\gamma$, we further split the first term resulting in the three-term bound
\begin{equation}
\begin{split}
        \mathbb E\left[
P_B^{\mathrm{ML}}(\cC) \right]
    &
    \leq
    \sum_{\substack{1\leq w<\zeta n}}
    \overline{N_{\cC}}(w)D^w
    +
    \sum_{\substack{(\zeta n\leq w<\gamma n) \lor ((1-\gamma)n< w\leq n )}}
    \overline{N_{\cC}}(w)D^w +
    2^{-nE_r(R+\Delta R_n)}.
\end{split}
\label{eq:union_split}
\end{equation}
The first term in~\eqref{eq:union_split} contains codewords with sublinear weight and can be controlled using a finite-length bound on the average weight enumerator.

For the second term in~\eqref{eq:union_split}, the asymptotic growth rate 
can be used.
For a normalized Hamming weight
    $\hat w=\frac{w}{n},$
the average weight distribution growth rate of the average weight distribution is defined as
$g(\hat{w}) \triangleq \lim_{n\rightarrow\infty} \frac{1}{n}\log \overline{N_{\cC}}(n\hat w)$  
where the limit is taken over blocklengths of the ensemble.
Thus, roughly speaking, the average number of codewords of normalized
weight $\hat w$ grows exponentially as $2^{n g(\hat w)}$. Hence,
\begin{equation}
    \overline{N_{\cC}}(n\hat w)
    D^{n\hat w}
    \doteq
    2^{n\left[g(\hat w)+\hat w\log_2 D\right]}.
\end{equation}
Hence, the second term vanishes whenever
$    g(\hat w)+\hat w\log_2D<0$
throughout the corresponding normalized-weight region $\zeta\leq\hat w<\gamma  \lor 1-\gamma<\hat w\leq1 $.
Finally, for the third term in~\eqref{eq:union_split}, using Stirling's approximation
\begin{equation}
    \lim_{n \to \infty} \frac{1}{n}\log_2
    \frac{\binom{n}{n\hat w}(M-1)}{2^n}
    =
    h_2(\hat w)-1+R,
\end{equation}
the asymptotic rate penalty entering the error exponent argument of~\eqref{eq:union_split} is
\begin{equation}
    \Delta R
    =
    \max_{\gamma \leq \hat w \leq 1-\gamma}
    \left\{
        g(\hat w)
        -
        \left(h_2(\hat w)-1+R\right)
    \right\}.
    \label{eq:asymptotic_deltaR}
\end{equation}

Thus, the ML analysis requires two different properties of the
spectrum: the very-small-weight contribution must vanish, while for
linearly scaling weights the growth rate must make the union-bound
term vanish on $U$ and must remain sufficiently close to the
random-code growth rate on $U^c$. Consequently, the maximum correctable channel crossover probability can be defined as
\begin{equation}
p_{\mathrm{ML}}^{\mathrm{LB}}
\triangleq
\sup
\left\{
\begin{aligned}
p\in(0,1/2]:\quad
&\exists\,0<\zeta<\gamma<\frac12
\text{ such that} \\[1mm]
&
\lim_{n\rightarrow \infty}\sum_{1\leq w<\zeta n}
\overline{N_{\cC}}(w)D(p)^w
= 0, \\
&
\sup_{ \substack{\zeta\leq\hat w<\gamma \lor 1-\gamma<\hat w\leq1 }}
\left[
g(\hat w)+\hat w\log_2D(p)
\right]
<0,\\
&
R+ 
\max_{\gamma\leq\hat w\leq1-\gamma}
\left\{
g(\hat w)-(h_2(\hat w)-1+R)
\right\}
<
1-h_2(p)
\end{aligned}
\right\}.
\label{eq:pML_LB}
\end{equation}
For Construction~A, these conditions must be verified separately for
$\cC_1$ and $\cC_2^\perp$, since the two codes
have different ensemble structures.

We first consider
$
\mathcal  C_1.
$
The weight-distribution behavior of regular LDPC ensembles was already characterized by Gallager \cite{gallager1963ldpc}, including both the very-small-weight regime and the asymptotic behavior for weights growing linearly with $n$. In particular, for an $\operatorname{LDPC}(n,d_v,d_c)$ with $d_v\geq3$, the contribution of sufficiently small-weight codewords satisfies
\begin{equation}
    \sum_{1\leq w <n\delta_{d_v,d_c}}
    \overline{N_{\mathcal  C_1}}(w)D^w
    \leq
    \frac{d_c-1}{2n^{d_v-2}}
    +
    o\!\left(\frac{1}{n^{d_v-2}}\right),
    \label{eq:ldpc_small_weight_bound}
\end{equation}
for some $\delta_{d_v,d_c}>0$, and therefore vanishes as $n\to\infty$. For weights that scale linearly with $n$, the ensemble is characterized by the known growth rate
$g_{\mathcal  C_1}(\hat w)$ summarized for completeness in Appendix~\ref{app:ldpc}. 
Thus, the complete spectral information needed for the two parts of the Miller--Burshtein bound is already available for $\cC_1$.

\begin{figure}[H]
    \centering
    \includegraphics[width=0.9
    \textwidth]{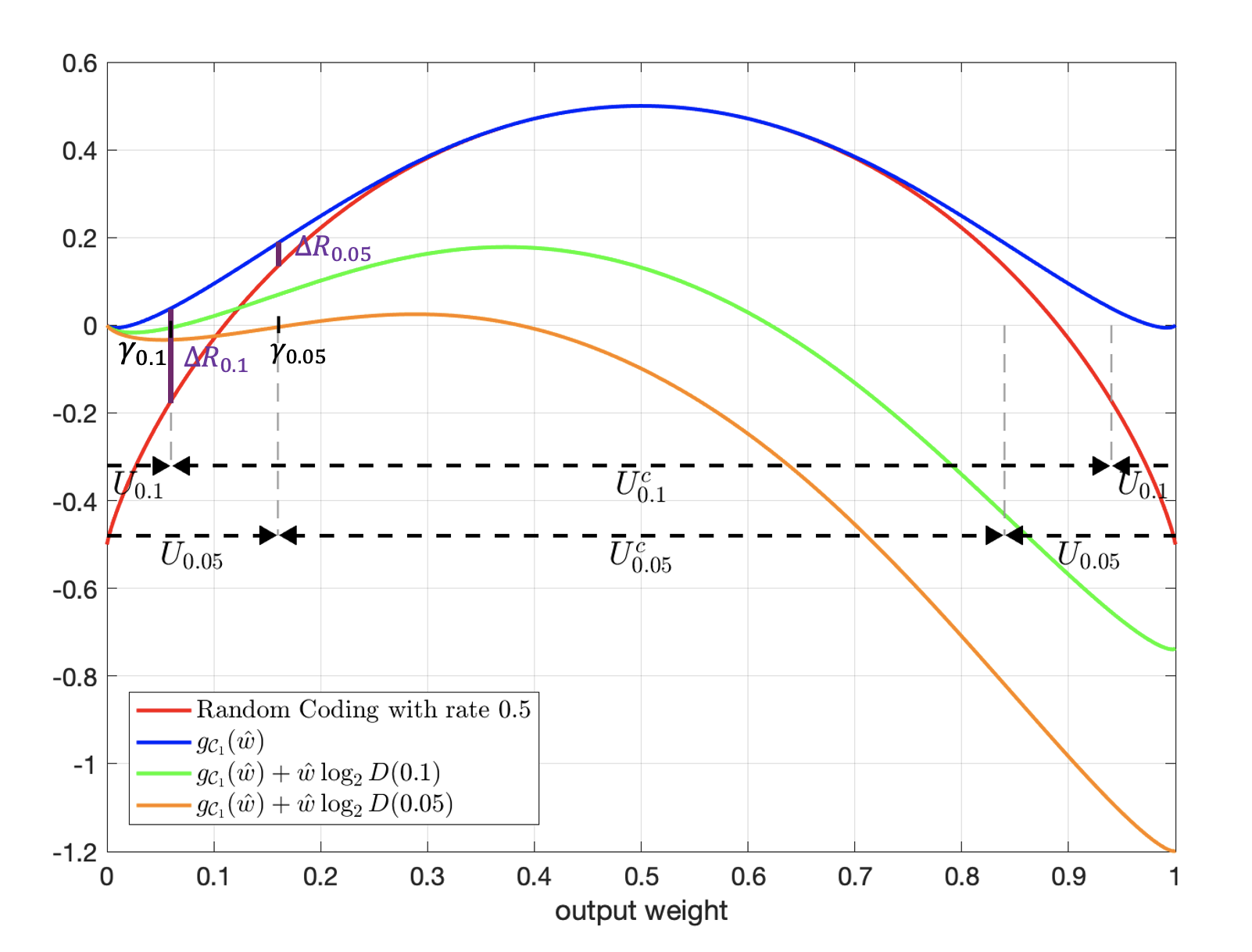}
    \caption{Illustration of the Miller--Burshtein partition for the
$\operatorname{LDPC}(3,6)$ ensemble.}
\label{fig:sec3.2_fig1}
\end{figure}

Figure~\ref{fig:sec3.2_fig1} illustrates how this characterization
is used in the ML analysis of $\cC_1$. For a BSC with crossover
probability $p$, consider the Bhattacharyya term
$    g_{\mathcal  C_1}(\hat w)
    +
    \hat w\log_2 D(p)$
(orange and green curves for two values of $p$). The normalized-weight regions in which this
quantity is negative are assigned to $U$, since the corresponding
terms in the Bhattacharyya union bound vanish exponentially with $n$.
The first positive zero crossing is denoted by $\gamma_p$; together
with the corresponding crossing on the high-weight side, it determines
the boundaries between $U$ and $U^c$. 

The remaining weights form $U^c$. Over this region, the LDPC growth
rate (blue curve) is compared with the random linear-code growth rate
$g_{\mathrm{rand}}(\hat w)=h_2(\hat w)-(1-R)$ (red curve).
The corresponding asymptotic rate penalty is 
the largest vertical separation between the two curves over $U^c$. Since the zero crossings
depend on $p$, the sets $U$ and $U^c$, and consequently
$\Delta R(p)$, also depend on the crossover probability. As $p$ decreases, the Bhattacharyya curve shifts downward, so its zero crossings move outward and the region over which the exponent is negative becomes larger.  Collecting the information in~\eqref{eq:asymptotic_deltaR}, \eqref{eq:ldpc_small_weight_bound} and Appendix~\ref{app:ldpc}, we can numerically evaluate $p_{\mathrm{ML}}^{\mathrm{LB}}$ for $\cC_1$.

We next consider $\cC_2^\perp$, which is obtained as the sum of
an independently drawn LDPC code and LDGM code. For weights that scale
linearly with $n$, the second and third terms in~\eqref{eq:union_split} can
be evaluated using the growth-rate upper bound derived in
the following lemma.

\begin{lemma}[Average weight distribution growth rate of the sum of two independent code ensembles]
\label{lem:sum_two_codes}

Let $\cC_a,\cC_b\subseteq\mathbb F_2^n$ be independently
drawn binary linear codes from two arbitrary ensembles.
The growth rate of the sum ensemble $\cC_a + \cC_b$ satisfies
\begin{equation}
g(\hat w)
\leq
\sup_{(\hat w_a,\hat w_b)\in\mathcal{D}_{\hat w}}
\Psi_{\hat w}(\hat w_a,\hat w_b),
\label{eq:sum_codes_growth}
\end{equation}
where
\begin{equation}
\mathcal{D}_{\hat w}
=
\left\{
(\hat w_a,\hat w_b)\in[0,1]^2:
\hat w\leq\hat w_a+\hat w_b\leq2-\hat w,
\;
|\hat w_a-\hat w_b|\leq\hat w
\right\},
\label{eq:sum_codes_feasible_region}
\end{equation}
and 
\begin{equation}
\Psi_{\hat w}(\hat w_a,\hat w_b)
=
\begin{cases}
\begin{aligned}
&g_a(\hat w_a)+g_b(\hat w_b)
+\hat w_a
h_2\!\left(
\frac{\hat w_a+\hat w_b-\hat w}{2\hat w_a}
\right)
\\[-0.2em]
&\quad
+(1-\hat w_a)
h_2\!\left(
\frac{\hat w_b-\hat w_a+\hat w}
{2(1-\hat w_a)}
\right)
-h(\hat w_b),
\end{aligned}
&
\begin{array}{l}
(\hat w_a,\hat w_b)\in\mathcal{D}_{\hat w},\\
0<\hat w_a<1,
\end{array}
\\[2.2em]
g_a(\hat w_a)+g_b(\hat w_b),
&
\begin{array}{l}
(\hat w_a,\hat w_b)\in\mathcal{D}_{\hat w},\\
\hat w_a\in\{0,1\}.
\end{array}
\end{cases}
\label{eq:sum_codes_pair_exponent}
\end{equation}
where $g_a(\hat w_a)$ and $g_b(\hat w_b)$ denote the asymptotic weight distribution growth rates of $\cC_a$ and $\cC_b$ respectively.
\end{lemma}
\begin{proof}
    See Appendix~\ref{app:Sum_LDPC_LDGM}.
\end{proof}

By substituting the growth
rates of the constituent ensembles, LDPC and LDGM in Appendix~\ref{app:ldpc}
, into
\eqref{eq:sum_codes_growth}, we obtain an upper bound on
$g_{\cC_2^\perp}(\hat w)$, which can be evaluated numerically.
This allows us to determine the zero crossings of
$    g_{\cC_2^\perp}(\hat w)
    +\hat w\log_2 D(p)$
and the corresponding rate penalty over $U^c$.

The remaining challenge is to bound the contributions
from the small-weight region $w<\zeta n$ and the
very large-weight region $w>(1-\zeta)n$ in the
first term of~\eqref{eq:union_split}.
Unlike the regular LDPC ensemble, for
$\cC_2^\perp$ we do not have an analytical characterization of
the average weight distribution in this regime, and therefore the
first condition in~\eqref{eq:pML_LB} cannot be established. 
We will perform a rigorous analysis of this term in Section~\ref{sec:analysis_css} in the context of a more elaborate construction. 
In the following, we present numerical results taking into account only the second and third terms of~\eqref{eq:union_split}.

We conclude this discussion by noting that, 
for sufficiently large but fixed degree parameters, the independently chosen codes $\cC_1^\perp$ and $\widetilde{\cC}$ have code rates approaching the stated design rates.
This result is established by noting that the exponential rate of growth of the intersection of the two codes can be bounded above by the quantity $\sup_{\hat{w}} \{ g_{\cC_1^\perp}(\hat{w})+g_{\tilde{\cC}}(\hat{w})-h_2(\hat{w}) \}$ (cf.~\eqref{eq:sum_codes_pair_exponent}).
As it turns out, using the LDPC results from Fact~\ref{app:ldpc}, as well as the LDGM weight-distribution bounds of~\cite{HsAn10} we can guarantee that the above quantity is non-positive which implies that $\Pr(R_{true}<R_{design}-r)\to 0$ as $n\to \infty$ for any $r>0$.
%

\subsection{Numerical Results}

We consider Pauli channels that are symmetric with respect to $X$ and $Z$ errors, i.e., $p_X=p_Z$. Consequently, $q_X=q_Z$ and the corresponding BSC capacities are equal, $C_X=C_Z$. This symmetry implies that $\cC_1$ and $\cC_2^\perp$ have the same rate
\begin{equation}
R_{\cC_1}
=
R_{\cC_2^\perp}
=
\frac{1+R_{CSS}}{2}.
\label{eq:classical_rates_css}
\end{equation}
We apply the procedure described above separately to $\cC_1$ and $\cC_2^\perp$ in Construction~A.
For each code component, two crossover probabilities are
reported. The quantity
$p_{cap}
=
h_2^{-1}(1-R)$
is the BSC crossover probability at which the channel capacity equals
the corresponding code rate. It therefore provides the
capacity-limit benchmark for a code of rate $R$. The quantity
$p_{\mathrm{ML}}^{\mathrm{LB}}$ is a lower
bound on the maximum correctable crossover probability obtained from
the numerical method described above.

\begin{table}[H]
\centering
\begin{tabular}{c|c|c|c|c|c|c}
\hline
$R_{\mathrm{CSS}}$
&
$\cC_1:\,(d_v,d_c)$
&
$\widetilde{\cC}:\,(\tilde d_v,\tilde d_c)$
&
$R_{\cC_1}$,$R_{\cC_2^\perp}$
&
$p_{1,cap}$,$p_{2^{\perp},cap}$
&
$p_{1,\mathrm{ML}}^{\mathrm{LB}}$
&
$p_{2^\perp,\mathrm{ML}}^{\mathrm{LB}}$
\\
\hline

0.25
& $(3,8)$
& $(9,12)$
& 0.625
& 0.07245
& 0.04167
& 0.07244
\\ \hline

0.$\overline{3}$
& $(3,9)$
& $(6,9)$
& 0.$\overline{6}$
& 0.06149
& 0.03360
& 0.06149
\\ \hline

0.5
& $(3,12)$
& $(3,6)$
& 0.75
& 0.04169
& 0.02029
& 0.04169
\\ \hline

0.$\overline{6}$
& $(3,18)$
& $(3,9)$
& 0.8$\overline{3}$
& 0.02462
& 0.01173
& 0.02462
\\ \hline

0.75
& $(3,24)$
& $(3,12)$
& 0.875
& 0.01713
& 0.00628
& 0.01712
\\ \hline

\end{tabular}

\caption{Lower bounds on the maximum correctable BSC crossover
probability under ML decoding. The pair $(d_v,d_c)$ specifies the
regular LDPC ensemble used for $\cC_1$, while
$(\tilde d_v,\tilde d_c)$ specifies the auxiliary LDPC ensemble
$\widetilde{\cC}$ appearing as a code component of
$\cC_2^\perp$.}
\label{tab:ML_results_tab}
\end{table}

The behavior of $\cC_1$ is consistent with the known ML
performance of regular LDPC ensembles. For fixed finite degrees,
regular LDPC ensembles are not, in general, capacity-achieving under
ML decoding; a nonzero gap to capacity remains because their average weight distribution growth rate differs from that of a random linear code. This limitation
was already identified in the analyses of regular LDPC
codes, where the gap decreases as the graph degrees increase
\cite{Miller2001}. Accordingly, for the first three rates in
Table~\ref{tab:ML_results_tab}, the lower bound
$p_{1,\mathrm{ML}}^{\mathrm{LB}}$ is strictly smaller than the
corresponding capacity-limit value $p_{1,cap}$.

A markedly different behavior is observed for
$\cC_2^\perp$. For every rate considered in
Table~\ref{tab:ML_results_tab}, the numerically obtained value of
$p_{2^\perp,\mathrm{ML}}^{\mathrm{LB}}$ coincides, to the displayed
precision, with the corresponding capacity-limit value
$p_{2^\perp,cap}$.

This is a notable feature of the construction. Unlike
$\cC_1$, which is a regular LDPC ensemble and exhibits thes
expected spectral penalty at several of the rates considered,
$\cC_2^\perp$ is obtained by combining the LDGM and LDPC code
components. Although both constituent codes are sparse and highly
structured, their combination produces an average weight distribution growth rate
that, over the region entering the second term of the
Miller--Burshtein bound, it closely matches the random-linear-code growth rate\footnote{This is a numerically demonstrated property. In Section~\ref{sec:constructionB} we will show conclusively that the random coding growth rate is a true upper bound.}.

For a representative example, consider $R_{CSS}=0.5$, for
which both code components have rate $0.75$. Figures
\ref{fig:LDPC(3,12)} and \ref{fig:sumofLDPC(3,6)LDGM(12,3)} compare
their average weight-distribution growth rates with that of a random
linear code of the same rate.

For $\cC_1$, the regular-LDPC growth rate differs noticeably
from the random-coding growth rate near the low- and high-weight
regions, and approaches the random-coding curve as $\hat w$ moves
toward the central part of the spectrum. This discrepancy is precisely
what produces a nonzero spectral penalty in the Miller--Burshtein
analysis and leads to a gap between
$p_{1,\mathrm{ML}}^{\mathrm{LB}}$ and $p_{1,cap}$.

For $\cC_2^\perp$, its
growth-rate upper bound follows the random-coding growth rate much more
closely over essentially the entire linearly scaling weight region.
Consequently, the supremum defining the rate penalty over $U^c$
is numerically negligible in this example. This explains why the
asymptotic part of the Miller--Burshtein analysis for
$\cC_2^\perp$ reaches the corresponding capacity-limit
crossover probability, whereas the regular-LDPC code component
$\cC_1$ remains the limiting code component of Construction~A.

\begin{figure}[H]
    \centering
\includegraphics[width=0.77\textwidth]{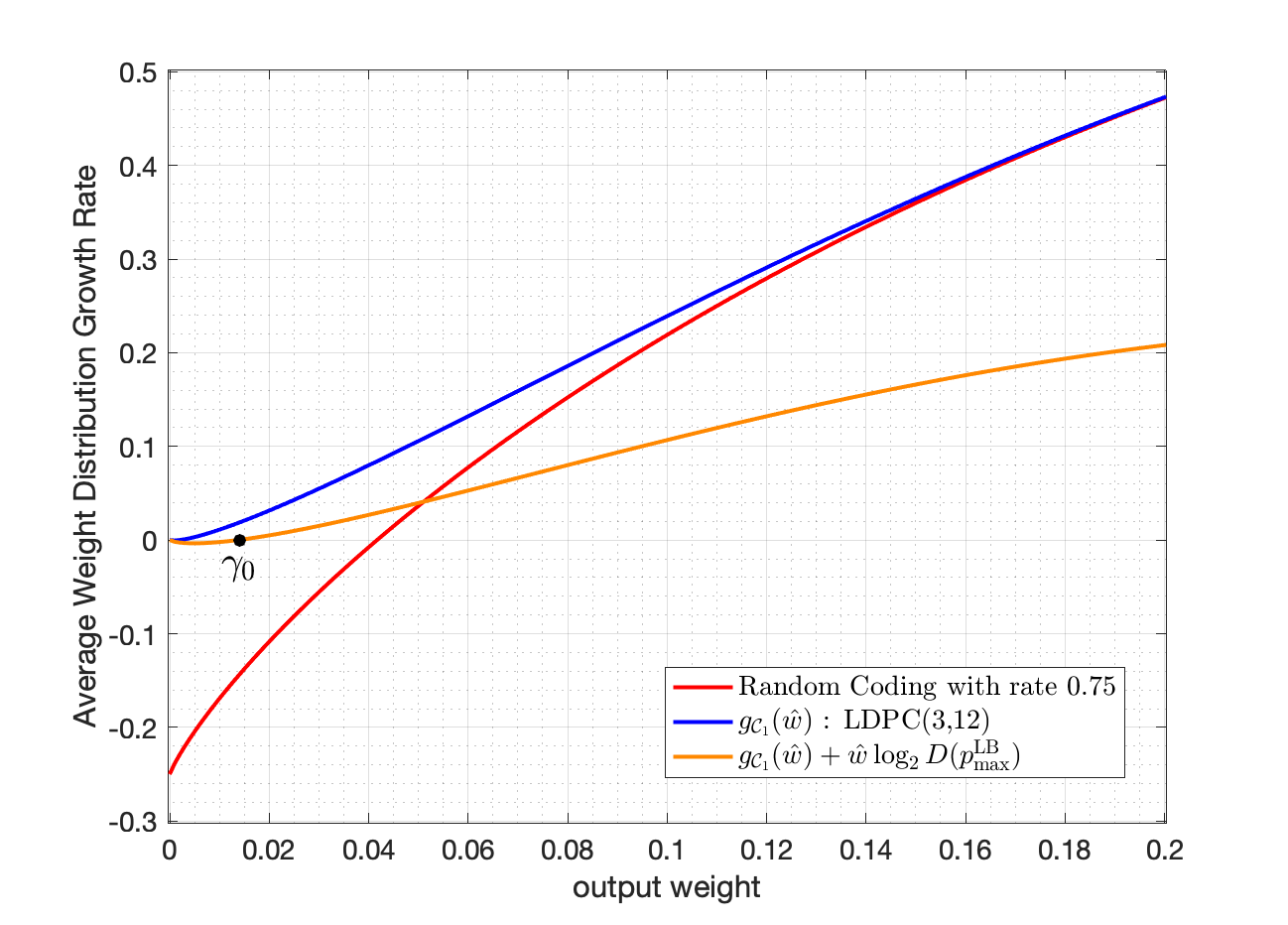}
    \caption{Average weight distribution growth rates for the regular
$\operatorname{LDPC}(3,12)$ ensemble and a random code of rate $0.75$.
The orange curve represents
$g_{\cC_1}(\hat w)+\hat w\log_2 D(p)$ at
$p=p_{\max}^{\mathrm{LB}}$, the lower bound on the maximum correctable
crossover probability.}
    \label{fig:LDPC(3,12)}
\end{figure}
\begin{figure}[H]
    \centering
    \includegraphics[width=0.77\textwidth]{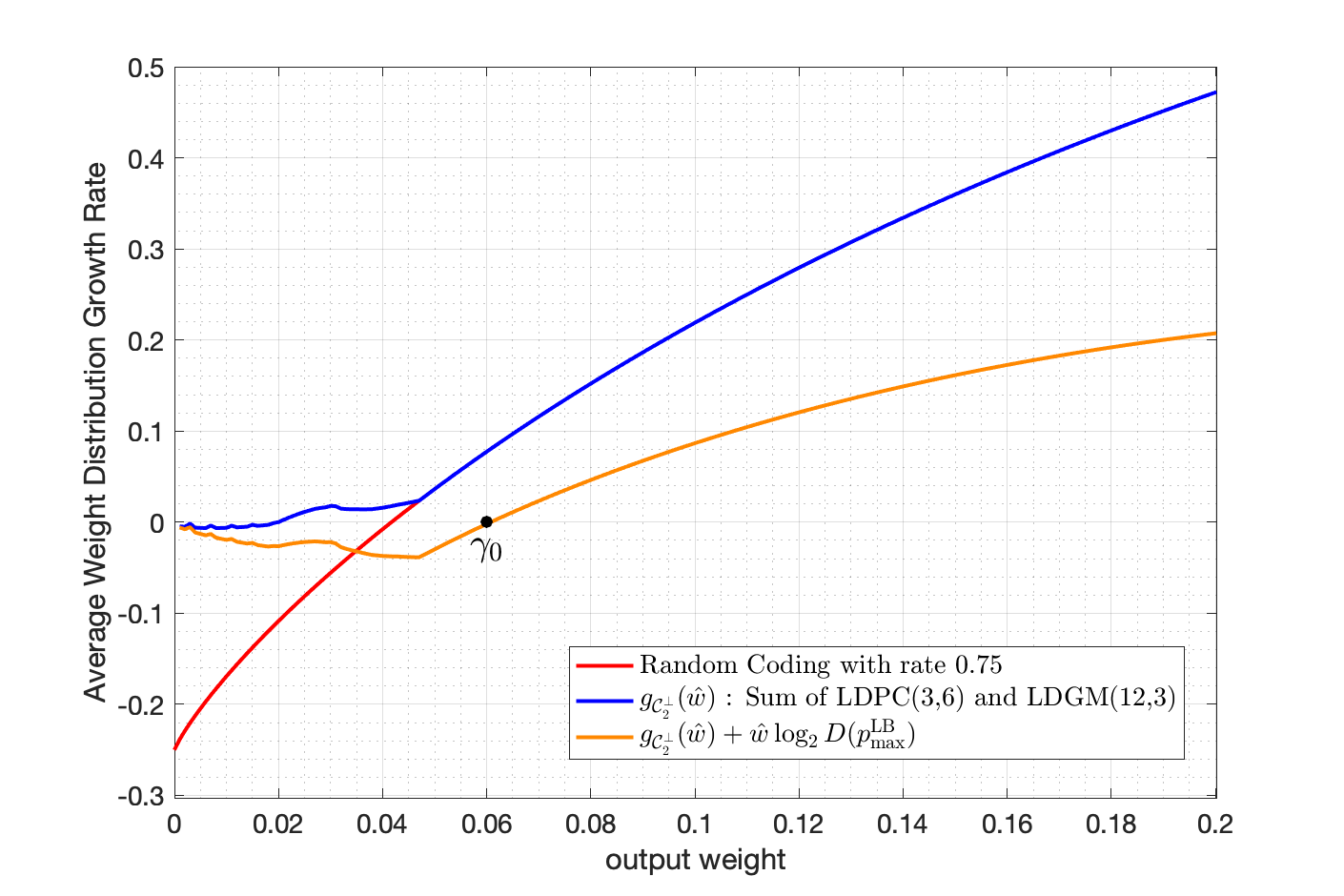}
    \caption{Average weight distribution growth rates for  $\operatorname{LDGM}(12,3) + \operatorname{LDPC}(3,6)$ ensemble and a random code of the same rate $0.75$.
The orange curve represents
$g_{\cC_2^\perp}(\hat w)+\hat w\log_2 D(p)$ at
$p=p_{\max}^{\mathrm{LB}}$, the lower bound on the maximum correctable
crossover probability.}
    \label{fig:sumofLDPC(3,6)LDGM(12,3)}
\end{figure}

Since reliable decoding of the CSS construction requires both
classical decoding problems to have vanishing error probability, the
crossover probability certified for Construction~A is determined by
the more restrictive of the two components:
\begin{equation}
p_{\mathrm{CSS}}^{\mathrm{LB}}
=
\min
\left\{
p_{1,\mathrm{ML}}^{\mathrm{LB}},
p_{2^\perp,\mathrm{ML}}^{\mathrm{LB}}
\right\}.
\end{equation}
Thus, for the first three rates in
Table~\ref{tab:ML_results_tab}, the overall ML guarantee is limited by
the regular LDPC component $\cC_1$, whereas
$\cC_2^\perp$ reaches the corresponding capacity-limit
crossover probability to the reported numerical precision.

This observation motivates the next construction. In particular, the
performance of Construction~A suggests replacing the regular LDPC
choice for $\cC_1$ with a stronger code ensemble,
while preserving the nested structure required by the CSS
construction. The goal is therefore to improve the ML performance of
the $\cC_1$ component so that it no longer determines the
overall decoding threshold.

\section{CSS Code Construction B}
\label{sec:constructionB}

The first construction has a particularly simple sparse graphical
structure and illustrates the main idea that will also be used in
Construction~B. However, ordinary LDPC ensembles do not retain bounded
graphical complexity as their rates approach capacity: approaching
capacity generally requires the average degrees of the parity-check
graph to increase~\cite{RichardsonUrbanke2008}. Therefore, although Construction~A is useful for
motivating the structure, it is not the most suitable choice when
capacity achievability together with bounded graphical complexity is
desired.

This observation leads to Construction~B. Here, we no longer require
the parity-check matrix of $\cC_1$ itself to be sparse.
Instead, a potentially dense parity-check matrix is represented as the
product of sparse constituent matrices. Our construction utilizes an LDPC/GM code for $\cC_1$ and, correspondingly, a GIRA
code for its dual. In this way, the sparse-representation
principle introduced above is applied to both components.

\subsection{Code Construction }

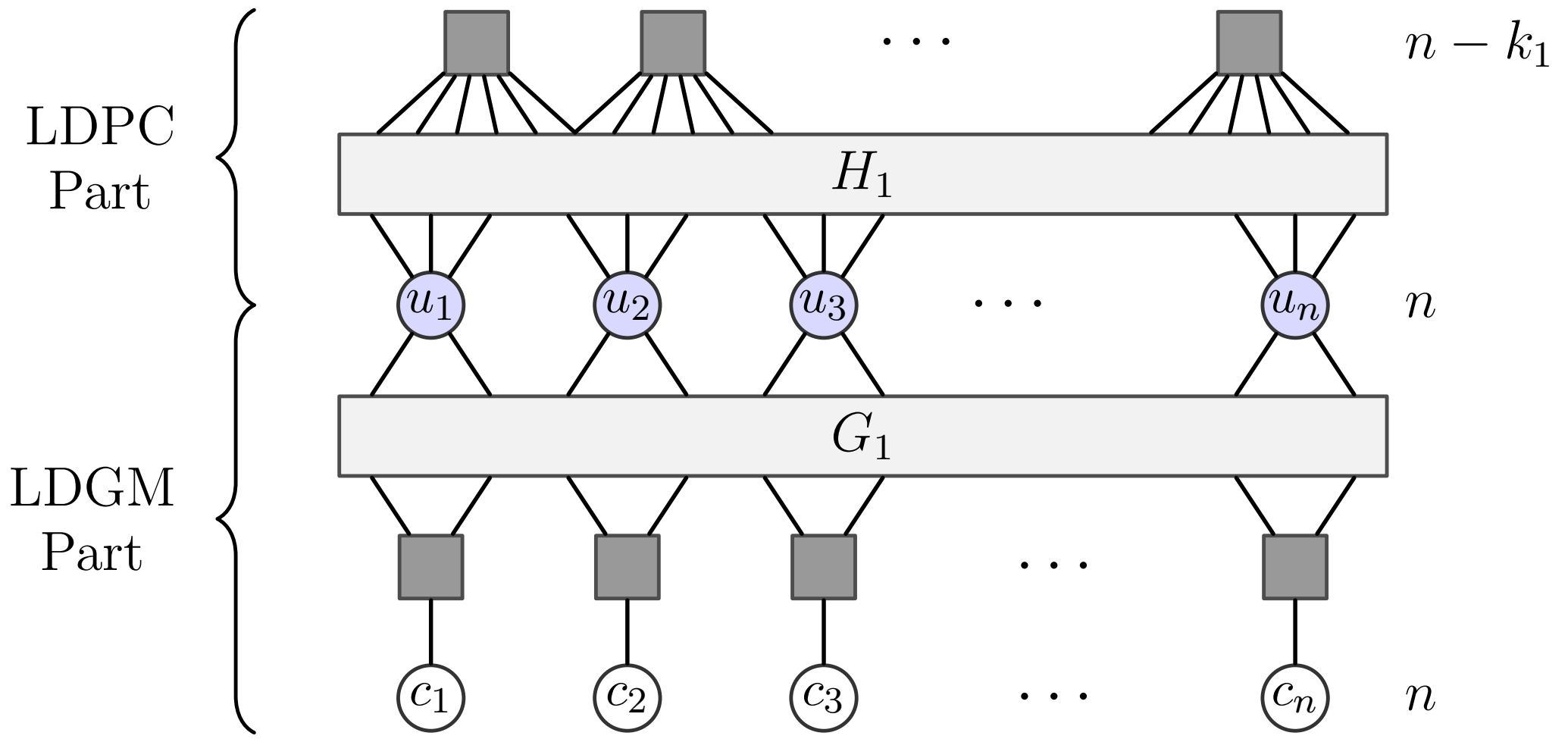
\begin{figure}[H]
    \centering
    \resizebox{0.8\textwidth}{!}{
        \begin{tikzpicture}[
    scale=1.0, transform shape,
    line cap=round,
    line join=round,
    edge/.style={draw=black, line width=0.9pt},
    block/.style={
        draw=black!70,
        fill=gray!10,
        line width=0.8pt,
        text=black,
        font=\Large
    },
    check/.style={
        draw=black!70,
        fill=gray!80,
        line width=0.8pt,
        text=black,
        font=\Large,
        minimum width=0.48cm,
        minimum height=0.48cm,
    },
    var/.style={
        circle,
        draw=black!80,
        fill=white,
        line width=0.8pt,
        minimum size=0.50cm,
        inner sep=0pt
    },
    state/.style={
        circle,
        draw=black!80,
        fill=blue!15,
        line width=0.8pt,
        minimum size=0.50cm,
        inner sep=0pt
    }
]


\def\xA{0.7}
\def\xB{2.2}
\def\xC{3.7}
\def\xD{7.3}


\def\yck{4.0}
\node[check] (ck1) at (\xA+0.35,\yck) {};
\node[check] (ck2) at (\xB+0.35,\yck) {};
\node[check] (ckm) at (\xD-0.35,\yck) {};
\node at (4.45,\yck) {\Large $\cdots$};

\def\yHo{3.0}
\node[
    block,
    minimum width=8.0cm,
    minimum height=0.5cm,
    text=black,
    font=\large
] (Ho) at (4,\yHo) {$H_1$};

\def\yx{2.0}
\node[state] (x1) at (\xA,\yx) {$u_1$};
\node[state] (x2) at (\xB,\yx) {$u_2$};
\node[state] (x3) at (\xC,\yx) {$u_3$};
\node[state] (xn) at (\xD,\yx) {$u_n$};
\node at (5.15,\yx) {\Large $\cdots$};

\def\yGi{1.0}
\node[
    block,
    minimum width=8.0cm,
    minimum height=0.50cm,
    text=black,
    font=\large
] (Gi) at (4,\yGi) {$G_1$};

\def\yckl{0.0}
\node[check] (ckl1) at (\xA,\yckl) {};
\node[check] (ckl2) at (\xB,\yckl) {};
\node[check] (ckl3) at (\xC,\yckl) {};
\node[check] (ckln) at (\xD,\yckl) {};
\node at (5.5,\yckl) {\Large $\cdots$};

\def\yc{-1.0}
\node[var] (c1) at (\xA,\yc) {$c_1$};
\node[var] (c2) at (\xB,\yc) {$c_2$};
\node[var] (c3) at (\xC,\yc) {$c_3$};
\node[var] (cn) at (\xD,\yc) {$c_n$};
\node at (5.5,\yc) {\Large $\cdots$};



\draw[edge] (ck1) -- ($(ck1.center |- Ho.north)+(0*0.3-0.75,0)$);
\draw[edge] (ck1) -- ($(ck1.center |- Ho.north)+(1*0.3-0.75,0)$);
\draw[edge] (ck1) -- ($(ck1.center |- Ho.north)+(2*0.3-0.75,0)$);
\draw[edge] (ck1) -- ($(ck1.center |- Ho.north)+(3*0.3-0.75,0)$);
\draw[edge] (ck1) -- ($(ck1.center |- Ho.north)+(4*0.3-0.75,0)$);
\draw[edge] (ck1) -- ($(ck1.center |- Ho.north)+(5*0.3-0.75,0)$);

\draw[edge] (ck2) -- ($(ck2.center |- Ho.north)+(0*0.3-0.75,0)$);
\draw[edge] (ck2) -- ($(ck2.center |- Ho.north)+(1*0.3-0.75,0)$);
\draw[edge] (ck2) -- ($(ck2.center |- Ho.north)+(2*0.3-0.75,0)$);
\draw[edge] (ck2) -- ($(ck2.center |- Ho.north)+(3*0.3-0.75,0)$);
\draw[edge] (ck2) -- ($(ck2.center |- Ho.north)+(4*0.3-0.75,0)$);
\draw[edge] (ck2) -- ($(ck2.center |- Ho.north)+(5*0.3-0.75,0)$);

\draw[edge] (ckm) -- ($(ckm.center |- Ho.north)+(0*0.3-0.75,0)$);
\draw[edge] (ckm) -- ($(ckm.center |- Ho.north)+(1*0.3-0.75,0)$);
\draw[edge] (ckm) -- ($(ckm.center |- Ho.north)+(2*0.3-0.75,0)$);
\draw[edge] (ckm) -- ($(ckm.center |- Ho.north)+(3*0.3-0.75,0)$);
\draw[edge] (ckm) -- ($(ckm.center |- Ho.north)+(4*0.3-0.75,0)$);
\draw[edge] (ckm) -- ($(ckm.center |- Ho.north)+(5*0.3-0.75,0)$);


\draw[edge] ($(x1.center |- Ho.south)+(0*0.45-0.45,0)$) -- (x1);
\draw[edge] ($(x1.center |- Ho.south)+(1*0.45-0.45,0)$) -- (x1);
\draw[edge] ($(x1.center |- Ho.south)+(2*0.45-0.45,0)$) -- (x1);
\draw[edge] ($(x2.center |- Ho.south)+(0*0.45-0.45,0)$) -- (x2);
\draw[edge] ($(x2.center |- Ho.south)+(1*0.45-0.45,0)$) -- (x2);
\draw[edge] ($(x2.center |- Ho.south)+(2*0.45-0.45,0)$) -- (x2);
\draw[edge] ($(x3.center |- Ho.south)+(0*0.45-0.45,0)$) -- (x3);
\draw[edge] ($(x3.center |- Ho.south)+(1*0.45-0.45,0)$) -- (x3);
\draw[edge] ($(x3.center |- Ho.south)+(2*0.45-0.45,0)$) -- (x3);
\draw[edge] ($(xn.center |- Ho.south)+(0*0.45-0.45,0)$) -- (xn);
\draw[edge] ($(xn.center |- Ho.south)+(1*0.45-0.45,0)$) -- (xn);
\draw[edge] ($(xn.center |- Ho.south)+(2*0.45-0.45,0)$) -- (xn);


\draw[edge] (x1) -- ($(x1.center |- Gi.north)+(0*0.90-0.45,0)$);
\draw[edge] (x1) -- ($(x1.center |- Gi.north)+(1*0.90-0.45,0)$);
\draw[edge] (x2) -- ($(x2.center |- Gi.north)+(0*0.90-0.45,0)$);
\draw[edge] (x2) -- ($(x2.center |- Gi.north)+(1*0.90-0.45,0)$);
\draw[edge] (x3) -- ($(x3.center |- Gi.north)+(0*0.90-0.45,0)$);
\draw[edge] (x3) -- ($(x3.center |- Gi.north)+(1*0.90-0.45,0)$);
\draw[edge] (xn) -- ($(xn.center |- Gi.north)+(0*0.90-0.45,0)$);
\draw[edge] (xn) -- ($(xn.center |- Gi.north)+(1*0.90-0.45,0)$);


\draw[edge] ($(ckl1.center |- Gi.south)+(0*0.90-0.45,0)$) -- (ckl1);
\draw[edge] ($(ckl1.center |- Gi.south)+(1*0.90-0.45,0)$) -- (ckl1);
\draw[edge] ($(ckl2.center |- Gi.south)+(0*0.90-0.45,0)$) -- (ckl2);
\draw[edge] ($(ckl2.center |- Gi.south)+(1*0.90-0.45,0)$) -- (ckl2);
\draw[edge] ($(ckl3.center |- Gi.south)+(0*0.90-0.45,0)$) -- (ckl3);
\draw[edge] ($(ckl3.center |- Gi.south)+(1*0.90-0.45,0)$) -- (ckl3);
\draw[edge] ($(ckln.center |- Gi.south)+(0*0.90-0.45,0)$) -- (ckln);
\draw[edge] ($(ckln.center |- Gi.south)+(1*0.90-0.45,0)$) -- (ckln);


\draw[edge] (ckl1.south) -- (c1.north);
\draw[edge] (ckl2.south) -- (c2.north);
\draw[edge] (ckl3.south) -- (c3.north);
\draw[edge] (ckln.south) -- (cn.north);


\draw[
    decorate,
    decoration={brace,mirror,amplitude=8pt},
    line width=0.8pt
]
(-0.65,0|- ck1.north) -- (-0.65,0|- x1)
node[midway,left=13pt,align=center,font=\large]
{LDPC\\Part};

\draw[
    decorate,
    decoration={brace,mirror,amplitude=8pt},
    line width=0.8pt
]
(-0.65,0|- x1) -- (-0.65,0|- c1.south)
node[midway,left=13pt,align=center,font=\large]
{LDGM\\Part};


\node[
    right,
    align=center,
    font=\large
] at (8.0,\yck) {$n-k_1$};

\node[
    right,
    align=center,
    font=\large
] at (8.0,\yx)
{$n$};

\node[
    right,
    align=center,
    font=\large
] at (8.0,\yc)
{$n$};

\end{tikzpicture}
    }
    \caption{Factor Graph of $\cC_1\sim\operatorname{LDPC/GM}(n,3,6,2)$.}
    \label{fig:ldpcgm_graph}
\end{figure}

We extend Construction~A by choosing the first component $\cC_1$ from the $\operatorname{LDPC/GM} (n,d_v^{H_1},d_c^{H_1},d^{G_1})$ ensemble of Definition~\ref{def:ldpcgm}. 
We independently draw an LDPC code from the $\operatorname{LDPC}(n,d_v^{H_1},d_c^{H_1})$ ensemble with parity-check matrix $H_1$, and a rate-one LDGM code from the $\operatorname{LDGM}(n,d^{G_1},d^{G_1})$
ensemble with generator matrix $G_1$. The corresponding design rate is $R_1=1-d_v^{H_1}/d_c^{H_1}$.
As established in Definition~\ref{def:ldpcgm}, the dual code $\cC_1^\perp$ admits a GIRA representation. The factor graph of $\cC_1$ is illustrated in Figure~\ref{fig:ldpcgm_graph}.

As in Construction~A, we independently choose the auxiliary code $\widetilde{\cC}$ from the $\operatorname{LDPC}(n,d_v^{\widetilde H},d_c^{\widetilde H})$ ensemble, with dimension $(k_1-k_2)$. 

Furthermore, the weight-distribution argument used for Construction~A shows that, for sufficiently large fixed degree parameters, the independently chosen components have an asymptotically negligible intersection with high probability. Furthermore, as established in \cite{HsAn10}, for sufficiently large fixed degree parameters, the LDPC/GM component $\cC_1$ incurs no rate loss relative to its underlying LDPC code with high probability. 
Consequently, Construction~B incurs no asymptotic rate loss due to the intersection of its components. 
The components of Construction~B are summarized as
\begin{equation}
\boxed{
\begin{aligned}
\cC_1
&\sim
\operatorname{LDPC/GM}
(n,d_v^{H_1},d_c^{H_1},d^{G_1}),
\
R_1=1-\frac{d_v^{H_1}}{d_c^{H_1}}\\
\widetilde{\cC}
&\sim
\operatorname{LDPC}
(n,d_v^{\widetilde H},d_c^{\widetilde H}), 
\ 
\tilde{R}=1-\frac{d_v^{\tilde{H}}}{d_c^{\tilde{H}}}\\
\cC_2^\perp
&=\cC_1^\perp+\widetilde{\cC},
\
R_2^\perp = 1-R_1+\tilde{R}.
\end{aligned}
}
\label{eq:constructionB_summary}
\end{equation}

Unlike Construction~A, where $\cC_1^\perp$ has an LDGM representation, Construction~B introduces a GIRA component through the additional rate-one LDGM transformation. The auxiliary LDPC component and the
algebraic structure enforcing CSS nesting remain unchanged. Thus,
$\cC_2^\perp$ is represented as the sum of a GIRA code and an LDPC code, while retaining a sparse factor-graph representation, as illustrated in
Figure~\ref{fig:config2_c2perp}.

\begin{figure}[H]
    \centering
    \resizebox{0.8\textwidth}{!}{
        \input{figures/constructionB_c2perp_graph.tex}
    }
    \caption{Factor Graph of $\cC_2^{\perp}\sim\operatorname{GIRA}(n,6,3,2)+\operatorname{LDPC}(n,3,4)$.}
    \label{fig:config2_c2perp}
\end{figure}

This decomposition is particularly useful for the subsequent
achievability analysis, since the asymptotic weight distribution of the GIRA component can first be studied separately and then combined with that of the auxiliary LDPC ensemble

\subsection{ML Performance Analysis}
\label{sec:analysis_css}
We now provide an analytical proof of the achievability of the proposed
CSS construction using the weight-distribution framework of
Miller and Burshtein \cite{Miller2001}. Rather than relying on numerical evaluation, we
verify directly the required spectral properties for the two constituent
ensembles, $\cC_1$ and $\cC_2$. 

\begin{fact}[{\cite[Theorem~2]{HsAn10}}]
\label{fact:ldpcgm}
There exist finite degree parameters $d_v^H\geq 4$, even $d_c^H$, and $d^G$ such that the  $\operatorname{LDPC/GM}(n,d_v^H,d_c^H,d^G)$ ensemble with design rate $R_{LDPC/GM}$ satisfies the following. For any $\mathrm{BSC}(p)$ with capacity $\cC= 1-h_2(p)>R_{LDPC/GM}$ and for every $\epsilon>0$, there exists $n_\epsilon$ such that, for every  $n>n_\epsilon$,
\begin{equation}
\Pr\!\left(P_B^{\mathrm{ML}}(\cC)\leq\epsilon\right)>1-\epsilon .
\end{equation}
\end{fact}

Fact~\ref{fact:ldpcgm} provides the required ML achievability result for the first component of Construction~B, with the degree parameters chosen as specified above.\footnote{We remark that although the precise statement of~\cite[Theorem~2]{HsAn10} is not as in Fact~\ref{fact:ldpcgm}, the essential result is the same.}

We therefore focus on $\cC_2^\perp$, which in our construction is given by
the sum of a GIRA ensemble and an LDPC ensemble.
%
%
Now, we will provide an analytical characterization of the average weight distribution of the GIRA ensemble and its asymptotic growth rate. This characterization allows us to study the spectral behavior of the GIRA ensemble directly through its degree parameters.

Note that throughout, the degree parameters are ordered as first variable-node degree and then check-node degree. The following results are established considering the GIRA ensemble as the dual of the LDPC/GM ensemble.

\begin{lemma}[Average weight distribution growth rate of GIRA ensemble]
\label{lem:gira}
The  ensemble $\operatorname{GIRA}(n,d_c^H,d_v^H,d^G)$ 
 satisfies
\begin{equation}
\overline N_{\cC}(w)
\leq
\sum_{\ell=0}^{n-k}
\binom{n-k}{\ell}\binom{n}{w}
\frac{
\mathrm{coef}\Big(
\left( 
\frac{(1+x)^{d_v^H}(1+y)^{d^G} + (1-x)^{d_v^H}(1-y)^{d^G}}{2}
\right)
^n,
 x^{\ell d_c^H} y^{w d^G}
\Big)
}{
\binom{n d_v^H}{\ell d_c^H}\binom{n d^G}{w d^G}
}.
\label{eq:C2_pair_bound_statement2}
\end{equation}

The output-weight growth rate is obtained by maximizing over the normalized input weight:
\begin{equation}
    g(\hat{w}) = \max_{\hat{\ell}\in[0,1]} g(\hat{\ell},\hat{w})
    \label{eq:max_dualodLDPCGM}
\end{equation}
where
\begin{equation}
 g(\hat{\ell},\hat{w})=
\frac{d_v^H}{d_c^H} h_2(\hat{\ell}) + h_2(\hat{w})-d_v^H h_2(\hat{\ell})
- d^G h_2(\hat{w})
+ \inf_{x>0,y>0}
  \log_2\left(
  \frac{
  (1+x)^{d_v^H}(1+y)^{d^G} + (1-x)^{d_v^H}(1-y)^{d^G}
  }{2 x^{\hat{\ell} d_v^H} y^{\hat{w} d^G}}
  \right).   
\end{equation}
\end{lemma}
\begin{proof}
    See Appendix~\ref{app:awg_gira}.
\end{proof}

Using this characterization, we next establish two properties of the GIRA ensemble that are essential for the achievability analysis. First, we show that, for sufficiently large degree parameters, its asymptotic growth rate is strictly negative in neighborhoods of zero and one, while being upper bounded by the random linear code growth rate in the remaining interval. Second, we show that the expected number of codewords with small linear and sublinear weights near either endpoint vanishes as the block length increases. The latter property is particularly important because the asymptotic growth rate alone does not guarantee the absence of low-weight codewords. Together, these results provide the spectral properties needed to analyze the sum ensemble.

\begin{lemma} 
\label{lem:lemma3}
There exist finite degree parameters $d_v^{H}$, $d_c^{H}$,  even $d^G>2$, and a positive real number 
$\delta < h^{-1}\bigl(1-d_v^H/d_c^H \bigr)$
such that the $\operatorname{GIRA}(n,d_c^H,d_v^H,d^G)$ ensemble satisfies the following:
\begin{enumerate}
    \item
\begin{equation}
g(\hat w) =
\begin{cases}
\le 0, & \hat w=0, \hat w=1 ,\\
<0, & \hat w\in(0,\delta]\cup [1-\delta,1),\\
\le h_2(\hat w)-(1-d_v^H/d_c^H), & \hat w\in(\delta,1-\delta).
\end{cases}
\label{eq:th_ldpcgm_dual}
\end{equation}
    \item 
\begin{equation}
\sum_{w=1}^{\lfloor\delta n\rfloor}
\overline N_{\cC}(w)
=
O\!\left(n^{-(\frac{d^G}{2}-1)}\right)
\qquad
\text{and}
\qquad
\sum_{w=\lceil(1-\delta)n\rceil}^{n-1}
\overline N_{\cC}(w)
=
O\!\left(n^{-(\frac{d^G}{2}-1)}\right).
\label{eq:C1perp_weight_sum}
\end{equation}

\end{enumerate}
\end{lemma}
\begin{proof}
    See Appendix~\ref{app:lemma3}.
\end{proof}

Lemma~\ref{lem:lemma3} establishes the required weight-distribution properties of the GIRA ensemble. We now extend this analysis to the sum ensemble formed by independently drawing a code from the GIRA ensemble and a code from the LDPC ensemble. The main objective is to establish that the sum ensemble inherits analogous weight-distribution properties, with the corresponding random linear code bound determined by the overall design rate $R=R_{GIRA}+R_{LDPC}= d_v^{H_1}/d_c^{H_1}+(1-d_v^{\tilde{H}}/d_c^{\tilde{H}})$.

\begin{theorem}
\label{thm:thm3}
There exist finite degree parameters $d_v^{H_1}$, $d_c^{H_1}$, even $d^{G_1}>2$, $d_v^{\tilde H}>2$, $d_c^{\tilde H}$ and a positive real number $\delta < h^{-1}\bigl(d_v^{\tilde H}/d_c^{\tilde H}-d_v^{H_1}/d_c^{H_1}\bigr)$ such that the sum ensemble obtained from independent draws from the $\operatorname{GIRA}(n,d_c^{H_1},d_v^{H_1},d^{G_1})$ and $\operatorname{LDPC}(n,d_v^{\tilde H},d_c^{\tilde H})$ ensembles satisfies the following:
\begin{enumerate}
    \item
\begin{equation}
g_{\cC_2^\perp}(\hat w) =
\begin{cases}
\le 0, & \hat w=0, \hat w=1 ,\\
<0, & \hat w\in(0,\delta]\cup [1-\delta,1),\\
\le h_2(\hat w)-\bigl(d_v^{\tilde H}/d_c^{\tilde H}-d_v^{H_1}/d_c^{H_1}\bigr), & \hat w\in(\delta,1-\delta).
\end{cases}
\label{eq:sum_ldgmpc__lpdc_claim}
\end{equation}

\item \begin{equation}
\overline N_{\cC_2^\perp}(w)
=
O\!\left(n^{-\min\left\{
\frac{d^{G_1}}{2}-1,\,
d_v^{\tilde H}-2
\right\}}\right)
\text{ for \ 
$w\in
\{1,\ldots,\lfloor\delta n\rfloor\}
\cup
\{\lceil(1-\delta)n\rceil,\ldots,n-1\}.$}
\label{eq:C2perp_weight_sum}
\end{equation}

\end{enumerate}
\end{theorem}
\begin{proof}
    See Appendix~\ref{app:ach_sum_gira_ldpc}.
\end{proof}


Theorem~\ref{thm:thm3} establishes that the sum ensemble satisfies the spectral properties required for the capacity-achievability analysis. 
These results allow us to apply the weight-distribution-based achievability framework of Miller and Burshtein.

\begin{theorem} [CSS Capacity Achievability of Construction B]
\label{thm:thm_sum_gira_ldpc_ach}
There exist finite degree parameters $d_v^{H_1}>4$, $d_c^{H_1}$,
$d_v^{\tilde H}>3$, $d_c^{\tilde H}>3$, and even $d^{G_1}>4$ such that the CSS ensemble of Construction~B with these
parameters satisfies the following. For any memoryless Pauli channel with
error probabilities $(p_X,p_Y,p_Z)$ satisfying
\begin{equation}
1-h_2(p_X+p_Y)>1-d_v^{H_1}/d_c^{H_1}
\quad\text{and}\quad
1-h_2(p_Z+p_Y)>1-d_v^{\tilde H}/d_c^{\tilde H}+d_v^{H_1}/d_c^{H_1}
\end{equation}
and for every $\epsilon>0$, there exists $n_\epsilon$ such that, for every
 $n>n_\epsilon$,
\begin{equation}
\Pr\!\left(P_B^{\mathrm{ML}}(\cC_1)\leq\epsilon,\;
P_B^{\mathrm{ML}}(\cC_2^{\perp})\leq\epsilon\right)>1-\epsilon .
\end{equation}
\end{theorem}

\begin{proof}
    See Appendix~\ref{app:thm_sum_gira_ldpc_ach}.
\end{proof}

We remark that Theorem~\ref{thm:thm_sum_gira_ldpc_ach} establishes simultaneous ML reliability of the two components when their design rates \(R_1=1-d_v^{H_1}/d_c^{H_1}\) and \(R_2^\perp=1-d_v^{\widetilde H}/d_c^{\widetilde H}+d_v^{H_1}/d_c^{H_1}\) lie strictly below their respective marginal channel capacities. Since \(R_{\mathrm{CSS}}=R_1+R_2^\perp-1\), and $C_{\mathrm{CSS}}=1-h_2(p_X+p_Y)-h_2(p_Y+p_Z)$, these conditions imply that 
the proposed construction can achieve rates approaching the CSS hashing bound, provided that the component codes asymptotically attain their design rates.
Indeed, the rank deficiencies of the underlying LDPC parity-check matrices are sublinear in $n$ with high probability, so the actual rates of the corresponding codes converge to their design rates \cite{MillerCohen2003}. 
Moreover, as discussed in Section~4.1 for Construction B, the intersection of $\cC_1^\perp$ and $\widetilde{\cC}$ has asymptotically negligible dimension with high probability, so the sum construction incurs no asymptotic rate loss.
Consequently, the actual rates of both $\cC_1$ and $\cC_2^\perp$ converge to their respective design rates, and the actual CSS rate approaches $R_1+R_2^\perp-1$ as $n\to\infty$.
Importantly, the construction retains a sparse graphical representation, providing an analytical foundation for the development of practical iterative decoding algorithms.

\section{Conclusion}
\label{sec:conclusion}

In this work, we investigated two constructions of quantum CSS codes based on classical linear codes with sparse graphical representations. We first introduced Construction~A, in which the first component is an LDPC code, while the second component is represented as the sum of its LDGM dual and an auxiliary LDPC code. Building on this framework, we developed Construction~B by replacing the LDPC component with an LDPC/GM code, whose dual admits a GIRA representation. This extension preserves the CSS nesting condition and the sum-of-codes structure while introducing additional flexibility in the design of the components.

Our main theoretical result establishes that Construction~B asymptotically achieves the CSS hashing bound over memoryless Pauli channels under maximum-likelihood (ML) decoding. We first proved that the sum ensemble of GIRA and LDPC, $\cC_2^\perp$, is capacity-achieving over binary symmetric channels (BSCs) under ML decoding, provided that the degree parameters are appropriately chosen and the design rate is below the channel capacity. Combining this result with the previously established ML capacity-achieving properties of the LDPC/GM ensemble $\cC_1$~\cite{HsAn10}, we showed that both components are simultaneously reliable under ML decoding with asymptotically high probability. Thus, the proposed construction establishes the asymptotic achievability of the CSS hashing bound while retaining a sparse graphical representation.

Although ML decoding generally entails exponential computational complexity and is therefore impractical for large block lengths, the capacity-achievability result provides a theoretical foundation for further investigation of the proposed construction under low-complexity iterative decoding.
In particular, the sparse factor-graph representation naturally motivates the application of belief-propagation (BP) decoding. An important direction for future work is therefore to investigate the performance of BP decoding for the proposed construction, characterize its decoding thresholds, and jointly optimize the degree distributions of both constituent codes. Such an investigation may help bridge the gap between the capacity-achieving performance established under ML decoding and the practical implementation of efficient iterative decoders.

A second, specifically quantum, challenge concerns syndrome extraction. The sparse extended factor graphs developed here do not imply that the effective parity-check matrices of $\cC_1$ and $\cC_2^\perp$ are sparse; indeed, approaching capacity generally requires their parity-check densities to grow. Directly identifying the rows of these dense matrices with stabilizer measurements would therefore produce high-weight checks and correspondingly dense syndrome-extraction circuits, potentially offsetting the implementation advantages of the sparse graphical representation. Realizing the full potential of the proposed construction will require syndrome-extraction methods that exploit its sparse constituent structure and auxiliary variables, rather than directly measuring the effective dense checks. 
Alternatively it may lead to fully quantum decoders without any required measurement, as those of~\cite{Gottesman1997,renes2017belief,Knill_2000,LiNakaharaPoonSzeTomita2012,PremakumarWangSaffmanJoynt2023,HeussenLocherMuller2024,VeroniMullerGiudice2024,ButtLocherBrechtelsbauerBuchlerMuller2025}.
Developing such indirect or extended measurement protocols, and analyzing their circuit depth, ancilla requirements, fault propagation, and compatibility with iterative decoding, is an important direction for future work.

\appendix

\section{ Facts about LDPC Codes} 
\label{app:ldpc}

\begin{fact}{ (Average Weight Distribution of a Regular LDPC Code).}
For the regular $\operatorname{LDPC}(n,d_v,d_c)$ ensemble, the ensemble-average
number of codewords of Hamming weight $w$ is given by
\begin{equation}
\overline{N_{\cC}}(w)
=
\binom{n}{w}
\frac{
\left[x^{w d_v}\right]
\left(
\frac{(1+x)^{d_c}+(1-x)^{d_c}}{2}
\right)^{n d_v/d_c}
}{
\binom{n d_v}{w d_v}
},
\label{eq:ldpc_average_we}
\end{equation}
where $[x^k]f(x)$ denotes the coefficient of $x^k$ in the polynomial
$f(x)$.

Moreover, letting $\hat w=w/n$ and taking the normalized logarithm as
$n\to\infty$, the corresponding ensemble-average weight distribution
growth rate is
\begin{subequations}
\begin{align}
g(\hat w)
&=
(1-d_v)h_2(\hat w)
+
\inf_{x>0}
\left\{
\frac{d_v}{d_c}
\log_2
\left(
\frac{(1+x)^{d_c}+(1-x)^{d_c}}{2}
\right)
-\hat w d_v\log_2 x
\right\}
\label{eq:ldpc}
\\
&\stackrel{(a)}{\leq}
h_2(\hat w)
+
\frac{d_v}{d_c}
\log_2
\left(
\frac{1+(1-2\hat w)^{d_c}}{2}
\right),
\label{eq:ldpc_ub}
\end{align}
\end{subequations}
where $h_2(\cdot)$
is the binary entropy function.
The expression in \eqref{eq:ldpc_average_we} is the
ensemble-average weight distribution of a regular LDPC ensemble derived
by Gallager \cite[Appendix A]{gallager1963ldpc}. The coefficient
extraction term counts the configurations of the check-node sockets
compatible with a codeword of weight $w$, while the two binomial terms
account for the choice of the $w$ nonzero variable nodes and the
corresponding placement of their $w d_v$ sockets.

The asymptotic expression in \eqref{eq:ldpc} follows by taking the
normalized logarithm of \eqref{eq:ldpc_average_we}, applying Stirling's
approximation to the binomial coefficients, and using the standard
coefficient bound
\begin{equation}
[x^m]f(x)
\leq
\inf_{x>0}\frac{f(x)}{x^m}.
\end{equation}

Finally, inequality~(a) follows from substituting $x=\frac{\hat{w}}{1-\hat{w}}$ in \eqref{eq:ldpc}.

\end{fact}

\begin{fact}{(Average Input--Output Weight Distribution of a Regular LDGM Code).}
Consider a regular $\operatorname{LDGM}(n,d_v,d_c)$ ensemble with $k$ input
nodes of degree $d_v$ and $n$ output nodes of degree $d_c$, where
$    k d_v = n d_c,
$ and $
    R_{LDGM}
    =
    \frac{k}{n}
    =
    \frac{d_c}{d_v}.$
Let $\overline{A}(\ell,w)$ denote the ensemble-average number of input
vectors of Hamming weight $\ell$ whose corresponding output has
Hamming weight $w$. Then
\begin{equation}
\overline{A}(\ell,w)
=
\binom{k}{\ell}\binom{n}{w}
\frac{
\left[x^{\ell d_v}\right]
f_-(x)^w f_+(x)^{n-w}
}{
\binom{n d_c}{\ell d_v}
},
\label{eq:ldgm_average_iowe}
\end{equation}
where
\begin{equation}
f_-(x)
=
\frac{(1+x)^{d_c}-(1-x)^{d_c}}{2},
\qquad
f_+(x)
=
\frac{(1+x)^{d_c}+(1-x)^{d_c}}{2}.
\end{equation}
where the functions $f_-(x)$ and $f_+(x)$ enumerate, respectively, the odd
and even numbers of active sockets incident on an output node.
Let
$    \hat\ell=\frac{\ell}{k},
    $ and $
    \hat w=\frac{w}{n}.$
By Stirling's approximation and the
standard coefficient bound,
the corresponding asymptotic input--output weight distribution growth
rate is
\begin{subequations}
\begin{align}
g(\hat\ell,\hat w)
&=
\left(\frac{d_c}{d_v}-d_c\right)h_2(\hat\ell)
+h_2(\hat w)+
\inf_{x>0}
\left\{
\hat w\log_2 f_-(x)
+(1-\hat w)\log_2 f_+(x)
-\hat\ell d_c\log_2 x
\right\}
\label{eq:ldgm_glw}
\\
&\stackrel{(a)}{\leq}\frac{d_c}{d_v}h_2(\hat\ell)
+h_2(\hat w)
+\hat w
\log_2
\frac{1-(1-2\hat\ell)^{d_c}}{2}
+(1-\hat w)
\log_2
\frac{1+(1-2\hat\ell)^{d_c}}{2}.
\end{align}
\end{subequations}
where inequality~(a) follows from substituting $x=\frac{\hat{\ell}}{1-\hat{\ell}}$ in \eqref{eq:ldgm_glw}.
Consequently, the output-weight growth rate is
\begin{equation}
    g(\hat w)
    =
    \max_{\hat\ell} g(\hat\ell,\hat w).
\label{eq:ldgm_output_growth}
\end{equation}

If $d_c$ is even,$(1-2\hat\ell)^{d_c}\geq0$, further implies
\begin{equation}
g(\hat\ell,\hat w)
\leq
\frac{d_c}{d_v}h_2(\hat\ell)
+h_2(\hat w)
+
\log_2
\frac{1+(1-2\hat\ell)^{d_c}}{2}.
\label{eq:ldgm_ub}
\end{equation}
The latter bound is exact at $\hat\ell=1/2$ and a nonzero coefficient in
\eqref{eq:ldgm_average_iowe} requires
\begin{equation}
\frac{\hat w}{d_c}
    \leq
    \hat\ell
    \leq
    1-\frac{\hat w}{d_c},    
    \label{eq:ldgm_feasible_region}
\end{equation}
and hence the maximization in~\eqref{eq:ldgm_output_growth} may be
restricted to this interval, which follows from the smallest and largest
possible powers of $x$ in $f_-(x)^w f_+(x)^{n-w}$.

\noindent
The exact ensemble-average input--output weight enumerator in
\eqref{eq:ldgm_average_iowe} is derived in
\cite[Sec.~3, Eqs.~(4)--(7)]{HsAn10}.
  
\end{fact}

    \section{ Proof of Lemma~\ref{lem:sum_two_codes}}
\label{app:Sum_LDPC_LDGM}
\begin{proof}
For a particular pair of codes, the weight distribution of the sum
code is
\begin{equation}
N_{\cC}(w)
=
\#\{c\in\cC:\operatorname{wt}(c)=w\}.
\end{equation}
Since a codeword $c\in\cC$ may admit more than one
representation as $c=c_a+c_b$, we have
\begin{equation}
N_{\cC}(w)
\leq
\#
\left\{
(c_a,c_b)\in\cC_a\times\cC_b:
\operatorname{wt}(c_a+c_b)=w
\right\}.
\label{eq:sum_codes_pair_bound}
\end{equation}

Condition on the weights of the two constituent codewords and define
\begin{equation}
\begin{aligned}
N(w\mid w_a,w_b)
=
\#\bigl\{
(c_a,c_b)\in\cC_a\times\cC_2:
\operatorname{wt}(c_a)=w_a,
\operatorname{wt}(c_b)=w_b,
\operatorname{wt}(c_a+c_b)=w
\bigr\}.
\end{aligned}
\end{equation}
Then
\begin{equation}
N_{\cC}(w)
\leq
\sum_{w_a=0}^n
\sum_{w_b=0}^n
N(w\mid w_a,w_b).
\label{eq:sum_codes_conditioned}
\end{equation}
Taking the ensemble average and using independence of
$\cC_a$ and $\cC_b$ gives
\begin{subequations}
\begin{align}
\overline N(w\mid w_a,w_b)
&=
\mathbb E_{\cC_a,\cC_b}
\left[
N(w\mid w_a,w_b)
\right]
\\
&=
\sum_{\substack{c_a,c_b\in\mathbb F_2^n\\
\operatorname{wt}(c_a)=w_a,\,
\operatorname{wt}(b_a)=w_b}}
\mathbf 1_{\{\operatorname{wt}(c_a+c_b)=w\}}
\Pr(c_a\in\cC_a)
\Pr(c_b\in\cC_b)
\\
&=
\frac{\overline N_{\cC_a}(w_a)}
{\binom n{w_a}}
\frac{\overline N_{\cC_b}(w_b)}
{\binom n{w_b}}
\sum_{\substack{c_a,c_b\in\mathbb F_2^n\\
\operatorname{wt}(c_a)=w_a,\,
\operatorname{wt}(c_b)=w_b}}
\mathbf 1_{\{\operatorname{wt}(c_a+c_b)=w\}}.
\label{eq:sum_codes_expectation}
\end{align}
\end{subequations}
where $\overline N_{\cC_i}(w_i)$ denote the ensemble-average
number of codewords of weight $w_i$ in $\cC_i$, $i=a,b$.

Let $t$ denote the number of positions in which both $a$ and $b$
are equal to one. Then
$\operatorname{wt}(c_a,c_b)
=
w_a+w_b-2t,$
and therefore
$t=\frac{w_a+w_b-w}{2}.$
For a fixed vector $c_a$ of weight $w_a$, the number of vectors $c_b$
of weight $w_b$ having exactly $t$ overlapping ones with $a$ is
$\binom{w_a}{t}
\binom{n-w_a}{w_b-t}.$
Since there are $\binom n{w_1}$ possible vectors $c_a$, it follows that
\begin{equation}
\overline N(w\mid w_a,w_b)
=
\overline N_{\cC_a}(w_a)
\overline N_{\cC_b}(w_b)
\frac{
\binom{w_a}{t}
\binom{n-w_a}{w_b-t}
}{
\binom n{w_b}
}.
\label{eq:sum_codes_conditional_average}
\end{equation}

Substituting \eqref{eq:sum_codes_conditional_average} into
\eqref{eq:sum_codes_conditioned} and taking the ensemble average
gives
\begin{equation}
\overline N_{\cC}(w)
\leq
\sum_{w_a=0}^n
\sum_{w_b=0}^n
\overline N_{\cC_a}(w_a)
\overline N_{\cC_b}(w_b)
\frac{
\binom{w_a}{t}
\binom{n-w_a}{w_b-t}
}{
\binom{n}{w_b}
},
\label{eq:sum_codes_finite}
\end{equation}
where the sum contains only pairs $(w_a,w_b)$ for which $t$ is an
integer satisfying
\begin{equation}
0\leq t\leq w_a,
\quad
0\leq w_b-t\leq n-w_a.
\label{eq:sum_codes_feasibility_finite}
\end{equation}

It remains to derive the asymptotic form.
Let
$\hat w=\frac wn,$
$\hat w_a=\frac{w_a}{n},$
$\hat w_b=\frac{w_b}{n}.$
The right-hand side of
\eqref{eq:sum_codes_finite} contains at most $(n+1)^2$ nonnegative
summands. Hence its logarithm is upper bounded by the logarithm of
its largest summand plus
$2\log_2(n+1).$
After normalization by $n$,
$\frac{2\log_2(n+1)}{n}
\longrightarrow0.$
Therefore only the largest exponential contribution remains.

The combinatorial factor in
\eqref{eq:sum_codes_conditional_average} is
$B_n(w_a,w_b,t)
=
\frac{
\binom{w_a}{t}
\binom{n-w_a}{w_b-t}
}{
\binom{n}{w_b}
}.$
Applying Stirling's approximation to each binomial coefficient, we obtain,
for $0<\hat w_a<1$,
\begin{equation}
\begin{aligned}
\frac{1}{n}\log_2 B_n(w_a,w_b,t)
={}
\hat w_a
h_2\left(\frac{\hat t}{\hat w_a}\right)
+(1-\hat w_a)
h_2\left(
\frac{\hat w_b-\hat t}{1-\hat w_a}
\right)-h_2(\hat w_b)
+O\left(\frac{\log n}{n}\right).
\end{aligned}
\label{eq:boundary_log_expression}
\end{equation}
which proves the first line of \eqref{eq:sum_codes_pair_exponent}.

Also, the finite-length feasibility conditions
\eqref{eq:sum_codes_feasibility_finite}, after normalization by $n$,
are equivalent to
\begin{equation}
\hat w
\leq
\hat w_a+\hat w_b
\leq
2-\hat w,
\qquad
|\hat w_a-\hat w_b|
\leq
\hat w,
\end{equation}
which gives the set $\mathcal{D}_{\hat w}$ in
\eqref{eq:sum_codes_feasible_region}. Maximizing over all feasible
pairs proves \eqref{eq:sum_codes_growth}.


It remains to establish the boundary cases
$\hat w_a\in\{0,1\}$. We first consider the case
$\hat w_a=0$. We investigate the behavior of \eqref{eq:boundary_log_expression} as
$\hat w_a\to 0$ using a directional approach.
Combining the feasible conditions,
$|\hat w_a-\hat w_b|\leq\hat w$
and
$\hat w\leq\hat w_a+\hat w_b,$
we have
$\hat w-\hat w_a\leq\hat w_b\leq\hat w+\hat w_a.$
Therefore, as $\hat w_a\to0$, the feasible values
of $\hat w_b$ necessarily converge to $\hat w$.
To characterize this limiting behavior, we
consider the directional approach
$\hat w_b=\hat w+\alpha\hat w_a,$
where
$-\frac{1}{\sqrt{2}}\leq\alpha
\leq\frac{1}{\sqrt{2}}$
follows directly from the feasibility conditions.
Substituting into \eqref{eq:boundary_log_expression} yields
\begin{equation}
\begin{aligned}
\frac{1}{n}\log_2 B_n
={}
\hat w_a h_2\left(\frac{1+\alpha}{2}\right)
+(1-\hat w_a)
h_2\left(
\frac{
\hat w-\frac{1-\alpha}{2}\hat w_a
}{
1-\hat w_a
}
\right)
-h_2(\hat w+\alpha\hat w_a)
+O\left(\frac{\log n}{n}\right).
\end{aligned}
\label{eq:boundary_directional_expression}
\end{equation}

Since $0\leq h_2(p)\leq1$, the first term vanishes as $\hat w_a\to0$.
Moreover,
$\frac{
\hat w-\frac{1-\alpha}{2}\hat w_a
}{
1-\hat w_a
}
\longrightarrow\hat w,$
and
$\hat w+\alpha\hat w_a
\longrightarrow\hat w.$
By continuity of the binary entropy function,
the remaining terms converge to
$h_2(\hat w)-h_2(\hat w)=0$.

Consequently,
\begin{equation}
\lim_{\hat w_a\to0}
\lim_{n\to\infty}
\frac{1}{n}\log_2 B_n(w_a,w_b,t)
=0.
\label{eq:boundary_limit_zero}
\end{equation}
The limit is taken along feasible integer
sequences with the prescribed normalized
weights, with the normalized-weight limit
performed after the large-blocklength limit.

At the exact boundary $\hat w_a=0$, feasibility
forces $\hat w_b=\hat w$ and $t=0$, so the
combinatorial factor equals one. At the other
boundary $\hat w_a=1$, feasibility requires
$\hat w_b=1-\hat w$ and $t=w_b$, and the
combinatorial factor again equals one.

Consequently, the combinatorial contribution
to the asymptotic exponent vanishes at both
boundaries. Therefore,
\begin{equation}
\Psi_{\hat w}(\hat w_a,\hat w_b)
=
g_a(\hat w_a)+g_b(\hat w_b),
\qquad
\hat w_a\in\{0,1\},
\end{equation}
which establishes the second branch of \eqref{eq:sum_codes_pair_exponent} .


\end{proof}

 \section{ Proof of Lemma~\ref{lem:gira}}
\label{app:awg_gira}
\begin{proof}

 Define
\begin{equation}
 A(\ell,w)\triangleq \#\{(c,x): c \in \mathbb F_2^n, \ x \in \mathbb F_2^{n-k}, \  w(c)=w, \ w(x)=\ell, \ xH+cG^T=0 \}.   
\end{equation}
which denotes the number of codewords with input weight \(\ell\) and output weight \(w\).
Therefore,
\begin{equation}
A(\ell,w)
=
\sum_{\substack{x:\,w(x)=\ell}}
\sum_{\substack{c:\,w(c)=w}}
\mathbf{1}\{xH+cG^T=0\}.    
\end{equation}
Taking the ensemble average gives
\begin{equation}
  \overline{A}(\ell,w)
=
\sum_{\substack{x:\,w(x)=\ell}}
\sum_{\substack{y:\,w(c)=w}}
P(xH+cG^T=0).  
\end{equation}
Since all vectors of the same weights are equivalent by symmetry,
\begin{equation}
 \overline{A}(\ell,w)
=
\binom{n-k}{\ell}\binom{n}{w}\,P(xH+cG^T=0),   
\end{equation}
where the probability depends only on \(\ell\) and \(w\).
\paragraph{Combinatorial interpretation of \(P(xH+cG^T=0)\).}

The event \(xH+cG^T=0\) means that the induced connections produce an even number of \(1\)'s at every check node. Hence,
\begin{equation}
 P(xH+cG^T=0)
=
\frac{\#\{\text{valid edge connections such that each check has even parity}\}}
{\binom{nd^G}{wd_c^G}\binom{nd_v^H}{\ell d_c^H}}.   
\end{equation}
The denominator counts all possible ways of choosing the active sockets, namely, \(wd_c^G\) active sockets from the \(nd^G\) sockets on the \(G^T\) side and \(\ell d_c^H\) active sockets from the \(nd_v^H\) sockets on the \(H\) side.

\paragraph{Counting valid configurations.}
The idea is to track simultaneously the number of active sockets induced by the vector $x$ on the $H$ side and by the vector $c$ on the $G^T$ side, and then enforce that the total number of active incident edges at each variable position is even. For one variable position, let $a$ denote the number of active sockets contributed from the $H$ side and let $b$ denote the number of active sockets contributed from the $G^T$ side. Without the parity constraint, the local generating function is
\begin{equation}
F(x,y)
= \sum_{a=0}^{d_v^H} \sum_{b=0}^{d^G}
\binom{d_v^H}{a}\binom{d^G}{b} x^a y^b
= (1+x)^{d_v^H}(1+y)^{d^G}.
\end{equation}
To keep only those local patterns for which the total parity is even, namely $a+b \equiv 0 \pmod 2$, we apply the standard parity projection and obtain
\begin{equation}
F_{\mathrm{+}}(x,y)
= \frac{F(x,y)+F(-x,-y)}{2}
= \frac{(1+x)^{d_v^H}(1+y)^{d^G} + (1-x)^{d_v^H}(1-y)^{d^G}}{2}.
\end{equation}
Since the $n$ variable positions contribute independently in the ensemble counting argument, the total number of valid configurations is obtained by extracting the coefficient of $x^{\ell d_c^H} y^{w d^G}$ from $F_{\mathrm{+}}(x,y)^n$. Thus,
\begin{equation}
\#{\text{valid configurations}} = \mathrm{coef}\Big(
F_{\mathrm{+}}(x,y)^n,
 x^{\ell d_c^H} y^{w d^G}
\Big).
\end{equation}
Therefore,
\begin{equation}
P(xH+cG^T=0) = \frac{
\mathrm{coef}\Big(
F_{\mathrm{+}}(x,y)^n,
 x^{\ell d_c^H} y^{w d^G}
\Big)
}{
\binom{n d_v^H}{\ell d_c^H}\binom{n d^G}{w d^G}
}.
\end{equation}
Hence the average input-output weight enumerator can be written as
\begin{equation}
\overline{A}(\ell,w)= \binom{n-k}{\ell}\binom{n}{w}
\frac{
\mathrm{coef}\Big(
F_{\mathrm{+}}(x,y)^n,
 x^{\ell d_c^H} y^{w d^G}
\Big)
}{
\binom{n d_v^H}{\ell d_c^H}\binom{n d^G}{w d^G}
}.
\label{eq:dual_ldpcgm_average_io}
\end{equation}

\paragraph{Asymptotic exponent}
Define the normalized quantities $\hat{\ell} = \frac{\ell}{n-k},$ $\hat{w} = \frac{w}{n}.$
Since for the regular parity-check ensemble
$
\frac{n-k}{n} = \frac{d_v^H}{d_c^H},
$
we have $\frac{\ell d_c^H}{n} = \hat{\ell} d_v^H,$ $\frac{w d^G}{n} = \hat{w} d^G.  $
We obtain the asymptotic growth rate
\begin{equation}
 g(\hat{\ell},\hat{w})
:= \lim_{n\to\infty} \frac{1}{n}\log_2 \overline{A}(\ell,w).   
\end{equation}

Applying Stirling's approximation and the saddle-point expression to the
coefficient in~\eqref{eq:dual_ldpcgm_average_io} yields
\begin{equation}
g(\hat\ell,\hat w)
=
(\frac{d_v^H}{d_c^H} -d_v^H)h_2(\hat\ell)
+(1-d^G)h_2(\hat w)+
\inf_{x>0,\,y>0}
\log_2\left(
\frac{F_+(x,y)}
{x^{\hat\ell d_v^H}y^{\hat w d^G}}
\right)
\label{eq:dual_ldpcgm_exp}
\end{equation}

Let $\overline N_{\cC}(w)$ be the average weight distribution in the dual ensemble. By counting pairs $(x,c)$ we can get,
\begin{equation}
\overline N_{\cC}(w)
\leq
\sum_{\ell=0}^{n-k}\overline A(\ell,w).
\end{equation}
Therefore, the output-weight average weight distribution growth rate satisfies
\begin{equation}
g(\hat w)
= \lim_{n\to\infty}\frac1n\log_2\overline N_{\cC}(w) \leq
\max_{\hat\ell\in[0,1]}g(\hat\ell,\hat w).
\end{equation}

\end{proof}

 \section{ Proof of Lemma~\ref{lem:lemma3}}
 \label{app:lemma3}

\subsection{Proof of Statement~1}

\begin{proof} 
Let $d_v^H,d_c^H,d^G$ be fixed finite degree parameters, with $R=d_v^H/d_c^H\in(0,1)$. For the existence assertion, we select $d_v^H$ and $d^G$ to be even, with $d_c^H>2$ and $d^G>2$.
Choose and fix $0<\delta<h_2^{-1}(1-R)$.

Fix an output weight $\hat w\in[0,1]$. We decompose the maximization in asymptotic growth rate derived in Lemma~\ref{lem:gira} over
$\hat \ell$ into  an interior region
and its complement:
\begin{equation}
\mathcal{B}(\gamma)=[\gamma,1-\gamma],
\qquad
\mathcal{B}^c(\gamma)=[0,\gamma)\cup(1-\gamma,1],
\end{equation}
where we now take $\gamma=(d_v^H)^{-1/2}$. 
More explicitly, the maximization in Lemma~\ref{lem:gira} can be decomposed as
\begin{equation}
\max_{\hat \ell\in[0,1]}
g(\hat \ell,\hat w)
=
\max\left\{
\sup_{\hat \ell\in\mathcal{B}(\gamma)}g(\hat \ell,\hat w),
\;
\sup_{\hat \ell\in\mathcal{B}^c(\gamma)}g(\hat \ell,\hat w)
\right\}.
\label{eq:dual_ldpcgm_region_split}
\end{equation}
We upper bound the two terms separately.

Using the growth rate expression derived in Lemma~\ref{lem:gira}, we can further upper bound $g(\hat{\ell},\hat{w})$ by evaluating the infimum at $x=\hat{\ell}/(1-\hat{\ell})$ and $y=\hat{w}/(1-\hat{w})$. This yields
\begin{equation}
\begin{aligned}
g(\hat{\ell},\hat{w})
&\leq {}
\frac{d_v^H}{d_c^H}h_2(\hat{\ell})
+h_2(\hat{w})
+\log_2\left(
1+(1-2\hat{\ell})^{d_v^H}
(1-2\hat{w})^{d^G}
\right)-1
\label{eq:dual_ldpcgm_inner_bound}
\\
&\triangleq {} \phi(\hat{\ell},\hat{w}).
\end{aligned}
\end{equation} 

\paragraph{Case 1: $\hat \ell \in\mathcal{B}(\gamma)$.}
We will prove this result by proving concavity of the function $\phi(\hat{\ell},\hat{w})$ wrt $\hat{\ell}$. 
The first derivative of $\phi(\hat \ell, \hat w)$ in \eqref{eq:dual_ldpcgm_inner_bound} with respect to $\hat \ell$ is
\begin{equation}
\frac{\partial \phi(\hat \ell,\hat w)}{\partial \hat \ell}
=
\frac{d_v^H}{d_c^H}
\log_2\left(\frac{1-\hat \ell}{\hat \ell}\right)
-
\frac{
2d_v^H(1-2\hat w)^{d^G}(1-2\hat \ell)^{d_v^H-1}
}{
\ln 2\left(
1+(1-2\hat \ell)^{d_v^H}(1-2\hat w)^{d^G}
\right)
}.
\label{eq:dual_ldpcgm_phi_first_derivative}
\end{equation}
At $\hat \ell=1/2$, both terms vanish. Hence
\begin{equation}
\left.
\frac{\partial \phi(\hat \ell,\hat w)}{\partial \hat \ell}
\right|_{\hat \ell=1/2}
=0.
\end{equation}
Therefore, $\hat \ell=1/2$ is a critical point.
The second derivative is
\begin{equation}
\begin{aligned}
\frac{\partial^2 \phi(\hat \ell,\hat w)}{\partial \hat \ell^2}
&=
-\frac{d_v^H}{d_c^H}
\frac{1}{\ln 2}
\frac{1}{\hat \ell(1-\hat \ell)}
\\
&\quad
+
\frac{
4d_v^H(1-2\hat w)^{d^G}
\left(d_v^H-1-(1-2\hat w)^{d^G}(1-2\hat \ell)^{d_v^H}\right)
(1-2\hat \ell)^{d_v^H-2}
}{
\ln 2
\left(
1+(1-2 \hat \ell)^{d_v^H}(1-2\hat w)^{d^G}
\right)^2
}.
\label{eq:dual_ldpcgm_phi_second_derivative}
\end{aligned}
\end{equation}
Because the selected degrees $d_v^H,d^G$ are even, it is true that $0\leq(1-2\hat w)^{d^G}\leq1$, and the denominator of the second term is at least $\ln 2$. For $\hat\ell\in[\gamma,1-\gamma]$ we therefore have
\begin{equation}
\frac{\partial^2\phi}{\partial\hat\ell^2}
\leq-\frac{4R}{\ln2}
+\frac{4(d_v^H)^2}{\ln2}(1-2\gamma)^{d_v^H-2}.
\label{eq:dual_ldpcgm_uniform_curvature}
\end{equation}
With $\gamma=(d_v^H)^{-1/2}$, the positive term tends to zero as $d_v^H$ increases. Thus there is a finite choice of $d_v^H$ for which the right-hand side is negative. Once this finite degree is fixed, $\phi(\hat\ell,\hat w)$ is strictly concave in $\hat\ell$ on $\mathcal{B}(\gamma)$, uniformly in $\hat w$. Its critical point $\hat\ell=1/2$ is consequently its maximizer, so
\begin{equation}
\sup_{\hat \ell\in\mathcal{B}(\gamma)}
 g(\hat \ell,\hat w)
\le
\phi\left(\frac12,\hat w\right)
=
R+h_2(\hat w)-1.   
\label{eq:dual_ldpcgm_interior_bound}
\end{equation}
\noindent

\paragraph{Case 2: $\hat \ell \in\mathcal{B}^c(\gamma)$.}
It remains to study the two boundary intervals where $\hat \ell$ is close to zero or
one.
We first evaluate the exponent at the endpoint $\hat{\ell}=0$. 
Using the growth rate expression derived in Lemma~\ref{lem:gira},  the factor
$x^{\hat{\ell}d_v^H}$ disappears when $\hat{\ell}=0$, and hence
\begin{equation}
\begin{aligned}
g(0,\hat w)
&=
h_2(\hat w)-d^G h_2(\hat w)
\\
&\quad+
\inf_{x>0,y>0}
\log_2
\left(
\frac{
(1+x)^{d_v^H}(1+y)^{d^G}
+
(1-x)^{d_v^H}(1-y)^{d^G}
}{
2y^{\hat w d^G}
}
\right).\\
&=
h_2(\hat w)-d^G h_2(\hat w)
+
\inf_{y>0}
\log_2
\left(
\frac{
(1+y)^{d^G}+(1-y)^{d^G}
}{
2y^{\hat w d^G}
}
\right),
\end{aligned}
\end{equation}
\noindent
since for fixed $y>0$, the infimum over $x$ is attained in the limit $x\to0^+$. This is the average weight distribution growth rate of the corresponding
rate-zero regular LDPC ensemble \cite{gallager1963ldpc}. Therefore,
\begin{equation}
g(0,\hat w)\leq0,
\qquad
\hat w\in[0,1],
\end{equation}
with
\begin{equation}
g(0,\hat w)=0
\qquad
\text{for }
\hat w\in\left\{0,\frac12,1\right\},
\end{equation}
and
\begin{equation}
g(0,\hat w)<0
\qquad
\text{for }
\hat w\in
(0,1)\setminus\left\{\frac12\right\}.
\label{eq:rate_zero_strict_negative}
\end{equation}

We now study the exponent for small positive values of $\hat{\ell}$.
The proof is divided according to the value of $\hat w$.

\textbf{Case 2(a): The corners $\hat w\in \{0,1\}$.} To analyze the corner $(\hat \ell, \hat w)=(0,0)$, it is sufficient to study the
directional derivative along an arbitrary ray. For a fixed interior direction \(\alpha,\beta>0\), let
\((x_t,y_t)\) denote the minimizing saddle point corresponding to
\((\hat{\ell},\hat{w})=(\alpha t,\beta t)\). The stationarity equations are
\begin{equation}
\alpha d_v^H t
=
\frac{x_t\,\partial_x F_+(x_t,y_t)}{F_+(x_t,y_t)},
\qquad
\beta d^G t
=
\frac{y_t\,\partial_y F_+(x_t,y_t)}{F_+(x_t,y_t)}.
\label{eq:corner_stationarity}
\end{equation}
Near the origin,
\begin{equation}
F_+(x,y)
=
1+\binom{d_v^H}{2}x^2
+d_c^H d^Gxy
+\binom{d^G}{2}y^2
+O\bigl((x+y)^4\bigr).
\label{eq:F_corner_expansion}
\end{equation}
Substituting \eqref{eq:F_corner_expansion} into
\eqref{eq:corner_stationarity} shows that, for every fixed
\(\alpha,\beta>0\),
\begin{equation}
x_t=\Theta(\sqrt{t}),
\qquad
y_t=\Theta(\sqrt{t}).
\label{eq:corner_saddle_scaling}
\end{equation} 
Consequently,
\begin{equation}
\phi_{\alpha,\beta}(t) \triangleq g(\alpha t,\beta t)
=
-K(\alpha,\beta)t\log_2\frac1t+O(t),
\end{equation}
where
\begin{equation}
K(\alpha,\beta)
=
d_v^H\left(\frac12-\frac1{d_c^H}\right)\alpha
+
\left(\frac{d^G}{2}-1\right)\beta.
\end{equation}
Therefore,
\begin{equation}
\phi_{\alpha,\beta}'(t)
=
\alpha\frac{\partial g}{\partial \hat \ell}
+
\beta\frac{\partial g}{\partial \hat w}
=
-K(\alpha,\beta)\log_2\frac1t+O(1).
\end{equation}
If $d_c^H>2$ and $d^G>2$, then
$K(\alpha,\beta)>0$ for every nonzero direction
$(\alpha,\beta)\in\mathbb{R}_+^2$. Hence,
\begin{equation}
\phi_{\alpha,\beta}'(t)\to-\infty
\qquad\text{as }t\to0^+.
\end{equation}
Thus, the exponent decreases from $(0,0)$ along every direction in the
nonnegative quadrant. And, there exists a neighborhood of $(0,0)$ in which
\begin{equation}
g(\hat\ell,\hat w)<0
\end{equation}
at every point other than the origin.

The same argument applies to the corner
$(\hat\ell,\hat w)=(0,1)$ after writing
\begin{equation}
\hat\ell=\alpha t,
\qquad
1-\hat w=\beta t.
\end{equation}
The corner cases with $\hat\ell=1$ follow by symmetry.

The preceding corner analysis is uniform over all directions. Hence, there
exist sufficiently small constant $\eta>0$ and large enough $d_v^H$ such that 
\begin{equation}
\sup_{\hat{\ell}\in\mathcal{B}^{c}(\gamma)}
g(\hat{\ell},\hat{w})
\begin{cases}
=0, & \hat{w}\in\{0,1\},\\[1mm]
<0, & \hat{w}\in(0,\eta]\cup[1-\eta,1).
\end{cases}
\label{eq:corner_bound}
\end{equation}

By decreasing $\eta$ if necessary, we may also assume that
$0<\eta<\delta.$

\textbf{Case 2(b): Normalized weight $\hat{w}$ bounded away from $0$ and $1$.}  

We next consider \(\hat w\in[\eta,1-\eta]\) where $\eta$ is fixed in Case 2(a).
By further upper bounding the combinatorial upper bound in \eqref{eq:dual_ldpcgm_inner_bound},  we have
\begin{equation}
g(\hat\ell,\hat w)\leq
h_2(\hat\ell)
+
h_2(\hat w)
+
\log_2\left(
1+(1-2\hat w)^{d^G}
\right)
-1.
\label{eq:boundary_combinatorial_bound}
\end{equation}
Define
\begin{equation}
B_{d^G}(\hat w)
\triangleq
h_2(\hat w)
+
\log_2\left(
1+(1-2\hat w)^{d^G}
\right)
-1.
\end{equation}
Since $d^G$ is even, $B_{d^G}(\hat w)$ is symmetric around $\hat w=1/2$. Hence, it is sufficient to consider $\hat w\in[0,1/2]$. 
Let
$t=1-2\hat w\in[0,1].$
Using
\begin{equation}
1-h_2(\hat w)\geq \frac{t^2}{2\ln 2}
\end{equation}
and $\ln(1+x)\leq x$, we obtain 
\begin{equation}
\begin{aligned}
B_{d^G}(\hat w)
&=
h_2(\hat w)-1+\log_2\left(1+t^{d^G}\right)\\
&\leq
-\frac{t^2}{2\ln 2}
+\frac{t^{d^G}}{\ln 2}\\
&=
\frac{t^2}{\ln 2}
\left(
t^{d^G-2}-\frac{1}{2}
\right).
\end{aligned}
\end{equation}

We will first consider $\eta\leq\hat w\leq\delta<1/2$, so we have
$0<1-2\delta\leq t\leq1-2\eta<1.$
We require the even degree $d^G>2$ to be sufficiently large that
$(1-2\eta)^{d^G-2}<\frac12.$
Then $t^{d^G-2} \leq(1-2\eta)^{d^G-2}<\frac12,$
and hence
\begin{equation}
B_{d^G}(\hat w)<0,
\qquad \eta\leq\hat w\leq\delta.
\end{equation}
Because $B_{d^G}$ is continuous on the compact interval $[\eta,\delta]$, its maximum on this interval is strictly negative. Thus, the quantity
$m_\delta
\triangleq
-\max_{\hat w\in[\eta,\delta]}
B_{d^G}(\hat w)$
is positive and satisfies
\begin{equation}
B_{d^G}(\hat w)\leq-m_\delta,
\qquad
\eta\leq\hat w\leq\delta.
\label{eq:B_uniform_negative}
\end{equation}

We next consider $\hat w\in[\delta,1-\delta]$.
Since $d^G$ is even and $|1-2\hat w|\leq1-2\delta<1,$ we have
\begin{equation}
\begin{aligned}
B_{d^G}(\hat w)
&\leq
h_2(\hat w)-1+
\log_2\left(1+(1-2\delta)^{d^G}\right).
\end{aligned}
\label{eq:B_middle_bound}
\end{equation}
We also require $d^G$ to be sufficiently large that
$L_\delta
\triangleq
\log_2\left(1+(1-2\delta)^{d^G}\right)
<R.$

Under these two degree conditions, define
$\varepsilon_*
=
\frac12
\min\left\{
m_\delta,\,
R-L_\delta
\right\}>0.$
Since $\gamma=(d_v^H)^{-1/2}$, we require $d_v^H$ sufficiently large that $h_2(\gamma)\leq\varepsilon_*.$
For $\hat\ell\in\mathcal B^c(\gamma)$, the symmetry and monotonicity of binary entropy give $h_2(\hat\ell)\leq h_2(\gamma) \leq\varepsilon_*.$
Consequently,
\begin{equation}
g(\hat\ell,\hat w)
\leq
\varepsilon_*+B_{d^G}(\hat w).
\label{eq:boundary_epsilon_star}
\end{equation}

For $\hat w\in[\eta,\delta]$,
\eqref{eq:B_uniform_negative} gives
\begin{equation}
\begin{aligned}
g(\hat\ell,\hat w)\leq\varepsilon_*-m_\delta \leq-\frac{m_\delta}{2}<0.
\end{aligned}
\label{eq:case2b_negative}
\end{equation}

On the other hand, for $\hat w\in[\delta,1-\delta]$, \eqref{eq:B_middle_bound} yields
\begin{equation}
\begin{aligned}
g(\hat\ell,\hat w)
&\leq
\varepsilon_*+
h_2(\hat w)-1+L_\delta\\
&=
h_2(\hat w)-(1-R)
-\left(R-L_\delta-\varepsilon_*\right)\\
&\leq
h_2(\hat w)-(1-R),
\end{aligned}
\label{eq:case2b_random_bound}
\end{equation}
where the last inequality follows from
$\varepsilon_*\leq\frac{R-L_\delta}{2}.$

By symmetry, the strict negative bound also holds for
$\hat w\in[1-\delta,1-\eta]$.
We conclude that
\begin{equation}
\sup_{\hat\ell\in\mathcal B^c(\gamma)}
g(\hat\ell,\hat w)
\begin{cases}
<0,
&
\hat w\in[\eta,\delta]\cup
[1-\delta,1-\eta],
\\[1mm]
\leq h_2(\hat w)-(1-R),
&
\hat w\in[\delta,1-\delta].
\end{cases}
\label{eq:boundary_bound}
\end{equation}

\textbf{Combining Cases 2(a) and 2(b).}
Combining \eqref{eq:corner_bound} and \eqref{eq:boundary_bound}, for sufficiently large $d^G$ even and $d_v^H$ 
we obtain
\begin{equation}
\sup_{\hat\ell\in\mathcal B^c(\gamma)}
g(\hat\ell,\hat w)
\begin{cases}
=0,
&
\hat w\in\{0,1\},
\\[1mm]
<0,
&
\hat w\in(0,\delta]\cup[1-\delta,1),
\\[1mm]
\leq h_2(\hat w)-(1-R),
&
\hat w\in(\delta,1-\delta).
\end{cases}
\label{eq:bound_case2}
\end{equation}


\paragraph{Combining Cases 1 and 2.} 
Fix $0<\delta<h_2^{-1}(1-R)$ and recall that $\gamma=(d_v^H)^{-1/2}$.
Choose finite degree parameters $d_v^H$ and even $d^G>2$ sufficiently large so that the interior estimate \eqref{eq:dual_ldpcgm_interior_bound} from Case~1 and the boundary estimates \eqref{eq:bound_case2} from Case~2 hold simultaneously.

For these fixed parameters, combining the two cases yields
\begin{equation}
g(\hat w)
\leq
\max\left\{
h_2(\hat w)-(1-R),
\; \sup_{\hat\ell\in\mathcal{B}^c(\gamma)}
g(\hat\ell,\hat w)
\right\},
\qquad
\hat w\in[0,1].
\label{eq:dual_ldpcgm_combined_bound}
\end{equation}
where
$\sup_{\hat\ell\in\mathcal{B}^c(\gamma)}
g(\hat\ell,\hat w)$
is given in \eqref{eq:bound_case2}. We now use these bounds to establish the three claims of Statement~1.

We first consider the endpoint weights. Since the ensemble is linear, the
all-zero codeword is always present and unique. Therefore,
$g(0)=0.$
Similarly, there can be at most one codeword of weight $n$, namely the
all-one vector, and hence
$g(1)\leq0.$
This proves the first line of the theorem.

Now consider $\hat w\in(0,\delta]\cup[1-\delta,1)$. 
Recall that $0<\delta<h_2^{-1}(1-R).$

We first consider $\hat w\in(0,\delta]$.
Since $h_2(x)$ is strictly increasing on $[0,1/2]$, we have
$h_2(\hat w)
\leq h_2(\delta)
<h_2\left(h_2^{-1}(1-R)\right)
=1-R.$
Therefore,
\begin{equation}
h_2(\hat w)-(1-R)<0,
\qquad 0<\hat w\leq\delta.
\label{eq:interior_negative_combined}
\end{equation}
Combining this inequality with the results from the boundary region in \eqref{eq:bound_case2}, both terms in the maximization in \eqref{eq:dual_ldpcgm_combined_bound} are strictly negative.
It follows that
\begin{equation}
\begin{aligned}
g(\hat w)
&=
\max\left\{
\sup_{\hat\ell\in\mathcal B(\gamma)}
g(\hat\ell,\hat w),
\sup_{\hat\ell\in\mathcal B^c(\gamma)}
g(\hat\ell,\hat w)
\right\}<0,
\qquad 0<\hat w\leq\delta.
\end{aligned}
\end{equation}

For $\hat w\in[1-\delta,1)$, we use the symmetry of binary entropy,
$h_2(\hat w)=h_2(1-\hat w),$
together with the corresponding symmetric
boundary estimates.

The same argument gives
\begin{equation}
g(\hat w)<0,
\qquad 1-\delta\leq\hat w<1.
\end{equation}

Therefore,
\begin{equation}
g(\hat w)<0,
\qquad
\hat w\in(0,\delta]\cup[1-\delta,1).
\end{equation}

This proves the second line of the theorem.

Finally, consider the region
$\delta<\hat w<1-\delta.$
The same upper bound holds over both the interior and boundary regions.
We obtain
\begin{equation}
\begin{aligned}
g(\hat w)
&=
\max\left\{
\sup_{\hat\ell\in\mathcal B(\gamma)}
g(\hat\ell,\hat w),
\sup_{\hat\ell\in\mathcal B^c(\gamma)}
g(\hat\ell,\hat w)
\right\}\\
&\leq
\max\left\{
h_2(\hat w)-(1-R),
h_2(\hat w)-(1-R)
\right\}\\
&=h_2(\hat w)-(1-R).
\end{aligned}
\end{equation}
Therefore,
\begin{equation}
g(\hat w)\leq h_2(\hat w)-(1-R),
\qquad \hat w\in(\delta,1-\delta).
\end{equation}

This establishes the third line of Statement~1.

\end{proof}

It remains to prove Statement~2.

\subsection{Proof of Statement~2}

\begin{proof}

Because $d_v^H$ is even,
\begin{equation}
\overline A(\ell,w)
=
\overline A(m-\ell,w).
\label{eq:C2_l_symmetry_statement2}
\end{equation}
Thus it is sufficient to consider
$0\leq\ell\leq\frac{m}{2}$
and multiply the resulting bound by two.

Choose a sufficiently small fixed constant $\varepsilon>0$ and define
$q:=\ell+w.$
We divide the sum into the two regions
$q\leq\varepsilon n$
and $ q>\varepsilon n.$
The second region will subsequently be divided according to the
size of $w$. Hence,
\begin{equation}
   \sum_{w=1}^{\lfloor\delta n\rfloor}
\overline N_{\cC}(w)\leq \sum_{\substack{
1\leq w\leq\delta n,\;
0\leq\ell\leq m/2\\
\ell+w\leq\varepsilon n}}
\overline A(\ell,w) + \sum_{\substack{
1\leq w\leq\delta n,\;
0\leq\ell\leq m/2\\
\ell+w >\varepsilon n}}
\overline A(\ell,w)
\end{equation}

\paragraph{Case 1: $q=\ell+w\leq\varepsilon n$.}
We first derive an upper bound on \(\overline{A}(\ell,w)\) in this
regime. To this end, we begin by upper-bounding the coefficient term
appearing in \(\overline{A}(\ell,w)\). Let
$E=\ell d_c^H+w d^G.$
Since \(F_+(s,t)^n\) has nonnegative coefficients, setting \(s=t=z\)
and collecting all terms of total degree \(E\) gives
\begin{subequations}
\begin{align}
\left[s^{\ell d_c^H}t^{w d^G}\right]F_+(s,t)^n
&\stackrel{(a)}\leq
\left[z^E\right]F_+(z,z)^n
\\
&\stackrel{(b)}\leq
\frac{F_+(z,z)^n}{z^E}
\\
&\stackrel{(c)}\leq
(1+Bz^2)^n z^{-E}\\
&\stackrel{(d)}\leq
e^{Bn z^2}z^{-E}.
\label{eq:bound_coef}
\end{align}
\end{subequations}
Here, (a) follows because
\(\left[z^E\right]F_+(z,z)^n\) contains the sum of all bivariate
coefficients whose total degree is \(E\), including the coefficient
corresponding to
\((\ell d_c^H,w d^G)\).
Inequality (2b) follows from the standard coefficient bound
$[z^E]f(z)\leq \frac{f(z)}{z^E},
 z>0,$
for a polynomial \(f\) with nonnegative coefficients. Inequality
(c) follows from the bound
$F_+(z,z)\leq 1+Bz^2,$
while (d) follows from \(1+x\leq e^x\).

We now optimize \eqref{eq:bound_coef} over \(z>0\). Choosing
$z^2=\frac{E}{2Bn}$
yields
\begin{equation}
e^{Bn z^2}z^{-E}
=
e^{E/2}
\left(\frac{2Bn}{E}\right)^{E/2}
=
\left(\frac{2eBn}{E}\right)^{E/2}.
\end{equation}
Therefore,
\begin{equation}
 \left[s^{\ell d_c^H}t^{w d^G}\right]F_+(s,t)^n
\leq
\left(\frac{2eBn}{E}\right)^{E/2}.
\label{eq:last_coef_bound}
\end{equation}

We next bound the binomial terms appearing in
\(\overline{A}(\ell,w)\). Using the standard inequalities
\begin{equation}
\binom{N}{k}\leq \left(\frac{eN}{k}\right)^k,
\qquad
\binom{N}{k}\geq \left(\frac{N}{k}\right)^k,    
\end{equation}
we obtain
\begin{equation}
 \frac{\binom{m}{\ell}}
{\binom{m d_c^H}{\ell d_c^H}}
\leq
\frac{\left(\frac{em}{\ell}\right)^\ell}
     {\left(\frac{m}{\ell}\right)^{\ell d_c^H}}
=
e^\ell
\left(\frac{\ell}{m}\right)^{\ell(d_c^H-1)}.   
\label{eq:bound_binomialHside}
\end{equation}
Similarly,
\begin{equation}
  \frac{\binom{n}{w}}
{\binom{n d^G}{w d^G}}
\leq
e^w
\left(\frac{w}{n}\right)^{w(d^G-1)}.  
\label{eq:bound_binomialGside}
\end{equation}

Combining \eqref{eq:last_coef_bound}, \eqref{eq:bound_binomialHside}, and \eqref{eq:bound_binomialGside} gives
\begin{align}
\overline{A}(\ell,w)
&\leq
e^{\ell+w}
\left(\frac{\ell}{m}\right)^{\ell(d_c^H-1)}
\left(\frac{w}{n}\right)^{w(d^G-1)}
\left(\frac{2eBn}{E}\right)^{E/2}.
\end{align}

Let
$q:=\ell+w,$
$\alpha:=\frac{d_c^H}{2}-1,$
$\beta:=\frac{d^G}{2}-1,$ and
$\mu:=\min\{\alpha,\beta\}.$
By collecting all constants independent of \(n,\ell,w\) into a
constant \(C>0\), the preceding bound can be written as
\begin{equation}
\overline{A}(\ell,w)
\leq
C^q
\left(\frac{q}{n}\right)^{\alpha\ell+\beta w}.    
\end{equation}
Since \(w\geq 1\) and \(q=\ell+w\),
$\alpha\ell+\beta w
=
\beta+\alpha\ell+\beta(w-1)
\geq
\beta+\mu(q-1).$
Therefore, for \(q\leq \varepsilon n\),
\begin{subequations}
\begin{align}
\overline{A}(\ell,w)
&\leq
C^q
\left(\frac{q}{n}\right)^{\beta+\mu(q-1)}
\\
&=
C n^{-\beta}q^\beta
\left[
C\left(\frac{q}{n}\right)^\mu
\right]^{q-1}.
\end{align}
\end{subequations}
Choose \(\varepsilon>0\) sufficiently small so that
$\rho:=C\varepsilon^\mu<1.$
Since \(q\leq\varepsilon n\), we have
\begin{equation}
C\left(\frac{q}{n}\right)^\mu\leq \rho,    
\end{equation}
and hence
\begin{equation}
 \overline{A}(\ell,w)
\leq
C n^{-\beta}q^\beta\rho^{q-1}.   
\end{equation}

For each fixed \(q\), there are at most \(q\) pairs
\((\ell,w)\) satisfying \(\ell\geq0\), \(w\geq1\), and
\(\ell+w=q\). Thus,
\begin{subequations}
 \begin{align}
\sum_{\substack{1\leq w\leq\delta n,\;0\leq\ell\leq m/2\\
                 \ell+w\leq\varepsilon n}}
\overline{A}(\ell,w)
&\leq
C n^{-\beta}
\sum_{q=1}^{\lfloor\varepsilon n\rfloor}
q^{\beta+1}\rho^{q-1}
\\
&\leq
C n^{-\beta}
\sum_{q=1}^{\infty}
q^{\beta+1}\rho^{q-1}\\
&=O(n^{-\beta})
\label{eq:case1_thm1}
\end{align}   
\end{subequations}
where the last equality follows from the fact that, since \(0<\rho<1\), the last series converges to a finite constant.

\paragraph{Case 2: \(q=\ell+w>\varepsilon n\).} 
In this regime, the total weight \(q\) is linear in \(n\). We further
divide the analysis according to the size of \(w\). The reason for
this split is that, since \(q=\ell+w>\varepsilon n\), if \(w\) is
small, then \(\ell\) must necessarily be linear in \(n\), whereas if
\(w\) is bounded away from zero, then \(w\) itself is linear in \(n\).
These two regimes allow us to apply different bounds.

Fix  \(0<\eta<\varepsilon/2\), and consider the following two subcases.

\textbf{Case 2(a): \(q>\varepsilon n\) and \(1\leq w<\eta n\).}
 Since \(q=\ell+w>\varepsilon n\),
\begin{equation}
 \ell=q-w>(\varepsilon-\eta)n>\frac{\varepsilon}{2}n.   
\end{equation}
Thus, although \(w\) may be small, the input weight \(\ell\) is necessarily
linear in \(n\). Let
$\hat\ell=\frac{\ell}{m},$
$\hat w=\frac{w}{n},$
and
$R=\frac{m}{n}.$
Then
$\hat\ell>\frac{\varepsilon}{2R}=:a,$
and
$0<\hat w<\eta.$
Since we restrict to \(\ell\leq m/2\), we have
$a\leq\hat\ell\leq\frac12.$

Using the entropy bounds for the binomial coefficients together with
the coefficient bound obtained by substituting
$s=\frac{\hat\ell}{1-\hat\ell},$
and
$t=\frac{\hat w}{1-\hat w},$
we obtain
\begin{equation}
\overline A(\ell,w)
\leq
(nd_v^H+1)(nd^G+1)
2^{\,n\phi(\hat\ell,\hat w)}, 
\label{eq:exp_bpund_c2a}
\end{equation}
where
\begin{equation}
\phi(\hat\ell,\hat w)
=
R h_2(\hat\ell)+h_2(\hat w)
+\log_2\!\left(
1+(1-2\hat\ell)^{d_v^H}
  (1-2\hat w)^{d^G}
\right)-1.    
\end{equation}

We now show that this exponent is uniformly negative in the present
region. Since
$a\leq\hat\ell\leq\frac12,$
we have
$0\leq1-2\hat\ell\leq1-2a=:\theta<1.$
Moreover, since \(0\leq\hat w<\eta<1/2\),
$0\leq(1-2\hat w)^{d^G}\leq1.$
Hence
\begin{subequations}
 \begin{align}
\phi(\hat\ell,\hat w)
&\leq
R h_2(\hat\ell)+h_2(\hat w)
+\log_2\!\left(1+\theta^{d_v^H}\right)-1
\\
&\leq
R+h_2(\eta)
+\log_2\!\left(1+\theta^{d_v^H}\right)-1.
\label{eq:phi_bpund}
\end{align}   
\end{subequations}

Since \(\theta<1\), by choosing \(d_v^H\) sufficiently large,
$\log_2\!\left(1+\theta^{d_v^H}\right)
\leq\frac{1-R}{4}.$
We then choose \(\eta>0\) sufficiently small, while still satisfying
\(\eta<\varepsilon/2\), such that
$h_2(\eta)\leq\frac{1-R}{4}.$
It follows from \eqref{eq:phi_bpund} that
\begin{equation}
\phi(\hat\ell,\hat w)
\leq
-\frac{1-R}{2}
=: -\kappa<0.    
\end{equation}
Therefore,
\begin{equation}
\overline A(\ell,w)
\leq
(nd_v^H+1)(nd^G+1)2^{-\kappa n}.    
\end{equation}

Finally, there are at most
$\left(\frac{m}{2}+1\right)\lfloor\eta n\rfloor$
pairs \((\ell,w)\) in this region. Consequently,
\begin{subequations}
\begin{align}
\sum_{\substack{q>\varepsilon n\\
                 1\leq w<\eta n\\
                 0\leq\ell\leq m/2}}
\overline A(\ell,w)
&\leq
\left(\frac{m}{2}+1\right)
\lfloor\eta n\rfloor
(nd_v^H+1)(nd^G+1)
2^{-\kappa n}
\\
&=e^{-\Omega(n)}
=o(n^{-\beta}).
\end{align}    
\end{subequations}

\textbf{Case 2(b): $\eta n\leq w\leq\delta n$.} 
In this case, \(w\) itself is linear in \(n\), and the corresponding
linear-weight bound can be applied directly.

Since \(g(\hat w)<0\) for every \(\hat w\in(0,\delta]\) and
\([\eta,\delta]\subset(0,\delta]\) is compact, continuity of \(g\)
implies that there exists \(\kappa_3>0\) such that
\begin{equation}
 g(\hat w)\leq-\kappa_3,
\qquad
\eta\leq\hat w\leq\delta.   
\end{equation}

Consequently, uniformly for
$\eta n\leq w\leq\delta n,$
the finite-length weight enumerator satisfies
\begin{equation}
\overline N_{\cC}(w)
\leq
\operatorname{poly}(n)\,2^{-\kappa_3n}.
\label{eq:C2_finite_from_statement2}
\end{equation}
Therefore
\begin{subequations}
 \begin{align}
\sum_{w=\lceil\eta n\rceil}^{\lfloor\delta n\rfloor}
\overline N_{\cC}(w)
&\leq
O(n)\operatorname{poly}(n)2^{-\kappa_3n}\\
&=
2^{-\Omega(n)}\\
&=
o(n^{-\beta}).
\label{eq:C2_S3}
\end{align}   
\end{subequations}

Combining Cases~2(a) and~2(b), we obtain
\begin{subequations}
\begin{align}
    \sum_{\substack{\ell+w>\varepsilon n\\
1\leq w\leq\delta n\\
0\leq\ell\leq m/2}}
\overline{A}(\ell,w) &= \sum_{\substack{q>\varepsilon n\\
                 1\leq w<\eta n\\
                 0\leq\ell\leq m/2}}
\overline A(\ell,w) + \sum_{\substack{q>\varepsilon n\\
    \eta n\leq w\leq\delta n\\
                 0\leq\ell\leq m/2}}
\overline A(\ell,w) \\
&= o(n^{-\beta}) + o(n^{-\beta}) \\
&= o(n^{-\beta})
\label{eq:case2_thm1}
\end{align}
\end{subequations}

\paragraph{Combining the two cases.}

From~\eqref{eq:case1_thm1}, and
\eqref{eq:case2_thm1},
\begin{subequations}
 \begin{align}
\sum_{w=1}^{\lfloor\delta n\rfloor}
\overline N_{\cC}(w)
&\leq
O(n^{-\beta})
+
o(n^{-\beta})
\nonumber\\
&=
O(n^{-\beta}).
\label{eq:C2_low_sum_final}
\end{align}   
\end{subequations}

Furthermore, since every $\overline N_{C}(w)$ is nonnegative, for every
$1\leq w\leq\delta n,$
$\overline N_{\cC_2}(w)
=
O(n^{-\beta}),
\qquad
1\leq w\leq\delta n.$
Finally, when $d^G$ is even,
$\mathbf 1_n G^T=0.$
Thus, if $c\in \cC_{GIRA}$, then $c+\mathbf 1_n\in \cC_{GIRA}$ as well.
This gives a weight-complementing bijection and therefore
$N_{\cC}(w)=N_{\cC}(n-w)$
for every realization of the ensemble. Taking expectations gives
$\overline N_{\cC}(w)
=
\overline N_{\cC}(n-w).$
Consequently, for every $n-\delta n\leq w\leq n-1,$
$\overline N_{\cC}(w)
=
O(n^{-\beta}).$ This completes the theorem.

\end{proof}


\section{Proof of Theorem~\ref{thm:thm3}}
\label{app:ach_sum_gira_ldpc}

\subsection{Proof of Statement~1}

\begin{proof}

We first decompose the feasible region $\mathcal{D}_a$ in \eqref{eq:sum_codes_feasible_region} into the part
lying inside the box
\begin{equation}
    \mathcal{B}(\gamma)
    =
    \{(x,y): x,y\in[\gamma,1-\gamma]\},
\end{equation}
and its complement $\mathcal{B}^c(\gamma)$, where $\gamma>0$ can be chosen
arbitrarily close to zero. More explicitly, the maximization in \eqref{eq:sum_codes_growth} can be decomposed as
\begin{equation}
\begin{aligned}
\max_{(x,y)\in\mathcal{D}_a}\Psi_a(x,y)
=
\max\Bigg\{
&\sup_{(x,y)\in\mathcal{D}_a\cap\mathcal{B}(\gamma)}
\Psi_a(x,y),
&\sup_{(x,y)\in\mathcal{D}_a\cap\mathcal{B}^c(\gamma)}
\Psi_a(x,y)
\Bigg\}.
\end{aligned}
\label{eq:Da_split_five_regions_24}
\end{equation}
We upper bound these two terms separately.
\paragraph{Case~1: $(x,y)\in \mathcal{D}_a \cap \mathcal{B}(\gamma)$.} For a fixed normalized output weight $a\in[0,1]$, $x$ and $y$ denote the normalized weights of GIRA and LDPC components respectively. In this region, both $x$ and $y$ are bounded away from end points 0 and 1. This allows us to analyze the maximum of $\kappa(x,y)$ given in \eqref{eq:kappa} over this region using its first and second order derivatives. For GIRA component, we use the bound \eqref{eq:dual_ldpcgm_inner_bound} derived in the previous section. For the LDPC component, we use the standard upper bound \eqref{eq:ldpc_ub}. Substituting them into
\eqref{eq:sum_codes_pair_exponent},
we obtain
\begin{equation}
\Psi_a(x,y)
\le
\max_{\ell\in[0,1]}f(\ell,x,y),
\label{eq:ub_sum_ldgmpc_ldpc}
\end{equation}
\noindent
where
\begin{equation}
\begin{aligned}
f(\ell,x,y)
&=
\frac{d_v^{H_1}}{d_c^{H_1}}h_2(\ell)
+
h_2(x)
+
\log_2\left(
1+(1-2\ell)^{d_v^{H_1}}(1-2x)^{d^{G_1}}
\right)
-1  
+
\frac{d_v^{\tilde H}}{d_c^{\tilde H}}
\log_2\left(
\frac{1+(1-2y)^{d_c^{\tilde H}}}{2}
\right) \\
&\quad
+
x h_2\left(\frac{x+y-a}{2x}\right)
+
(1-x)h_2\left(\frac{-x+y+a}{2(1-x)}\right)
\\
& \leq \frac{d_v^{H_1}}{d_c^{H_1}}h_2(\ell)
+
h_2(x)
+
\log_2\left(
\frac{1+(1-2x)^{d^{G_1}}}{2}
\right) 
+
\frac{d_v^{\tilde H}}{d_c^{\tilde H}}
\log_2\left(
\frac{1+(1-2y)^{d_c^{\tilde H}}}{2}
\right) \\
&\quad
+
x h_2\left(\frac{x+y-a}{2x}\right)
+
(1-x)h_2\left(\frac{-x+y+a}{2(1-x)}\right)
\end{aligned}
\label{eq:f_dual_ldpcgm_ldpc}
\end{equation}
where the upper bound follows from $-1<(1-2\ell)^{d_v^{H_1}}<1 $. Hence in the resulting upper bound the only term with $\ell$ is $h_2(\ell)$ which is maximized at $\ell = 1/2$. Then,
\begin{equation}
\begin{aligned}
\Psi_a(x,y)
\le
\kappa(x,y)= \frac{d_v^{H_1}}{d_c^{H_1}} + E(x,y)+\theta_{G_1}(x)+\theta_{\tilde H}(y),
\label{eq:kappa}
\end{aligned}
\end{equation}
where
\begin{subequations}
\begin{equation}
E(x,y) = 
h_2(x)
+
x h_2\left(\frac{x+y-a}{2x}\right)
+
(1-x)h_2\left(\frac{-x+y+a}{2(1-x)}\right),  
\end{equation}
\begin{equation}
\theta_{G_1}(x)
=
\log_2\left(
\frac{1+(1-2x)^{d^{G_1}}}{2}
\right),
\label{eq:theta_ldgmpc}
\end{equation}
\begin{equation}
\theta_{\tilde H}(y)
=\frac{d_v^{\tilde H}}{d_c^{\tilde H}}
\log_2\left(
\frac{1+(1-2y)^{d_c^{\tilde H}}}{2}
\right) .
\label{eq:theta_ldpc}
\end{equation}
\end{subequations}
Then the Hessian of $\kappa(x,y)$ can be written as
\begin{equation}
\nabla^2 \kappa(x,y)
=
\nabla^2 E(x,y)
+
\begin{bmatrix}
\theta_{G_1}''(x) & 0\\
0 & \theta_{\tilde H}''(y)
\end{bmatrix}.
\label{eq:hessian_sumldgmpc_ldpc}
\end{equation}

We first analyze the Hessian of $E(x,y)$. Define
\begin{equation}
p_1=\frac{x+y-a}{2},
\qquad
p_2=\frac{a-x+y}{2},
\end{equation}
\begin{equation}
p_3=\frac{a+x-y}{2},
\qquad
p_4=\frac{2-a-x-y}{2}.
\end{equation}
On the feasible region $\mathcal{D}_a$, the quantities
$p_1,p_2,p_3,p_4$ are nonnegative and satisfy $p_1+p_2+p_3+p_4=1.$ Therefore, they define a valid probability distribution.
Then, 
\begin{equation}
E(x,y)
=
-\sum_{i=1}^4 p_i\log_2 p_i.
\end{equation}
Calculating the Hessian we obtain
\begin{subequations}
\begin{align}
\nabla^2 E(x,y)
& =
-\frac{1}{\ln 2}
\sum_{i=1}^4
\frac{1}{p_i}
\nabla p_i\nabla p_i^T.&&\\
& =
-\frac{1}{4\ln 2}
\left[
\left(\frac1{p_1}+\frac1{p_4}\right)
\begin{bmatrix}
1 & 1\\
1 & 1
\end{bmatrix}
+
\left(\frac1{p_2}+\frac1{p_3}\right)
\begin{bmatrix}
1 & -1\\
-1 & 1
\end{bmatrix}
\right].
\end{align}
\end{subequations}
\noindent
A direct evaluation of the eigenvalues of $\nabla^2 E(x,y)$ results in
\begin{equation}
\lambda_+(x,y)
=
-\frac{1}{2 \ln2}
\left(
\frac1{p_1}+\frac1{p_4}
\right),
\quad
\lambda_-(x,y)
=
-\frac{1}{2 \ln2}
\left(
\frac1{p_2}+\frac1{p_3}
\right),
\end{equation}

\noindent
Since $p_1+p_4=1-a$, we have
\begin{subequations}
\begin{equation}
\frac1{p_1}+\frac1{p_4}
\ge
\frac{4}{p_1+p_4}
=
\frac{4}{1-a}.
\end{equation}
Similarly, since $p_2+p_3=a$, we have
\begin{equation}
\frac1{p_2}+\frac1{p_3}
\ge
\frac{4}{a}.
\end{equation}
\end{subequations}
Hence,
\begin{subequations}
\begin{equation}
\lambda_+(x,y)
= -
\frac{1}{2 \ln2}
\left(
\frac1{p_1}+\frac1{p_4}
\right)
\leq -
\frac{2}{(1-a) \ln2},
\end{equation}
and
\begin{equation}
\lambda_-(x,y)
=
-\frac{1}{2 \ln 2}
\left(
\frac1{p_2}+\frac1{p_3}
\right)
\leq
-\frac{2}{a \ln2}.
\end{equation}
\end{subequations}

\noindent
Since $\frac{2}{\ln2}\min\left\{\frac1a,\frac1{1-a}\right\} \ge \frac{4}{\ln2}$, we have 
\begin{equation}
\nabla^2 E(x,y)\preceq -\frac{4}{\ln2}I.
\label{eq:ub_hessian_e_sum_ldgm_pc}
\end{equation}

Next we show that the remaining part of the Hessian, coming from asymptotic weight distributions of components, $\theta_{G_1}(x)$ and
$\theta_{\tilde H}(y)$, can be made uniformly small by choosing the check degrees $d_c^{\tilde H}$ and $d^{G_1}$ sufficiently
large. Consider the term $\theta_{G_1}(x)$ in \eqref{eq:theta_ldgmpc}, the second derivative of this term is 
\begin{equation}
\theta''_{G_1}(x)
=
\frac{
4d^{G_1}(1-2x)^{d^{G_1}-2}
\left(d^{G_1}-1-(1-2x)^{d^{G_1}}\right)
}{
\left(1+(1-2x)^{d^{G_1}}\right)^2\ln 2
} 
\end{equation}

\noindent
In the considered region we have (for even $d^{G_1}$)
\begin{align*}
x \in[\gamma,1-\gamma] &\Rightarrow |1-2x| \leq 1-2\gamma \\
&\Rightarrow (1-2x)^{d^{G_1}-2} \leq (1-2\gamma)^{d^{G_1}-2}.
\end{align*}

\noindent
Since $0<1-2\gamma<1$, the quantity
$|1-2x|^{d^{G_1}-2}$, and thus $\theta''_{G_1}(x)$,
decays exponentially fast to zero as $d^{G_1}\to\infty$. Thus there exists large enough $d^{G_1}$ such that \begin{equation}
\theta''_{G_1}(x)< \frac{2}{\ln 2}.
\label{eq:ub_hessian_thetas_ldgmpc_ldpc}
\end{equation}

\noindent
A similar argument holds for $\theta_{\tilde H}(y)$.
Then considering \eqref{eq:ub_hessian_e_sum_ldgm_pc} and \eqref{eq:ub_hessian_thetas_ldgmpc_ldpc}, the Hessian of $\kappa$ in \eqref{eq:hessian_sumldgmpc_ldpc} satisfies

\begin{equation}
\nabla^2 \kappa(x,y)
 \preceq -\frac{2}{\ln2}I .
\end{equation}
Therefore, in the range
of interest
the function $\kappa(x,y)$ is strictly concave for sufficiently large check degrees $d_c^{\tilde H}$ and $d^{G_1}$. 

The next step is to show that $(1/2,1/2)$ is the global maximum over this region. 
We first verify that $(1/2,1/2)$ is a critical point, and then by concavity on this region, we conclude that this local maximum is in fact the global maximum.
The first partial derivatives are
\begin{subequations}
\begin{equation}
\begin{aligned}
\frac{\partial \kappa}{\partial x}
&=
\log_2\left(\frac{1-x}{x}\right)
-
\frac{
2d^{G_1}(1-2x)^{d^{G_1}-1}
}{
\left(1+(1-2x)^{d^{G_1}}\right)\ln 2
} \\
&\quad
+
\frac12
\log_2
\left(
\frac{x^2(a-x+y)(2-a-x-y)}
{(1-x)^2(x+y-a)(a+x-y)}
\right),
\end{aligned}
\end{equation}
and
\begin{equation}
\begin{aligned}
\frac{\partial \kappa}{\partial y}
&=
-
\frac{
2d_v^{\tilde H}(1-2y)^{d_c^{\tilde H}-1}
}{
\left(1+(1-2y)^{d_c^{\tilde H}}\right)\ln 2
} \\
&\quad
+
\frac12
\log_2
\left(
\frac{(a+x-y)(2-a-x-y)}
{(a-x+y)(x+y-a)}
\right).
\end{aligned}
\end{equation}
\end{subequations}
Substituting $(x,y)=(1/2,1/2)$ verifies that it is a critical point. Since $\kappa$ is strictly concave, our critical point is also the unique global maximum in this region. 

Consequently,

\begin{subequations}
\begin{align}
\sup_{(x,y)\in\mathcal{D}_a\cap\mathcal{B}(\gamma)}
\Psi_a(x,y)
&\le
\kappa\left(\frac12,\frac12\right) \\
&=
\frac{d_v^{H_1}}{d_c^{H_1}}+
h_2(a)
-
\frac{d_v^{\tilde H}}{d_c^{\tilde H}}\\
& = h_2(a)
- ( 1- R_{\cC_2^\perp}).
\end{align}
\end{subequations}

\paragraph{Case 2: $(x,y)\in \mathcal{D}_a \cap \mathcal{B}^c(\gamma)$.} 
It remains to study the part of the complement region in \eqref{eq:Da_split_five_regions_24} which can be split into four boundary regions, corresponding to the cases where either $x$ or $y$ is close to $0$ or $1$.
Specifically,
\begin{equation}
\begin{aligned}
&\sup_{(x,y)\in\mathcal{D}_a\cap\mathcal{B}^c(\gamma)}
\Psi_a(x,y)&=
\max\Bigg\{
\sup_{\substack{(x,y)\in\mathcal{D}_a\\0\leq x<\gamma}}
\Psi_a(x,y),
\sup_{\substack{(x,y)\in\mathcal{D}_a\\1-\gamma<x\leq1}}
\Psi_a(x,y),
& 
\sup_{\substack{(x,y)\in\mathcal{D}_a\\0\leq y<\gamma}}
\Psi_a(x,y),
\sup_{\substack{(x,y)\in\mathcal{D}_a\\1-\gamma<y\leq1}}
\Psi_a(x,y)
\Bigg\}.
\end{aligned}
\label{eq:four_boundary_regions_sum_ldgmpc_ldpc}
\end{equation}

The entropy
terms satisfy the basic concavity bound
\begin{equation}
x h_2\left(\frac{x+y-a}{2x}\right)
+
(1-x)h_2\left(\frac{-x+y+a}{2(1-x)}\right)
\le h_2(y).
\label{eq:2case_bound_ldgmpc_ldpc}
\end{equation}
Therefore,  applying this bound to \eqref{eq:sum_codes_pair_exponent}, we obtain
\begin{equation}
\Psi_a(x,y)\leq g_{\cC_1^\perp}(x)+g_{\widetilde{\cC}}(y).
\end{equation}
where $g_{\cC_1^\perp}(.)$ and $g_{\widetilde{\cC}}(.)$ are asymptotic growth rates for $\operatorname{GIRA}(n,d_c^{H_1},d_v^{H_1},d^{G_1})$ and $\operatorname{LDPC}(n,d_v^{\tilde H},d_v^{\tilde H})$ respectively.

We consider the first term in \eqref{eq:four_boundary_regions_sum_ldgmpc_ldpc}, i.e., when $0<x<\gamma$.
In this region we have
\begin{equation}
x\in(0,\gamma),
\qquad
y\in(a-\gamma,a+\gamma).
\end{equation}
The required small-weight and endpoint behavior
of \(g_{\cC_1^\perp}\) follows from the result established in
the previous section. In particular, there exists \(\gamma_1>0\) such that for sufficiently large $d^{G_1}>2$ and $d_v^{H_1}$, $g_{\cC_1^\perp}(x)<0$ is satisfied for all $0<x<\gamma_1.$
Thus, by choosing \(\gamma<\gamma_1\), we obtain
\begin{equation}
\sup_{0\leq x<\gamma}
g_{\cC_1^\perp}(x)
=
0.
\end{equation}
Since \(g_{\widetilde{\cC}}\) is continuous at \(a\), for every
\(\varepsilon_2>0\), there exists \(\gamma_{\varepsilon_2}>0\) such that
$\left|
g_{\widetilde{\cC}}(y)-g_{\widetilde{\cC}}(a)
\right|
<
\varepsilon_2$
whenever
$|y-a|<\gamma_{\varepsilon_2}.$
Therefore, by choosing \(\gamma<\gamma_{\varepsilon_2}\), we obtain
\begin{equation}
 \sup_{a-\gamma<y<a+\gamma}
g_{\widetilde{\cC}}(y)
\leq
g_{\widetilde{\cC}}(a)+\varepsilon_2.
\end{equation}
Combining the two bounds gives
\begin{equation}
\sup_{\substack{(x,y)\in\mathcal{D}_a\\0\leq x<\gamma}}
\Psi_a(x,y)
\leq
g_{\mathrm{\widetilde{\cC}}}(a)+\varepsilon_2.
\end{equation}
Hence, by choosing \(\gamma\) sufficiently small, the contribution of
this boundary region can be made arbitrarily close to
\(g_{\widetilde{\cC}}(a)\).

Next, We consider the third term in \eqref{eq:four_boundary_regions_sum_ldgmpc_ldpc}, i.e., when $0<y<\gamma$.
In this region we have
\begin{equation}
y\in(0,\gamma),
\qquad
x\in(a-\gamma,a+\gamma).
\end{equation}
In the region of interest, we have
\begin{equation}
\begin{aligned}
\sup_{\substack{(x,y)\in\mathcal{D}_a\\0\leq y<\gamma}}
\Psi_a(x,y)
\leq
\sup_{\substack{0\leq x\leq1\\|x-a|<\gamma}}g_{\cC_1^\perp}(x)
+
\sup_{\substack{0\leq y<\gamma}} g_{\widetilde{\cC}}(y).
\end{aligned}
\end{equation}
The next step is to control the two terms on the right-hand side separately. We now use the known small-weight behavior of regular LDPC ensembles given in~\cite[Appendix A]{gallager1963ldpc}. 
Specifically, there exists a $\gamma_2>0$ such that for all sufficiently large check degree $d_c^{\tilde H}$, $g_{\widetilde{\cC}}(w)<0$ is satisfied for all $0<w<\gamma_2$.
Therefore, by choosing $\gamma>0$ small enough so that $\gamma<\gamma_2$,
we obtain
\begin{equation}
\sup_{0\leq y<\gamma}g_{\widetilde{\cC}}(y)=0.
\end{equation}
Since $g_{\cC_1^\perp}$ is continuous at $a$, for every $\epsilon_1>0$ there exists a $\gamma_{\epsilon_1} \ge \gamma$ such that $|g_{\cC_1^\perp}(x)-g_{\cC_1^\perp}(a)|<\epsilon_1$ whenever $|x-a|<\gamma_{\epsilon_1}$. Therefore, by choosing $\gamma>0$ small enough so that $\gamma<\gamma_{\epsilon_1}$ we obtain
\begin{equation}
 \sup_{a-\gamma<x<a+\gamma} g_{\cC_1^\perp}(x)
\leq g_{\cC_1^\perp}(a)+\epsilon_1 .   
\end{equation}
Combining this with the previous bound gives
\begin{equation}
\sup_{\substack{(x,y)\in\mathcal{D}_a\\0\leq y<\gamma}}
\Psi_a(x,y)
\leq g_{\cC_1^\perp}(a)+\varepsilon_1.
\end{equation}
Then, by choosing $\gamma$ sufficiently small, the upper bound can be made arbitrarily close to $g_{\cC_1^\perp}(a)$.

For even check degrees ,$d^{G_1}$, $d_v^{H_1}$ and $d_c^{\tilde H}$, the all-one vector belongs to each component code, so
its weight distribution is symmetric. Then the corresponding strips near \(x=1\) and \(y=1\) satisfy the same
bounds. Hence,
\begin{equation}
\sup_{(x,y)\in\mathcal{D}_a\cap\mathcal{B}^{c}(\gamma)}
\Psi_a(x,y)
\leq
\max\left\{
g_{ \cC_1^\perp}(a)+\varepsilon_1,\,
g_{\widetilde{\cC}}(a)+\varepsilon_2
\right\}.
\end{equation}

\paragraph{Combining the two cases.}
Combining the interior estimate from Case~1 with the boundary-region
estimate from Case~2 gives, for \(a\in(0,1)\),
\begin{equation}
g_{\cC_2^\perp}(a)
\leq
\max\left\{
h_2(a)-(1-R_{\cC_2^\perp}),\,
g_{\cC_1^\perp}(a)+\varepsilon_1,\,
g_{\widetilde{\cC}}(a)+\varepsilon_2
\right\}.
\label{eq:ldgmpc_ldpc_combined_bound}
\end{equation}

Figure~\ref{fig:combining_sumGIRA_LDPC} illustrates the three terms appearing in
\eqref{eq:ldgmpc_ldpc_combined_bound}. In particular, the both LDPC and GIRA component average weight distribution growth rates are negative before their first positive zero crossings, while the
random-coding term changes sign at the Gilbert--Varshamov point
$h_2^{-1}(1-R_{\cC_2^\perp})$. By choosing the check degrees $d^{G_1}$, $d_v^{H_1}$ and $d_c^{\tilde H}$ sufficiently  large, these zero crossings can be shifted to the right, allowing us to control the boundary terms in the upper bound on $g_{\cC_2^\perp}(a)$ for $a\in (0,1)$.
This is precisely the behavior used below.

We are now ready to prove \eqref{eq:sum_ldgmpc__lpdc_claim}.
First, at the end points, we argue directly from the weight distribution rather than
from the maximization formula. Since the sum ensemble is linear, the zero
codeword is always present and $\bar N_C(0)=1,$
and therefore
\begin{equation}
 g_{\cC_2^\perp}(0)
=
\lim_{n\to\infty}\frac{1}{n}\log_2 \bar N_{\cC_2^\perp}(0)
=
0.   
\end{equation}
Similarly, there can be at most one codeword of weight $n$, namely the all-one
vector. Therefore,
$ \bar N_{\cC_2^\perp}(n)\leq 1, $  
which implies
\begin{equation}
 g_{\cC_2^\perp}(1)
=
\lim_{n\to\infty}\frac{1}{n}\log_2 \bar N_{\cC_2^\perp}(n)
\leq 0.   
\end{equation}
This proves the first line of the theorem.

It remains to consider \(a\in(0,1)\). Define
\begin{equation}
\delta_{\mathrm{GV}}
=
h_2^{-1}(1-R_{\cC_2^\perp}).
\end{equation}
Since
\begin{equation}
R_{\cC_1^\perp}<R_{\cC_2^\perp},
\qquad
R_{\widetilde{\cC}}<R_{\cC_2^\perp},
\end{equation}
we have
\begin{equation}
h_2^{-1}\!\left(1-R_{\cC_1^\perp}\right)
>
\delta_{\mathrm{GV}},
\qquad
h_2^{-1}\!\left(1-R_{\widetilde{\cC}}\right)
>
\delta_{\mathrm{GV}}.
\end{equation}
By the result established in the previous section, the parameters of the GIRA component can be chosen so that
\(g_{\cC_1^\perp}(a)<0\) throughout an interval extending beyond
\(\delta_{GV}\). Similarly, by choosing the LDPC check degree sufficiently large, the first positive zero of
\(g_{\widetilde{\cC}}(a)\) can be placed strictly to the right of
\(\delta_{GV}\).
Choose any
$0<\delta<\delta_{GV}$
smaller than the corresponding positive zero crossings of both
component average weight distribution growth rates. Then, for every \(a\in(0,\delta]\),
\begin{equation}
h_2(a)-(1-R_{\cC_2^\perp})<0,
\qquad
g_{\cC_1^\perp}(a)<0,
\qquad
g_{\widetilde{\cC}}(a)<0.
\end{equation}
Letting \(\varepsilon_1,\varepsilon_2\downarrow0\) in
\eqref{eq:ldgmpc_ldpc_combined_bound} therefore gives
\begin{equation}
g_{\cC_2^\perp}(a)<0,
\qquad
a\in(0,\delta].
\end{equation}
By symmetry of the component average weight distribution growth rates and
\(h_2(a)=h_2(1-a)\), we similarly obtain
\begin{equation}
g_{\cC_2^\perp}(a)<0,
\qquad
a\in[1-\delta,1).
\end{equation}
This proves the second line of the theorem.

Finally, consider \(a\in(\delta,1-\delta)\). By the GIRA result
proved in the previous section and the standard regular-LDPC
average weight distribution growth rate result, the component degrees can be chosen sufficiently
large so that
\begin{equation}
g_{\cC_1^\perp}(a)
\leq
h_2(a)-(1-R_{\cC_2^\perp})
\end{equation}
and
\begin{equation}
g_{\widetilde{\cC}}(a)
\leq
h_2(a)-(1-R_{\cC_2^\perp})
\end{equation}
throughout this interval. Consequently, by choosing
\(\varepsilon_1\) and \(\varepsilon_2\) sufficiently small, the component terms can be kept below the random-coding term in this region. Then, 
\eqref{eq:ldgmpc_ldpc_combined_bound} yields
\begin{equation}
g_{\cC_2^\perp}(a)
\leq
h_2(a)-(1-R_{\cC_2^\perp}),
\qquad
a\in(\delta,1-\delta).
\end{equation}
This proves the third line of the first statement in the theorem. It remains to prove Statement~2.

\begin{figure}[H]
    \centering
    \includegraphics[width=0.8\textwidth]{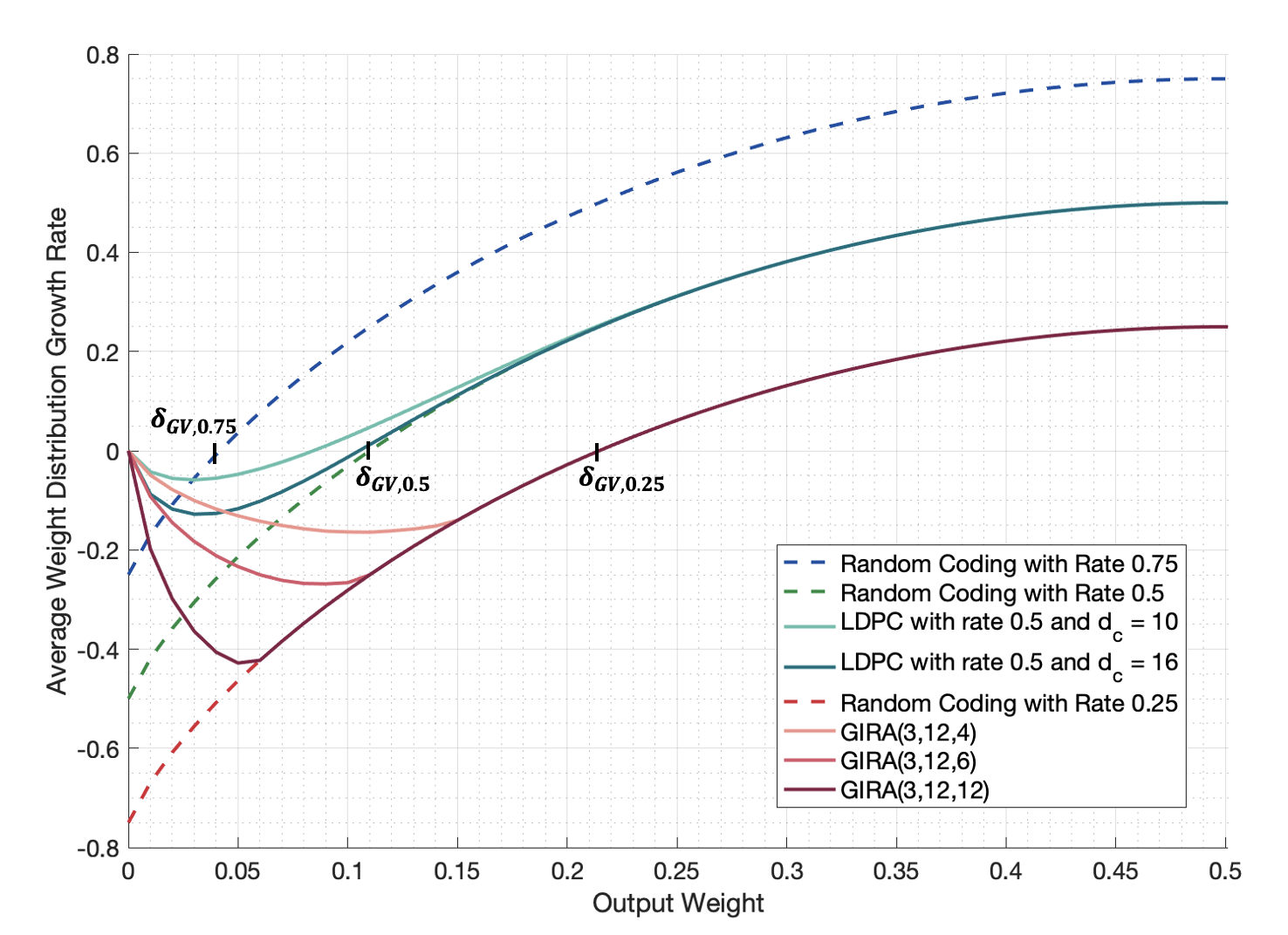}
    \caption{ Asymptotic Weight Distributions for the Component Ensembles.}
    \label{fig:combining_sumGIRA_LDPC}
\end{figure}

\end{proof}
\subsection{Proof of Statement~2}
\begin{proof}
Let \(\cC_1^\perp\) denote the GIRA component and \(\widetilde{\cC}\) the
LDPC component, and let
\(\overline N_{\cC_1^\perp}(w_1)\) and \(\overline N_{\widetilde{\cC}}(w_2)\)
denote their respective average weight enumerators. Using \eqref{eq:sum_codes_finite} 
\begin{equation}
 \overline N_{\cC_2^\perp}(w)
\leq
\sum_{w_1=0}^n\sum_{w_2=0}^n
\overline N_{\cC_1^\perp}(w_1)\overline N_{\widetilde{\cC}}(w_2)
\frac{
\binom{w_1}{t}
\binom{n-w_1}{w_2-t}
}{
\binom{n}{w_2}
},   
\end{equation}
where 
$t=\frac{w_1+w_2-w}{2},$
whenever \(t\) is admissible, and is zero otherwise.

We first separate the terms for which one of the two component
codewords is the all-zero codeword. Since
\(\overline N_{\cC_1^\perp}(0)=\overline N_{\widetilde{\cC}}(0)=1\), these terms are nonzero only
when the other component has weight \(w\). Hence,
\begin{equation}
\begin{aligned}
 \overline N_{\cC_2^\perp}(w)
&\leq
\overline N_{\cC_1^\perp}(w)+\overline N_{\widetilde{\cC}}(w)
+
\sum_{w_1=1}^{n}\sum_{w_2=1}^{n}
\overline N_{\cC_1^\perp}(w_1)\overline N_{\widetilde{\cC}}(w_2)
\frac{
\binom{w_1}{t}\binom{n-w_1}{w_2-t}
}{
\binom{n}{w_2}
}.   
\end{aligned}
\label{eq:sum_gira_ldpc}   
\end{equation}

Imposing the restriction of feasible region in \eqref{eq:sum_codes_feasible_region}, the sum in right hand side becomes
\begin{equation}
\begin{aligned}
\sum_{w_1=1}^{n}\sum_{w_2=1}^{n}
\overline N_{\cC_1^\perp}(w_1)\overline N_{\widetilde{\cC}}(w_2)
\frac{
\binom{w_1}{t}\binom{n-w_1}{w_2-t}
}{
\binom{n}{w_2}
} &= \sum_{w_1=1}^{\delta_0n}\sum_{w_2=1}^{(\delta_0+\delta)n}
\overline N_{\cC_1^\perp}(w_1)\overline N_{\widetilde{\cC}}(w_2)
\frac{
\binom{w_1}{t}\binom{n-w_1}{w_2-t}
}{
\binom{n}{w_2}
} \\ &+ \sum_{w_1=\delta_0n}^{(1-\delta_0)n}\sum_{w_2=(\delta_0-\delta)n}^{(1-\delta_0+\delta)n}
\overline N_{\cC_1^\perp}(w_1)\overline N_{\widetilde{\cC}}(w_2)
\frac{
\binom{w_1}{t}\binom{n-w_1}{w_2-t}
}{
\binom{n}{w_2}
}\\ &
+ \sum_{w_1=(1-\delta_0)}^{n}\sum_{w_2=(1-\delta_0-\delta)n}^{n}
\overline N_{\cC_1^\perp}(w_1)\overline N_{\widetilde{\cC}}(w_2)
\frac{
\binom{w_1}{t}\binom{n-w_1}{w_2-t}
}{
\binom{n}{w_2}
} .
\label{eq:sum_thm2_statement2}
\end{aligned}
\end{equation}
where $\delta_0$ is chosen such that 
\begin{equation}
    \sum_{u=1}^{\lfloor \delta_0n\rfloor}
\overline N_{{\widetilde{\cC}}}(u)
=
O(n^{-(d_v^{\tilde H}-2)})
\label{eq:ldpc_bound}
\end{equation}
which is proved in \cite[Appendix~A]{gallager1963ldpc}.

For the first term in \eqref{eq:sum_thm2_statement2}
$1\leq w_1<\delta_0 n$
and
$w_2\leq w_1+\delta n
<
(\delta_0+\delta)n
$, both \(w_1\) and \(w_2\) are small. Since $0\leq
\frac{
\binom{w_1}{t}\binom{n-w_1}{w_2-t}
}{
\binom{n}{w_2}
}
\leq1,$
\begin{equation}
    \sum_{w_1=1}^{\delta_0n}\sum_{w_2=1}^{(\delta_0+\delta)n}
\overline N_{\cC_1^\perp}(w_1)\overline N_{\widetilde{\cC}}(w_2)
\frac{
\binom{w_1}{t}\binom{n-w_1}{w_2-t}
}{
\binom{n}{w_2}
} \leq \left(
\sum_{w_1=1}^{\lfloor\delta_0n\rfloor}
\overline N_{\cC_1^\perp}(w_1)
\right)
\left(
\sum_{w_2=1}^{\lfloor(\delta_0+\delta)n\rfloor}
\overline N_{\widetilde{\cC}}(w_2)
\right)
\label{eq:theorem2_sum_bound}
\end{equation}
To bound the first factor in \eqref{eq:theorem2_sum_bound}, we further divide the range of
\(w_1\) into a small-weight region and a linear-weight region. Let
\(\delta'>0\) be a fixed constant, independent of \(n\), chosen
sufficiently small so that Statement~2 of Theorem~1 applies on
\(1\leq w_1\leq\delta'n\). We also choose \(\delta'\) such that
$0<\delta'<\delta_0+\delta.$
Then
\begin{equation}
\sum_{w_1=1}^{\lfloor(\delta_0+\delta)n\rfloor}
\overline N_{\cC_1^\perp}(w_1) \leq 
\sum_{w_1=1}^{\lfloor\delta'n\rfloor}
\overline N_{\cC_1^\perp}(w_1)
+
\sum_{w_1=\lceil\delta'n\rceil}^{\lfloor(\delta_0+\delta)n\rfloor}
\overline N_{\cC_1^\perp}(w_1)    
\end{equation}

By Statement~2 of Lemma~\ref{lem:lemma3},
\begin{equation}
    \sum_{w_1=1}^{\lfloor\delta'n\rfloor}
\overline N_{\cC_1^\perp}(w_1)
=
O\!\left(n^{-(\frac{d_v^{G_1}}{2}-1)}\right).
\label{eq:n1_small}
\end{equation}

For the second term, \(w_2\) is linear in \(n\), since
$\frac{w_1}{n}\in[\delta',\delta_0+\delta].$
By the preceding choice of \(\delta_0\) and \(\delta\), this interval
lies in the region where Statement~1 of Lemma~\ref{lem:lemma3} gives
\(g_{\cC_1^\perp}(\hat w_1)<0\). Since
\([\delta',\delta_0+\delta]\) is compact, there exists
\(\kappa>0\) such that
\begin{equation}
g_{\cC_1^\perp}(\hat w_1)\leq-\kappa,
\qquad
\delta'\leq\hat w_1\leq\delta_0+\delta.    
\end{equation}
Hence,
\begin{equation}
 \overline N_{\widetilde{\cC}}(w_1)
\leq
\operatorname{poly}(n)2^{-\kappa n}   
\end{equation}
uniformly over this range, and therefore
\begin{equation}
 \sum_{w_1=\lceil\delta'n\rceil}^{\lfloor(\delta_0+\delta)n\rfloor}
\overline N_{\cC_1^\perp}(w_1)
=
2^{-\Omega(n)}.   
\label{eq:n1_linear}
\end{equation}

Then combining \eqref{eq:ldpc_bound} and \eqref{eq:n1_linear}, the first term in \eqref{eq:sum_thm2_statement2} becomes
\begin{equation}
       \sum_{w_1=1}^{\delta_0n}\sum_{w_2=1}^{(\delta_0+\delta)n}
\overline N_{\cC_1^\perp}(w_1)\overline N_{\widetilde{\cC}}(w_2)
\frac{
\binom{w_1}{t}\binom{n-w_1}{w_2-t}
}{
\binom{n}{w_2}
} = O\!\left(n^{-(\frac{d_v^{G_1}}{2}-d_v^{\tilde H}-3)}\right).
\label{eq:thm2_2_1}
\end{equation}

We next consider the second term in \eqref{eq:sum_thm2_statement2}. In this region,
$\delta_0 n\leq w_1\leq(1-\delta_0)n,$
and
$(\delta_0-\delta)n\leq w_2\leq(1-\delta_0+\delta)n.$
Since \(\delta<\delta_0\), both component weights are linear in \(n\)
and bounded away from \(0\) and \(n\). 
Hence, this term lies entirely in the linear-weight region considered
in the proof of Statement~1.
 For \(1\leq w\leq\delta n\), we have
$0<\frac{w}{n}\leq\delta,$
and Statement~1 gives
$g_{\cC_2^\perp}\!\left(\frac{w}{n}\right)<0.$
Moreover, the linear-weight estimate used in its proof is uniform over
the above compact ranges of \(w_1/n\) and \(w_2/n\). Hence, there
exists a constant \(\kappa>0\) such that the contribution of this
region satisfies
\begin{equation}
\sum_{w_1=\lceil\delta_0n\rceil}^{\lfloor(1-\delta_0)n\rfloor}
\sum_{w_2=\lceil(\delta_0-\delta)n\rceil}^{\lfloor(1-\delta_0+\delta)n\rfloor}
\overline N_{\cC_1^\perp}(w_1)\overline N_{\widetilde{\cC}}(w_2)
\frac{
\binom{w_1}{t}\binom{n-w_1}{w_2-t}
}{
\binom{n}{w_2}
}
\leq
\operatorname{poly}(n)\,2^{-\kappa n}.    
\end{equation}
Therefore,
\begin{equation}
\sum_{w_1=\lceil\delta_0n\rceil}^{\lfloor(1-\delta_0)n\rfloor}
\sum_{w_2=\lceil(\delta_0-\delta)n\rceil}^{\lfloor(1-\delta_0+\delta)n\rfloor}
\overline N_{\cC_1^\perp}(w_1)\overline N_{\widetilde{\cC}}(w_2)
\frac{
\binom{w_1}{t}\binom{n-w_1}{w_2-t}
}{
\binom{n}{w_2}
}=2^{-\Omega(n)}.   
\label{eq:thm2_2_2}
\end{equation}

 We finally consider the third term in \eqref{eq:sum_thm2_statement2}. In this region,
$(1-\delta_0)n\leq w_1\leq n,$
and
$(1-\delta_0-\delta)n\leq w_2\leq n.$
Without loss of generality, we restrict attention to the case in which
\(d^{G_1}\), \(d_v^{H_1}\), and \(d_c^{\tilde H}\) are even. Under this
assumption, the all-one vector belongs to both component codes, and hence
their weight enumerators are symmetric:
$\overline N_{\cC_1^\perp}(w)=\overline N_{\cC_1^\perp}(n-w)$ and
$\overline N_{\widetilde{\cC}}(w)=\overline N_{\widetilde{\cC}}(n-w).$
Therefore, the high-weight region can be treated by the same argument as
the corresponding low-weight region after replacing
\(w_i\) by \(n-w_i\).

Consequently,
\begin{equation}
 \sum_{w_1=(1-\delta_0)n}^{n}
\sum_{w_2=(1-\delta_0-\delta)n}^{n}
\overline N_{\cC_1^\perp}(w_1)\overline N_{\widetilde{\cC}}(w_2)
\frac{
\binom{w_1}{t}\binom{n-w_1}{w_2-t}
}{
\binom{n}{w_2}
}
= O\!\left(n^{-(\frac{d_v^{G_1}}{2}-d_v^{\tilde H}-3)}\right).   
\label{eq:thm2_2_3}
\end{equation}

Combining \eqref{eq:thm2_2_1}, \eqref{eq:thm2_2_2}, and \eqref{eq:thm2_2_3}, \eqref{eq:sum_thm2_statement2} becomes
\begin{equation}
   \sum_{w_1=1}^{n}\sum_{w_2=1}^{n}
\overline N_{\cC_1^\perp}(w_1)\overline N_{\widetilde{\cC}}(w_2)
\frac{
\binom{w_1}{t}\binom{n-w_1}{w_2-t}
}{
\binom{n}{w_2}
} = O\!\left(n^{-(\frac{d_v^{G_1}}{2}-d_v^{\tilde H}-3)}\right).
\label{eq:double_sum}
\end{equation}

We lastly consider the first two terms in \eqref{eq:sum_gira_ldpc}
appearing outside the double sum. 
Recall that the component rates satisfy
$R_{\cC_1^\perp}<R_{\cC_2^\perp},$
and
$R_{\widetilde{\cC}}<R_{\cC_2^\perp},$
where \(R_{\cC_2^\perp}\) is the design rate of the sum ensemble.

To see the significance of this rate ordering, consider the corresponding
random-coding curves
$h_2(a)-(1-R_{\cC_1^\perp}),$
$h_2(a)-(1-R_{\widetilde{\cC}}),$ and
$h_2(a)-(1-R_{\cC_2^\perp}).$
Since \(R_{\cC_1^\perp}<R_{\cC_2^\perp}\) and \(R_{\widetilde{\cC}}<R_{\cC_2^\perp}\), we have
$1-R_{\cC_1^\perp}>1-R_{\cC_2^\perp},$ $1-R_{\widetilde{\cC}}>1-R_{\cC_2^\perp},$
and therefore
$h_2(a)-(1-R_{\cC_1^\perp})<h_2(a)-(1-R_{\cC_2^\perp})$, $h_2(a)-(1-R_{\widetilde{\cC}})<h_2(a)-(1-R_{\cC_2^\perp}).$
Thus, the random-coding curves associated with both component ensembles
lie strictly below that of the sum ensemble. Consequently, their first
positive zero crossings occur strictly to the right of the first positive
zero crossing associated with the sum rate \(R\).

Let \(\delta\) be the constant used for the sum ensemble. Since
\(\delta\) lies strictly to the left of the first positive zero crossing
of \(h_2(a)-(1-R_{\cC_2^\perp})\), it also lies strictly to the left of the corresponding
zero crossings of both component ensembles.

For the first component, \(\cC_1^\perp\), which is the GIRA ensemble, the above
observation allows us to apply Lemma~\ref{lem:lemma3} on an interval containing
\((0,\delta]\). More precisely, since the zero crossing corresponding
to \(R_1\) lies strictly to the right of \(\delta\), the degrees in the
GIRA ensemble can be chosen sufficiently large so that the small-weight
region in Lemma~\ref{lem:lemma3} contains the interval \((0,\delta]\). Statement~2
of Lemma~\ref{lem:lemma3} then yields
$\sum_{u=1}^{\lfloor\delta n\rfloor}
\overline N_{\cC_1^\perp}(u)
=
O\!\left(
n^{-\left(\frac{d^{G_1}}{2}-1\right)}
\right).$
Since every term in this sum is nonnegative, for each
\(1\leq w\leq\delta n\),
\begin{equation}
 \overline N_{\cC_1^\perp}(w)
\leq
\sum_{u=1}^{\lfloor\delta n\rfloor}
\overline N_{\cC_1^\perp}(u)
=
O\!\left(
n^{-\left(\frac{d^{G1}}{2}-1\right)}
\right).   
\end{equation}

We next consider the LDPC component \(\widetilde{\cC}\). Let \(\delta_{0,2}\)
denote the first positive zero crossing of the asymptotic growth rate of
the regular LDPC ensemble. For a fixed LDPC rate \(R_{\widetilde{\cC}}\), the Gallager
ensemble satisfies
$\sum_{u=1}^{\lfloor\delta_{0,2}n\rfloor}
\overline N_{\widetilde{\cC}}(u)
=
O\!\left(n^{-(d_v^{\tilde H}-2)}\right).$

It remains to verify that the interval \(0<w\leq\delta n\) is contained
in this small-weight region. Since \(R_{\widetilde{\cC}}<R_{\cC_2^\perp}\), the random-coding curve
$h_2(a)-(1-R_{\widetilde{\cC}})$
lies below
$h_2(a)-(1-R_{\cC_2^\perp}),$
and hence its first positive zero crossing lies strictly to the right of
that associated with the sum rate. In particular, because \(\delta\)
was chosen to lie to the left of the latter crossing, it also lies to
the left of the random-coding zero corresponding to \(R_{\widetilde{\cC}}\).

By the LDPC weight-distribution result of Gallager, the zero crossing
\(\delta_{0,2}\) of the LDPC growth rate can be pushed arbitrarily close
to the corresponding random-coding zero by taking the check degree
\(d_c^{\tilde H}\) sufficiently large. We may therefore choose
\(d_c^{\tilde H}\) large enough so that
$\delta_{0,2}>\delta.$
Consequently,
\begin{equation}
\sum_{u=1}^{\lfloor\delta n\rfloor}\overline N_{\widetilde{\cC}}(u)
\leq
\sum_{u=1}^{\lfloor\delta_{0,2}n\rfloor}\overline N_{\widetilde{\cC}}(u)
=
O\!\left(n^{-(d_v^{\tilde H}-2)}\right),    
\end{equation}
and hence, uniformly for \(1\leq w\leq\delta n\),
\begin{equation}
\overline N_{\widetilde{\cC}}(w)
=
O\!\left(n^{-(d_v^{\tilde H}-2)}\right).    
\end{equation}

Therefore,
\begin{equation}
 \overline N_{\cC_1^\perp}(w)+\overline N_{\widetilde{\cC}}(w)
=
O\!\left(
n^{-\min\left\{
\frac{d^{G_1}}{2}-1,\,
d_v^{\tilde H}-2
\right\}.}
\right),
\qquad
1\leq w\leq\delta n.
\label{eq:single_n1n2}
\end{equation}

Consequently by combining \eqref{eq:single_n1n2} and \eqref{eq:double_sum}, we obtain
\begin{equation}
    \overline N_{\cC_2^\perp}(w)
=  O\!\left(n^{-\min\left\{
\frac{d^{G_1}}{2}-1,\,
d_v^{\tilde H}-2,\,\frac{d_v^{G_1}}{2}-d_v^{\tilde H}-3
\right\}}\right)=  O\!\left(n^{-\min\left\{
\frac{d^{G_1}}{2}-1,\,
d_v^{\tilde H}-2
\right\}}\right) \qquad
1\leq w\leq\lfloor\delta n\rfloor.
\end{equation}

Since there are at most $n$ terms, the sum over terms up to $\lfloor\delta n\rfloor$ becomes
\begin{equation}
  \sum_{w=1}^{\lfloor\delta n\rfloor}
\overline N_{\cC_2^\perp}(w)
=  O\!\left(n^{-\min\left\{
\frac{d^{G_1}}{2}-2,\,
d_v^{\tilde H}-3
\right\}}\right).
\end{equation}
This completes the proof.

\end{proof}

\section{ Proof of Theorem~\ref{thm:thm_sum_gira_ldpc_ach} }
\label{app:thm_sum_gira_ldpc_ach}

\begin{proof}

Fix finite degree parameters
$d_v^{H_1}>4$, $d_c^{H_1}$,
$d_v^{\widetilde H}>3$,
$d_c^{\widetilde H}>3$, and even
$d^{G_1}>4$, satisfying the achievability conditions of Theorem~\ref{thm:thm3} and Fact~\ref{fact:ldpcgm}.
The corresponding component design rates are
$R_1
=1-\frac{d_v^{H_1}}{d_c^{H_1}},$ and
$R_2^\perp
=1-\frac{d_v^{\widetilde H}}{d_c^{\widetilde H}}
+\frac{d_v^{H_1}}{d_c^{H_1}}.$

Consider any memoryless Pauli channel with error probabilities $p_X$, $p_Y$, and $p_Z$ satisfying the conditions of Theorem~\ref{thm:thm_sum_gira_ldpc_ach}.
Let
$q_X=p_X+p_Y,$ and
$q_Z=p_Z+p_Y,$
and denote the corresponding BSC capacities by
$C_X=1-h_2(q_X),$ and
$C_Z=1-h_2(q_Z).$
By the assumptions of the theorem,
$R_1<C_X,$ and 
$R_2^\perp<C_Z.$
Thus, both component design rates lie strictly below their respective BSC capacities.
By construction,
$\cC_2^\perp
=\cC_1^\perp+
\widetilde{\cC},$
so that
$\cC_1^\perp\subseteq
\cC_2^\perp$ and, equivalently,
$\cC_2\subseteq\cC_1$.
Hence, the CSS nesting condition is satisfied.

We now establish that the two 
components are simultaneously reliable under
ML decoding.
For $i=1,2$, let $\mu_{i,n}$ denote the ensemble-averaged ML block error probability of the corresponding component.


\paragraph{ML achievability of $\cC_2^\perp$.}
We first consider $\cC_2^\perp$ over $\mathrm{BSC}(q_Z)$.

We apply the Miller--Burshtein bound on the ensemble-averaged ML block
error probability. Let
$U
=
\left\{
w\in\{1,\ldots,n\}:
\frac{w}{n}\in(0,\delta]\cup[1-\delta,1]
\right\},$
and let \( U^c\) denote its complement. Then
\begin{equation}
\mu_{2,n}
=
\mathbb E\left[
P_B^{\mathrm{ML}}(\cC_2^\perp;q_Z)
\right]
\leq
\sum_{w\in U}
N_{\cC_2^\perp}(w)D_Z^w
+
2^{-nE_{r,Z}\left(
R_2^\perp+\frac{\log_2\alpha_n}{n}
\right)},   
\label{eq:pb_bound}
\end{equation}
where
\begin{equation}
\alpha_n
=
\max_{w\in U^c}
\frac{N_{\cC_2^\perp}(w)}{2^{nR_2^\perp}-1}
\frac{2^n}{\binom{n}{w}},    
\end{equation}
$E_{r,Z}(\cdot)$ denotes the random-coding exponent of $\mathrm{BSC}(q_Z)$, and
\( D_Z=2\sqrt{q_Z(1-q_Z)}<1\) is its Bhattacharyya parameter.

We analyze the two terms in \eqref{eq:pb_bound} separately.

\textit{First term: contribution from $U$.}
Since \(0\leq D_Z<1\)
\begin{subequations}
\begin{align}
\sum_{w\in U}N_{\cC_2^\perp}(w)D_Z^w
&\leq
\sum_{w=1}^{\lfloor\delta n\rfloor}N_{\cC_2^\perp}(w)
+
\sum_{w=\lceil(1-\delta)n\rceil}^{n-1}N_{\cC_2^\perp}(w)
+ N_{\cC_2^\perp}(n)D_Z^n\\
&\leq
O\!\left(n^{-\min\left\{
\frac{d^{G_1}}{2}-2,\,
d_v^{\tilde H}-3 \right\}}\right)
+D_Z^n
\label{eq:exponent_thm2}
\\
&=o(1).
\label{eq:firstterm}
\end{align}    
\end{subequations}
The second inequality follows from Statement 2
of Theorem~\ref{thm:thm3} and the fact that
$N_{\cC_2^\perp}(n)\leq1$.
Indeed, since the sum code is a binary linear
code of length $n$, there is at most one distinct
codeword of weight $n$, namely the all-one vector. 
Consequently,
$N_{\cC_2^\perp}(n)
=
\Pr(\mathbf{1}_n\in\cC)
\leq1.$
Since $d^{G_1}>4$ and $d_v^{\widetilde H}>3$, the exponent in \eqref{eq:exponent_thm2} is strictly positive.
Furthermore, since $D_Z<1$, this contribution vanishes
exponentially with $n$.

\textit{Second term: contribution from $U^c$.}

We next consider the random-coding term in~\eqref{eq:pb_bound}. For
\(w=\hat w n\) with
\(\hat w\in(\delta,1-\delta)\), the Stirling approximation
$\frac{1}{n}\log_2\binom{n}{\hat w n}
=
h_2(\hat w)+o(1)$
together with the Statement~1 of Theorem~\ref{thm:thm3} gives
\begin{subequations}
\begin{align}
\limsup_{n\to\infty}
\frac{1}{n}\log_2\alpha_n
&\leq
\sup_{\hat w\in(\delta,1-\delta)}
\left\{
g_{\cC_2^\perp}(\hat w)-h_2(\hat w)+(1-R_2^\perp)
\right\}\\
&\leq0.
\end{align}  
\label{eq:css_alpha_bound}
\end{subequations}

Since the channel satisfies $R_2^\perp<C_Z$, we may choose a constant
$\kappa_Z>0$ such that
$R_2^\perp+\kappa_Z<C_Z.$
By \eqref{eq:css_alpha_bound},
for all sufficiently large $n$,
\begin{equation}
 R_2^\perp+
\frac{\log_2\alpha_n}{n}
\leq R_2^\perp+\kappa_Z<C_Z.   
\end{equation}
The random-coding exponent is strictly
positive for rates below channel capacity
and is nonincreasing in rate. Therefore,
\begin{equation}
   2^{-nE_{r,Z}\left(
R_2^\perp+\frac{\log_2\alpha_n}{n}
\right)} \le 2^{-nE_{r,Z}(R_2^\perp+\kappa_Z)}
=o(1).
\label{eq:2ndterm}
\end{equation}

Combining \eqref{eq:firstterm} and \eqref{eq:2ndterm} in \eqref{eq:pb_bound} yields
\begin{equation}
\mu_{2,n}
=
\mathbb E\left[
P_B^{\mathrm{ML}}(\cC_2^\perp;q_Z)
\right]
\leq
o(1)
+o(1)
\longrightarrow0.   
\end{equation}

Thus, for the fixed component design rate $R_2^\perp$ and any channel satisfying $C_Z>R_2^\perp$, the ensemble-averaged ML block error probability of $\cC_2^\perp$ vanishes.

\paragraph{ML achievability of $\cC_1$.}
We next consider $\cC_1$, which belongs to the LDPC/GM ensemble.
Its design rate is fixed by the degree parameters as
$R_1
=
1-\frac{d_v^{H_1}}{d_c^{H_1}}.$
By the channel condition
$R_1<C_X.$
By Fact~\ref{fact:ldpcgm}, the LDPC/GM ensemble admits ML reliability over MBIOS channels when its design rate lies strictly below the channel capacity and the degree parameters satisfy the corresponding achievability conditions.

Consequently, for the joint ensemble of Construction B, we have
$\mu_{1,n}\to 0$ and $\mu_{2,n}\to 0,$ as $n\to\infty$.

\paragraph{Simultaneous reliability.} We now show that both components are simultaneously reliable with probability approaching one over the joint ensemble of Construction B.

Fix any $\epsilon>0$.
By Markov's inequality,
\begin{equation}
\Pr\left(
P_B^{\mathrm{ML}}(\cC_1;q_X)
>\epsilon
\right)
\leq
\frac{\mu_{1,n}}{\epsilon},
\label{eq:css_markov1}
\end{equation}
and similarly,
\begin{equation}
\Pr\left(
P_B^{\mathrm{ML}}(\cC_2^\perp;q_Z)
>\epsilon
\right)
\leq
\frac{\mu_{2,n}}{\epsilon}.
\label{eq:css_markov2}
\end{equation}

Applying the union bound, we obtain
\begin{equation}
\begin{aligned}
\Pr\Big(
P_B^{\mathrm{ML}}(\cC_1;q_X)
\leq\epsilon,
\,
P_B^{\mathrm{ML}}(\cC_2^\perp;q_Z)
\leq\epsilon
\Big)
&\geq
1-
\Pr\left(
P_B^{\mathrm{ML}}(\cC_1;q_X)
>\epsilon
\right)
-
\Pr\left(
P_B^{\mathrm{ML}}(\cC_2^\perp;q_Z)
>\epsilon
\right)\\
&\quad\geq
1-\frac{\mu_{1,n}+\mu_{2,n}}
{\epsilon}.
\end{aligned}
\label{eq:css_joint}
\end{equation}

Since
$\mu_{1,n}\to0$ and
$\mu_{2,n}\to0$,
for every $\epsilon>0$ there exists $n_\epsilon$ such that, for all
$n>n_\epsilon$,
$\mu_{1,n}+\mu_{2,n}<\epsilon^2.$
Substituting this into \eqref{eq:css_joint} gives 

\begin{equation}
\Pr\left(
P_B^{\mathrm{ML}}(\cC_1)
\leq\epsilon,\,
P_B^{\mathrm{ML}}(\cC_2^\perp)
\leq\epsilon
\right)>1-\epsilon.
\label{eq:css_joint_final}
\end{equation}

This establishes the simultaneous
ML reliability claimed in the theorem.
Notice that this argument does not
require independence of the two
component error probabilities.

\end{proof}

\bibliographystyle{IEEEtran}
\bibliography{achilleas18abrv,references}

\end{document}